\documentclass{aa}
\usepackage{natbib,twoopt}
\usepackage[dvipsnames,svgnames,x11names]{xcolor}
\usepackage{subcaption}
\usepackage[breaklinks=true, colorlinks=true, linkcolor=blue, citecolor=blue]{hyperref} %% to avoid \citeads line fills
\bibpunct{(}{)}{;}{a}{}{,}             %% natbib format for A&A and ApJ
\makeatletter
\newcommandtwoopt{\citeads}[3][][]{\href{http://adsabs.harvard.edu/abs/#3}%
    {\def\hyper@linkstart##1##2{}%
     \let\hyper@linkend\@empty\citealp[#1][#2]{#3}}}
  \newcommandtwoopt{\citepads}[3][][]{\href{http://adsabs.harvard.edu/abs/#3}%
    {\def\hyper@linkstart##1##2{}%
     \let\hyper@linkend\@empty\citep[#1][#2]{#3}}}
  \newcommandtwoopt{\citetads}[3][][]{\href{http://adsabs.harvard.edu/abs/#3}%
    {\def\hyper@linkstart##1##2{}%
     \let\hyper@linkend\@empty\citet[#1][#2]{#3}}}
  \newcommandtwoopt{\citeyearads}[3][][]%
    {\href{http://adsabs.harvard.edu/abs/#3}
    {\def\hyper@linkstart##1##2{}%
     \let\hyper@linkend\@empty\citeyear[#1][#2]{#3}}}
\makeatother
\usepackage{graphicx}
\usepackage{multirow}
\usepackage{siunitx}  %unit
\usepackage{txfonts}
\usepackage{graphicx}	% Including figure files
\usepackage{amsmath}	% Advanced maths commands
\usepackage{amssymb}	% Extra maths symbols

\usepackage{multirow}
\usepackage{color}
\usepackage{verbatim}
\usepackage{hyperref}
\usepackage{breakurl}
\usepackage{float}
\usepackage[multidot]{grffile} %% to allow for dots in fig basename
\usepackage{xspace}

\usepackage{nicefrac}

\def\emi{EM}

\newcommand{\simgt}{\,\hbox{\lower0.6ex\hbox{$\sim$}\llap{\raise0.6ex\hbox{$>$}}}\,}
\newcommand{\simlt}{\,\hbox{\lower0.6ex\hbox{$\sim$}\llap{\raise0.6ex\hbox{$<$}}}\,}

\newcommand{\xrism}{\textit{XRISM}\xspace}
\newcommand{\athena}{\textit{Athena}\xspace}

\definecolor{coral}{rgb}{0.9, 0.46, 0.03}
\definecolor{coral2}{rgb}{0.8, 0, 0}
\definecolor{midnightblue}{rgb}{0.1, 0.1, 0.44}

\begin{document}

\title{The velocity structure of the intracluster medium during a major merger: simulated microcalorimeter observations}

\author{Veronica~Biffi\inst{1,2}\thanks{e-mail:         
    \href{mailto:veronica.biffi@inaf.it}{\tt veronica.biffi@inaf.it}}
    \and
    John~A.~ZuHone\inst{2}
    \and
    Tony~Mroczkowski\inst{3}
    \and
    Esra~Bulbul\inst{4}
    \and
    William Forman\inst{2}
    }
    
    \titlerunning{ICM velocity sructure in galaxy cluster major merger}
    \authorrunning{V. Biffi et al.}

    \institute{INAF --- Osservatorio Astronomico di Trieste, via Tiepolo 11, I-34143 Trieste, Italy
    \and
    Center for Astrophysics $\vert$ Harvard \& Smithsonian, 60 Garden St., Cambridge, MA 02138, USA
    \and
    ESO --- European Southern Observatory, Karl-Schwarzschild-Stra{\ss}e 2, D-85748 Garching, Germany
    \and      
    Max-Planck-Institut f\"ur extraterrestrische Physik, Gießenbachstraße 1, D-85748 Garching, Germany
    }

   \date{Received ; accepted}

% \abstract{}{}{}{}{} 
% 5 {} token are mandatory

  \abstract
      {Major mergers between galaxy clusters can produce large turbulent
   and bulk flow velocities in the intra-cluster medium (ICM) and thus
   imprint useful diagnostic features in X-ray spectral emission lines
   from heavy ions. As successfully achieved by \textit{Hitomi} in
   observations of the Perseus cluster, measurements of gas velocities
   in clusters from high-resolution X-ray spectra will be achievable
   with upcoming X-ray calorimeters like those on board \xrism,
   \athena, or a \textit{Lynx} like mission.  An interesting
   application to clusters involves detecting multiple velocity
   components or velocity gradients from diagnostic observations of
   specific interesting locations across the cluster.  To explore this
   possibility in the case of a major head-on cluster merger, we
   perform velocity analyses of a cluster-cluster merger from a
   hydrodynamical simulation by means of X-ray synthetic spectra with
   spectral resolution of order of a few eV. We observed the system
   along two extreme line-of-sight directions: 1) perpendicular to the
   plane of the merger and 2) along the merger axis. In these
   geometrical configurations, we find that clear non-Gaussian shapes
   of the iron He-like K$_\alpha$ line at $6.7$\,keV are expected.
   While the velocity dispersion predicted from the simulations can be
   retrieved for the brightest 100\,ks pointings with \xrism Resolve,
   some discrepancy with respect to the expected value is noted and
   can be attributed to the complex non-Gaussian line shapes.
   Measurements in low surface brightness regions, especially when
   multiple velocity components are present along the line of sight,
   require high S/N and the larger collecting area of the \athena
   X-IFU calorimeter is therefore required.  With the latter, we also
   investigate the ICM temperature and velocity gradient across the
   merger bow shock edge, from $20''$-wide annuli extracted from a
   single 1\,Ms X-IFU observation.  For both temperature and velocity
   dispersion, we find best-fit values that are consistent with
   predictions from the simulations within $1$-$\sigma$.  The
   uncertainties on the inferred velocity dispersion are however too
   large to place any stringent constraints on the shallow gradient
   downstream of the shock.  Additionally, we present simulated images
   of the thermal and kinetic Sunyaev-Zeldovich effects from this
   merging system, using the above viewing configurations, and compare
   the results at angular resolutions appropriate for future
   observatories such as CMB-S4 and the Atacama Large Aperture
   Submillimeter Telescope (AtLAST).}
  % aims heading (mandatory)   {}
  % methods heading (mandatory)   {}
  % results heading (mandatory)   {}
  % conclusions heading (optional), leave it empty if necessary    {}


   \maketitle
%
%-------------------------------------------------------------------

\section{Introduction}\label{sec:intro}

The diffuse intra-cluster medium (ICM) in galaxy clusters retains a wealth of information about the interaction with member galaxies, energetic feedback processes from stellar sources and black holes, chemical enrichment, and the system accretion history and merging events with other groups and clusters of galaxies. 
All these processes not only leave an imprint on the thermal structure of the ICM, but also exhibit kinematic signatures. The velocity diagnostics of the ICM provide, therefore, a complementary picture of cluster physics~\cite[][]{simionescu2019} to the thermal and chemical pictures.
Gas motions of non-thermal origin, driven for instance by accreting streams of gas from filaments, or in the form of motions caused by infalling substructures and mergers~\cite[][]{lau2009,fang2009,biffi2011,zuhone2016b}, are crucial to measure because they can provide substantial additional pressure support to the system. This, in turn, can bias X-ray and Sunyaev-Zeldovich (SZ) mass estimates based on the assumption of hydrostatic equilibrium~\cite[HE; e.g.][]{pratt2019}, as investigated by several simulation studies~\cite[][]{rasia2006,meneghetti2010,suto2013,lau2013,nelson2014,shi2015,biffi2016,oppenheimer2018,vazza2018,pearce2020,angelinelli2020,gianfagna2021}.

Even though residual gas motions can be present in relaxed clusters as well, they are particularly important in disturbed systems. In fact, cluster mergers can develop significant bulk motions in the ICM, with large gas velocities up to thousands of km/s~\cite[e.g.][]{dupke2005,simionescu2019}.
Furthermore, disturbances of the ICM velocity field can also persist after the merging event, contributing to a deviation from HE~\cite[][]{nelson2012,nelson2014,ghirardini2018}.
If, on the one hand, merging systems are evidently not suitable for measuring HE-based masses, they provide, on the other hand, the possibility to measure large gas velocities, and map the velocity field to constrain the merger state and geometry.
Observations have already singled out many systems showing clear signatures of merging activity, including extreme examples. 
These include cases where the merging occurs in (or close to) the plane of the sky, 
as well as galaxy clusters merging along the l.o.s.\ direction, as suggested for instance from multiple peaks in the redshift distribution of their galaxies or from the central off-set between the galaxy spatial distribution and the X-ray emission.
Mergers in the plane of the sky include well-known systems like the Bullet cluster~\cite[][]{markevitch2002}, A3266~\cite[][]{degrandi1999,henriksen2000,henriksen2002,finoguenov2006,sanders2021}, or the ``Sausage''~\cite[CIZA~J2242.8+5301;][]{kocevski2007,vanweeren2011,akamatsu2015} and ``Toothbrush''~\cite[1RXS~J0603.3+4214;][]{vanweeren2012,ogrean2013,rajpurohit2018} clusters showing famous giant radio relics.
Instead, candidates for l.o.s. mergers are for instance Cl~0024+17~\cite[][]{czoske2001,czoske2002,ota2004,zhang2005,jee2010}, Abell~576~\cite[][]{dupke2007}, the ``Cheshire Cat'' group~\cite[][]{irwin2015},
the Frontier Fields cluster MACS~J1149.6+2223~\cite[][]{ogrean2016}, and complex merging systems like Abell~697 from the Reionization Lensing Cluster Survey (RELICS)~\cite[][]{cibirka2018}.
Extreme geometrical configurations of cluster mergers can be particularly interesting to pose limits on the detectability of the gas velocity features during and after the dynamical interaction.
Furthermore, given the large gas velocities involved, they offer the conditions to test the capabilities of future high-resolution X-ray instruments in reconstructing the ICM velocity field.

ICM velocities can be in principle measured directly from X-ray observations from the Doppler shift and broadening of emission lines in the ICM X-ray spectrum~\cite[see][for a recent review]{simionescu2019}.
Early direct constraints on ICM bulk motions have been pursued by mapping redshift (i.e. velocity) differences across the system extent, from spatially resolved X-ray observations (e.g.\ early \textit{ASCA} measurements on the Centaurus and Perseus clusters by~\citealt{dupke2001a,dupke2001b}; see also more recent results for \textit{Suzaku} and \textit{Chandra} by~\citealt{tamura2014,liu2015,ota2016,liu2016}).
Direct measurements of the ICM non-thermal motions from line broadening are more challenging.
In principle, this can be achieved in the case of emission lines from heavy ions, such as the Fe-K complexes at $6.4$--$6.9$\,keV. In such cases, the line broadening due to thermal motions is sub-dominant compared to the broadening and shape distortions caused by bulk and turbulent gas motions.
Measurements of these line features requires very high spectral resolution~\cite[][]{inogamov2003}.
Given the typical resolutions of X-ray instruments available, so far most of the ICM velocity measurements actually involved indirect constraints, for instance from 
measurements of resonant scattering~\cite[e.g.][]{churazov2004,sanders2006,werner2009,deplaa2012,zhuravleva2013,hitomi2018_resscatt}, or
pressure and surface brightness fluctuations~\cite[e.g.][]{schuecker2004,churazov2012,sanders2012,gaspari2013,zhuravleva2015,zhuravleva2018}.
Estimates of ICM motions have also been derived from observations of the SZ~\cite[][]{SZ1969,ZS1969,tSZ1972} effect~\cite[e.g.][]{khatri2016,mroczkowski2019}, especially from the kinetic SZ signal~(kSZ, \citealt{kSZ1980}; e.g.\ \citealt{adam2017}). The latter can be particularly strong for merging galaxy clusters, given the large velocities developed~\cite[][]{sayers2013,adam2017}. 
\citealt{sayers2019} employed observations of the kSZ effect towards a sample of 10 merging clusters to constrain the ICM bulk l.o.s.\ velocity and global RMS gas motions within $r_{2500}$, which further allowed differentiating between l.o.s.\ and plane-of-sky mergers.
%%%%%%
In the X-rays, a few successful attempts were able to place direct constraints on the non-thermal (turbulent) gas velocities in cluster cores by exploiting the Reflection Grating Spectrometer (RGS) aboard \textit{XMM-Newton}~\cite[][]{sanders2010,sanders2011,bulbul2012,sanders2013,pinto2015}. %
More recently, an alternative strategy has been presented by~\cite{sanders2020} and \cite{gatuzz2021} exploiting the \textit{XMM-Newton} EPIC-pn detector. By studying the spatial and temporal variation of fluorescent instrumental background lines present in all observations, they could improve significantly the calibration accuracy and successfully map bulk motions in Coma, Perseus, and Virgo.

So far, the best and only example of direct ICM velocity measurement from high-resolution spectroscopy 
has been obtained from the observation of the Perseus galaxy cluster by the \textit{Hitomi} Soft X-ray Spectrometer (SXS)~\cite[][]{hitomi2018}. The high-resolution spectra allowed detailed modeling of the iron He-like K$_{\alpha}$ complex in the core of the Perseus Cluster, yielding typical ICM bulk velocities and velocity dispersion values of order $100$--$200$\,km/s. 
The unprecedented success obtained with \textit{Hitomi} confirms the power of high-resolution X-ray spectroscopy,
anticipating the further progress that will be achieved with the microcalorimeters on board the next-generation X-ray telescopes,
like \xrism~\cite[][]{xrism2020} and \athena~\cite[][]{athena}. These are expected within the next $\sim 15$ years, with \xrism launching already by early 2023 and \athena in the early 2030s. 
Like \textit{Hitomi}, these will allow for high spectral resolution, namely $7$\,eV for \xrism Resolve, and $2.5$\,eV, for \athena X-IFU.
If a \textit{Lynx}-type\footnote{\burl{https://wwwastro.msfc.nasa.gov/lynx/};\\ \url{https://www.lynxobservatory.com/}}~\cite[][]{lynx2019} mission with sub-arcsecond angular resolution is developed and includes a microcalorimeter, this will provide even further progress, with average spectral resolutions of few eV and arrays optimized to reach fractions of an eV in the soft band.
Hydrodynamical simulations are essential to predict the performance of high resolution X-ray calorimeters in the task of mapping ICM velocities, by producing realistic X-ray synthetic observations of simulated clusters~\cite[][]{zuhone2009,biffi2013,zuhone2016a,zuhone2016b,Roncarelli:2018,Cucchetti:2019,Bulbul2019}.

In this study, we aim at demonstrating in particular the capabilities of the \xrism Resolve~\cite[][]{xrismResolve2018} and \athena X-IFU~\cite[][]{xifu2018} to constrain interesting features of the ICM velocity field in extreme merger configurations, via numerical hydrodynamical simulations and synthetic X-ray observations derived from them.
Expanding on previous investigations, we also extend our focus beyond the bright central regions of the clusters to fainter regions, such as peripheral locations outside the bright shock fronts formed during the first core passage.
We focus on a merging pair of galaxy clusters, with a mass ratio of 1:3 and ICM temperatures of $\sim 5$\,keV and $\sim 2.5$\,keV for the main halo and the secondary cluster. This represents a major merger, similar to the case of the ``El Gordo''~\cite[ACT-CL J0102-4915;][]{marriage2011,menanteau2012,zhang2015,asencio2021} but with lower absolute masses of the two components.
The structure of the paper is organized as follows. In Sec.~\ref{sec:sims} we describe the simulations employed and, in Sec.~\ref{sec:mocks}, the approach to generate mock X-ray observations of simulated cluster mergers. Results are presented and discussed in Sec.~\ref{sec:results}, where we investigate the gas velocity field in the merging system after the first core passage, at interesting locations in the two main projections (perpendicular to the l.o.s.\ and aligned with it). In this section, we also complement the direct simulation analysis with mock X-ray observations for \xrism Resolve and \athena X-IFU. We exploit the larger collecting area and higher spatial resolution of X-IFU to explore also the velocity field across the main shock front.
Also, we show predictions for thermal and kinetic SZ effect imaging at the resolutions of CMB-S4 and the Atacama Large Aperture Submillimeter Telescope \citep[AtLAST;][]{Klaassen2020}. 
We summarize our results and draw final conclusions in Sec.~\ref{sec:concl}.

\section{Numerical simulations}\label{sec:sims}

\paragraph{Code}
The simulations have been performed with the AREPO
code~\cite[][]{arepo}\footnote{\url{https://arepo-code.org}}, and include the treatment of gravity and
hydrodynamics only. The AREPO code employs a finite-volume Godunov method on
an unstructured moving-mesh to solve the hydrodynamical equations, and a Tree-PM solver for solving for the gravitational force.

The simulations we analyse in this work contain three %types of
Lagrangian mass elements, i.e.\ gas, DM and stars.
The gas elements are simulated using the moving-mesh Voronoi
tessellation method. The gas is modeled as an ideal fluid with $\gamma=
5/3$ and mean molecular weight $\mu = 0.6$, i.e. primordial abundances
of H and He with trace amounts of metals.
DM and stellar particles only interact via gravity, with each other
and with the gas elements. In particular, star particles are included
only to serve as tracers of the stellar component associated with the BCG in the cluster center and the average stellar density throughout the cluster. These play no role for the analysis presented in this paper.

For the purpose of the present analysis, no radiative processes --
such as cooling, star formation or feedback processes -- are included. For the short timescales and and high temperatures involved in the analysis of a merging cluster, these effects are not significant.

The simulations analysed are non-cosmological, i.e., the equations of motion have no dependence on redshift. For the purpose of the following calculations and mock observations, we assume a $\Lambda$CDM Universe with $\Omega_{m,0}= 0.27$, $\Omega_{\Lambda,0}=0.73$ and $H_0=71$\,km\,s$^{-1}$\,Mpc$^{-1}$.

\subsection{Merging galaxy clusters}

For the current analysis we investigate the properties of a merging
system, where two clusters in a 1-to-3 mass ratio collide with zero
impact parameter. These are the same initial conditions as in \cite{zuhone2011}, with some changes as outlined here.

The two clusters are initially %set to be 
spherically symmetric and in virial and hydrostatic equilibrium, 
with parameters consistent with observed
relaxed clusters and cluster scaling relations (see Section 2.2 of
\citealt{zuhone2011} for details). The parameters which we use 
%to set up the clusters 
are given in Table~\ref{tab0}.

For the initial total mass distribution of gas, stars, and DM we choose an NFW
\citep{nfw97} profile:
\begin{equation}
\rho_{\rm tot}(r) = \frac{\rho_s}{{r/r_s{(1+r/r_s)}^2}}.
\end{equation}
The scale density $\rho_s$ and scale radius $r_s$ are determined by the
virial radius $r_{200c}$\footnote{According to the standard definition, $r_{200c}$ corresponds to the radius enclosing an average overdensity of $200$ times the critical density of the Universe.}
and concentration parameter $c_{200c}$ using:
\begin{align}
\begin{split}
r_s &= \frac{r_{200c}}{c_{200c}} 
\\
\rho_s &= \frac{200}{3}c_{200c}^3 \, \rho_{\rm crit}(z){\left[{\rm log}(1+r/r_s) - \frac{r/r_s}{{1 + r/r_s}}\right]}^{-1}
\end{split}
\end{align}
where we assume $z = 0.1$ to scale the critical density. The NFW functional form
is carried out to $r_{200c}$. Since the NFW mass profile does not converge as $r \rightarrow \infty$, we ``taper'' the density profile after this radius. For radii $r > r_{200c}$ the mass density follows an
exponential profile:
\begin{equation}
\rho_{\rm tot}(r) = \frac{\rho_{\rm s}}{{c_{200c}(1+c_{200c})^2}}
{\left(\frac{r}{r_{200c}}\right)^\kappa}{\rm exp}
{\left(- \frac{r-r_{200c}}{r_{\rm decay}}\right)}
\end{equation}
where $\kappa$ is set such that the density and its first derivative are
continuous at $r = r_{200c}$ and $r_{\rm decay} = 0.1r_{200c}$.

The spatial distribution of the stellar mass is divided into two components: a
BCG at the cluster center, and a smooth component over the entire cluster
representing the stars in the satellite galaxies and the intracluster light. For
the BCG component, we use an analytical approximation to a deprojected
S\'{e}rsic profile given by~\citet{merritt2006}:
\begin{align}
\begin{split}
\rho_{\rm}(r) &= \rho_e\exp \{ -d_{n_s} \left[(r/r_e)^{1/{n_s}}-1 \right]\}\label{eqn:sersic} 
\\
d_{n_s} &\approx 3n_s - 1/3 + 0.0079/n_s,\quad{\rm for}~n_s \gtrsim 0.5
\end{split}
\end{align}
which is a good representation of the stellar mass density profile of elliptical
galaxies. We set $n_s = 6$, and use the relationships for BCG masses and radii
with their cluster counterparts from \cite{Kravtsov2018} to
determine the other parameters for each cluster. We add the contribution to the
stellar density from the other galaxies and intracluster light by assuming an
NFW profile with the same scale radius as the DM profile and a scale density
which is 2\% of that of the DM profile.

For the gas, we begin by assuming it has an entropy profile of the form
\citep{Voit2005,Cavagnolo2008,Cavagnolo2009}:
\begin{equation}
S(r) = S_0 + S_1\left(\frac{r}{0.1r_{200c}}\right)^{\alpha}
\end{equation} 
where $S_0$ is the core entropy of the cluster and $S_1$ is the entropy at
$0.1r_{200c}$. We initialize our clusters with small core entropies by requiring
$S_1/S_0$ = 20, to make our models consistent with relaxed,
``cool-core'' galaxy clusters, and set $\alpha = 1.1$.

To use this equation to derive the other thermodynamic profiles from
the condition of hydrostatic equilibrium, we use the gas mass fraction taken
from scaling relations (see Table~\ref{tab0}) and impose $T(r_{200c}) =
\frac{1}{2}T_{200c}$~\citep[as in][]{Poole2006}, where
\begin{equation}
k_\textsc{b} T_{200c} \equiv \frac{GM_{200c}{\mu}m_p}{2r_{200c}}
\end{equation}
is the so-called ``virial temperature'' of the cluster. Once we have the gas
density and temperature profiles, we can also derive the dark matter density
profile via $\rho_{\rm DM} = \rho_{\rm tot} - \rho_g - \rho_*$.

These radial profiles for the gas, DM, and stars can now be used to generate the
properties of the particles/cells for the two clusters. To generate positions,
we randomly generate a number in the range $0 < u < 1$ and the function $u =
M_i(r)/M_i(r_{\rm max})$ is inverted to give the radius of the particle
from the center of the halo, where $i~\in~[{\rm DM, *}]$ and $r_{\rm max}$ =
5~Mpc is the maximum radius of the halo. The gas cell velocities are set to
zero in each cluster's rest frame initially. For the dark matter and stellar
particle velocities, we choose to directly calculate the distribution function
via the Eddington formula~\citep{Eddington1916}:
\begin{equation}
\mathcal{F}(\mathcal{E}) = \frac{1}{\sqrt{8}\pi^2}\left[\int^{\mathcal{E}}_{0} \frac{d^2\rho}{ d\Psi^2}
\frac{d\Psi}{\sqrt{\mathcal{E} - \Psi}} + \frac{1}{\sqrt{\mathcal{E}}}\left(\frac{d\rho}{d\Psi}\right)_{\Psi=0} \right]
\end{equation}
where $\Psi = -\Phi$ is the relative potential and $\mathcal{E} = \Psi -
\frac{1}{2}v^2$ is the relative energy of the particle. We tabulate the function
$\mathcal{F}$ in intervals of $\mathcal{E}$ and interpolate to solve for the
distribution function at a given energy. Particle speeds are chosen from this
distribution function using the acceptance-rejection method. Once particle radii
and speeds are determined, positions and velocities are determined by choosing
random unit vectors in $\Re^3$. The masses of the simulated 
resolution elements
are: $M_{\rm gas} = 9.22 \times 10^6~\rm M_{\odot}$, $M_{\rm DM} = 6.77 \times 10^7~\rm M_{\odot}$, and $M_{\rm star} = 6.91 \times 10^7~\rm M_{\odot}$.

The two clusters in the merging system have a virial mass ratio
$M_{200c,1}/M_{200c,2}=3$, with $M_{200c,1}=6\times 10^{14}\,{\rm M}_\odot$ for
the main cluster and $M_{200c,2}=2\times 10^{14}\,{\rm M}_\odot$ for the smaller
infalling subhalo. We set them up within a cubical computational domain of side
$L = 40$\,Mpc. The distance between the cluster centers is given by $d
= r_{200c,1}+r_{200c,2} \approx 2.8$\,Mpc. \citet{Vitvikska2002} demonstrated from
cosmological simulations that the average infall velocity for merging clusters
is $v_{\rm in}(r_{200c}) = 1.1V_c$, where $V_c = \sqrt{GM_{200c}/r_{200c}}$ is the circular velocity at $r_{200c}$ for the
primary cluster. This is then the relative velocity of the two clusters at the
beginning of the simulation, $v_{\rm rel} \approx 1500$\,km/s. Given that
the merger is head-on, the resulting velocity is only in the $x$-direction,
and is given for each cluster in the simulation frame by:
\begin{eqnarray}
v_{x,1} &=& v_{\rm rel}\frac{M_{200c,2}}{M_{200c,1}+M_{200c,2}} \\
v_{x,2} &=& -v_{\rm rel}\frac{M_{200c,1}}{M_{200c,1}+M_{200c,2}} 
\end{eqnarray}
The gas properties for the cells in the two clusters (which have some overlap in
their outskirts) are set by summing the densities and pressures from the two
profiles interpolated to the cell positions, and by summing the momenta from
each cluster as: 
\begin{eqnarray}
\rho(\textbf{r}) &=& \rho_1(\textbf{r}) + \rho_2(\textbf{r}) \\
P(\textbf{r}) &=& P_1(\textbf{r}) + P_2(\textbf{r}) \\
\rho{v_x}(\textbf{r}) &=& \rho_1(\textbf{r}){v_{x,1}}+\rho_2(\textbf{r}){v_{x,2}}
\end{eqnarray}
where $(\textbf{r})$ is the 3D coordinate of the gas cell.

\begin{table*}
\centering
\caption{Initial Cluster Parameters}
\label{tab0}
\begin{tabular}{cccccccc}
\hline
Cluster & $M_{200}$          & $r_{200}$ & $c_{200}$ & $f_g(r_{500})$  \\
        & ($\rm M_{\odot}$)      & (kpc)     &           &        \\ \hline
Main    & $6 \times 10^{14}$ & 1639.59   & 4.8       & 0.1056 \\ 
Subhalo & $2 \times 10^{14}$ & 1136.83   & 5.0       & 0.0879 \\ \hline
\end{tabular}
\end{table*}
  
Our primary goal is to use this merging configuration to
explore the features of the gas velocity field across the merging system
and the primary shock front caused by the 
subcluster infalling through the main halo. 
We therefore choose a snapshot soon after the first core passage,
corresponding to $1.6$\,Gyr since the beginning of the simulation.
This allows us to investigate particularly extreme conditions connected to the
geometry of the merger and their impact on the gas velocity distribution, shaping the non-thermal profile of gas emission lines in X-ray spectra. 
Given that no degeneracies due to additional baryonic processes nor to complex chemical distributions are present
in the non-radiative run considered here, this enables us to directly explore the possibility to reconstruct such imprints from mock X-ray observations simulating high-resolution X-ray microcalorimeters.

\subsection{Numerical estimates}\label{sec:num_estimate}

Direct estimates of average l.o.s.\ velocity and velocity
dispersion for the gas selected within a given region in the
simulation are given~by:
\begin{equation}\label{eq:shift-sigma}
  \mu_w = \frac{\Sigma_i w_i v_{\rm{los},i}}{\Sigma_i w_i}\,,
  \qquad
  \sigma_w = \frac{\Sigma_i w_i (v_{\rm{los},i} - \mu_w)^2}{\Sigma_i w_i}\,,
\end{equation}
where the summation is performed over the gas elements selected, and
$w_i$ is the weighing quantity (e.g. gas mass, emission measure,
Bremsstrahlung-like emissivity or Fe-line emissivity) for the $i$-th
gas element.
Values estimated according to Eq.~\ref{eq:shift-sigma} are used in
the following Sections as expectations and compared against
the quantities reconstructed from the analysis of the synthetic X-ray spectra.
Typically, we will report average numerical values weighted by Fe-line emissivity, to compare more directly with estimates reconstructed from the diagnostics of the Fe-K complex in high-resolution mock X-ray spectra.
In these computations, the Fe-line emissivity is an ideal estimate, computed for every gas element (given its thermodynamical properties) from the interpolation of emissivity tables, assuming an APEC model with fixed metallicity ($Z=0.3Z_\odot$), for the $[6.5$--$7.1]$\,keV rest-frame energy band including the $6.7$\,keV iron complex.

\section{Mock X-ray observations}\label{sec:mocks}

We generate synthetic X-ray observations from the hot gas properties in the simulated clusters via the
pyXSIM Python package~\cite[][]{pyXSIM}\footnote{pyXSIM (\url{http://hea-www.cfa.harvard.edu/~jzuhone/pyxsim/}) critically depends on the yt Project~\cite[][\url{http://yt-project.org}]{yt}, to link the hydrodynamical simulations to the algorithm for generating the X-ray ideal photons.}, which in turn is based on the PHOX algorithm~\cite[][]{biffi2012,biffi2013}. While referring the interested reader to the aforementioned papers for further details, we briefly recall here the main features of the approach.

We consider every gas element in the simulation as a single-temperature X-ray emitting plasma, for which the spectral emission follows a thermal Bremsstrahlung continuum with emission lines from heavy ions, implemented as an APEC~\cite[][]{apec} model in the X-ray analysis XSPEC package\footnote{\url{https://heasarc.gsfc.nasa.gov/xanadu/xspec/}}~\cite[][]{xspec}. Given that no stellar evolution nor chemical enrichment models are treated in our simulations, we assume a uniform metallicity for the ICM equal to $Z=0.3\,Z_\odot$ and adopt the solar reference abundances of~\cite{angr1989}.  The simulation snapshot is artificially positioned at a fiducial redshift of $z=0.057$, for which the angular scale is $\approx 1.1$\,kpc\,arcsec$^{-1}$ for the $\Lambda$CDM cosmological model adopted here.  In the initial phase of generating the ideal sets of photons, we assume artificially large values of collecting area and exposure time to allow for a statistically representative Monte Carlo sampling for later realistic exposures.
We also include a Galactic foreground absorption component by using the Tuebingen-Boulder ISM absorption model~\cite[tbabs,][]{tbabs} and assuming a uniform fiducial Galactic column density of $N_{\rm  H}=4\times 10^{20}$\,cm$^2$.
At a second stage, we project the photons along a chosen l.o.s.\ direction
and Doppler-shift their energies depending on the l.o.s.\ component of
the velocity associated with the emitting gas element.
The final event files are produced by convolving the ideal
photon lists with the specific characteristics of the 
\xrism Resolve and \athena X-IFU instruments,
accounting for the effective area, detector response,
vignetting, and point-spread function (PSF) of the chosen telescope. 
This step is performed by means of the
SOXS\footnote{\url{http://hea-www.cfa.harvard.edu/soxs/}} and SImulation of X-ray TElescopes\footnote{\url{https://github.com/thdauser/sixte}}
\cite[SIXTE, provided by ECAP/Remeis observatory;][]{sixte} packages, for \xrism and \athena respectively. 
Specific details of the synthetic observations we perform are given in the following sections, along with the results.

\section{Results}\label{sec:results}

In this Section we compare and discuss the features of the ICM velocity field predicted by the simulations and reconstructed from the synthetic X-ray observations therefrom derived. We concentrate on two l.o.s.\ directions, perpendicular to the plane of the merger (Sec.~\ref{sec:z-proj}) and along the merger axis (Sec.~\ref{sec:x-proj}). 
We intend to use these limit cases to constrain the merging signatures in the gas velocity field,
at particularly interesting locations in the system.

\subsection{Merger in the plane of the sky}\label{sec:z-proj}

\begin{figure*}
  \centering
\includegraphics[width=0.43\textwidth]{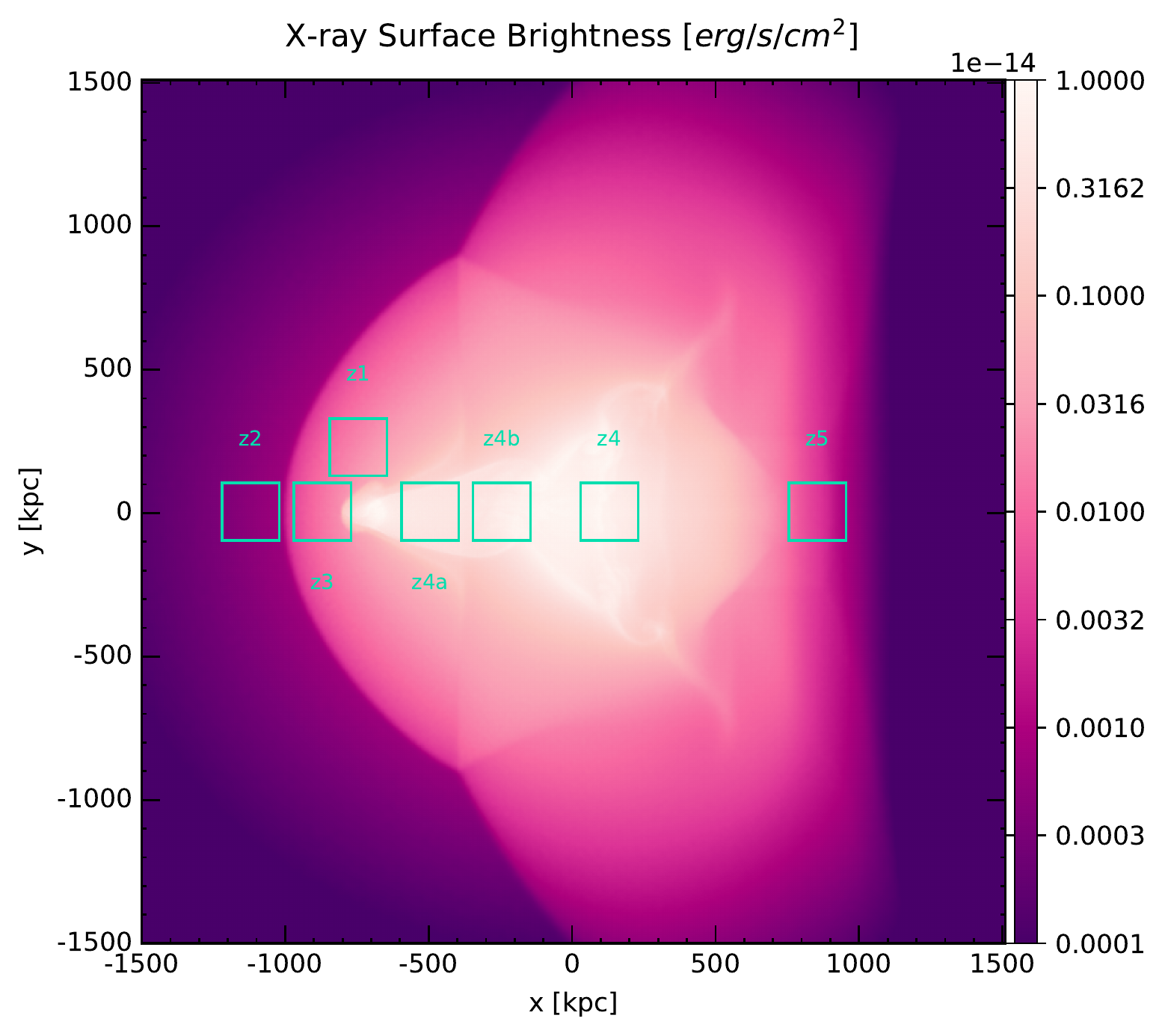}
\qquad
\includegraphics[width=0.43\textwidth]{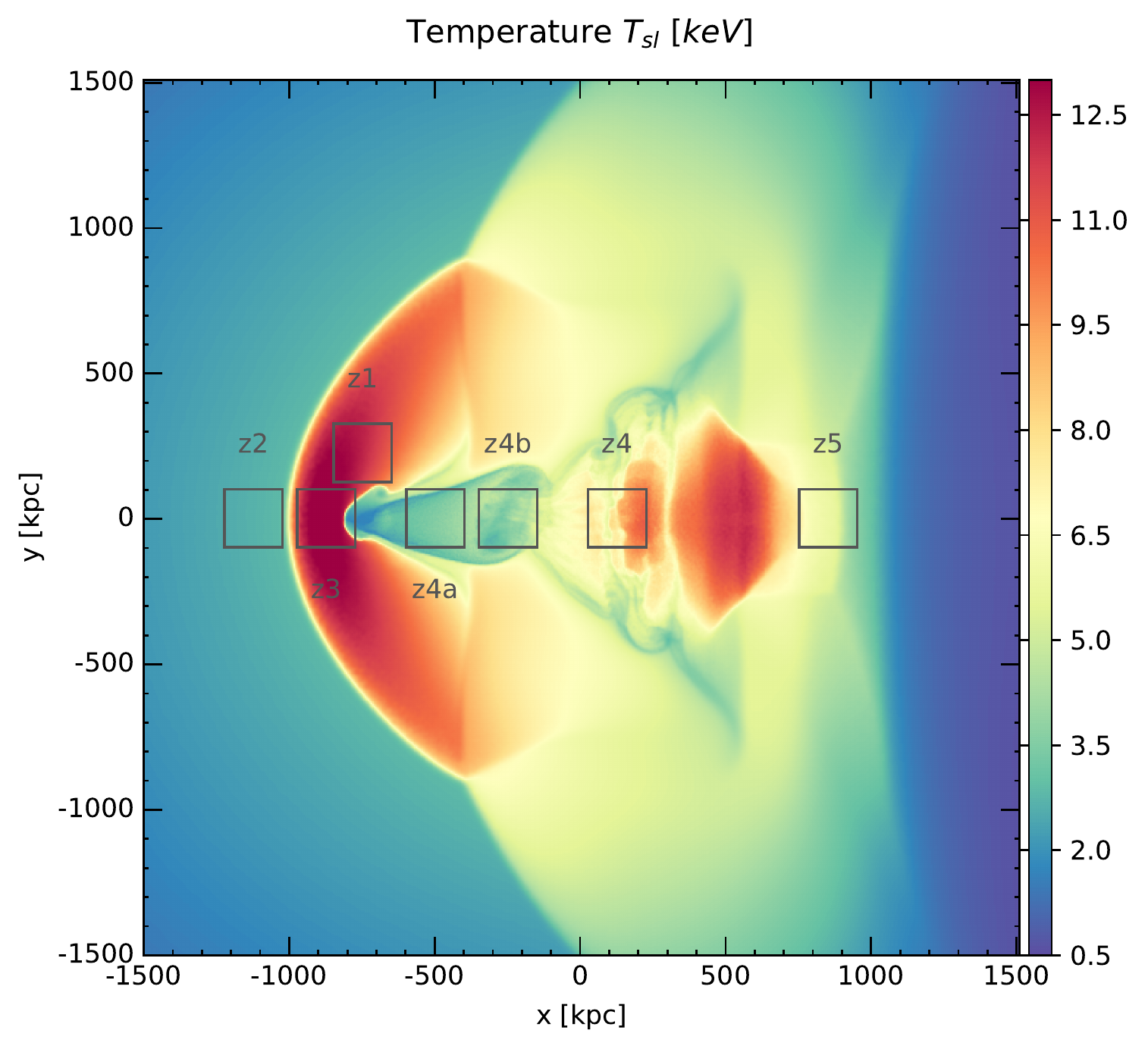}
\\
\includegraphics[width=0.43\textwidth]{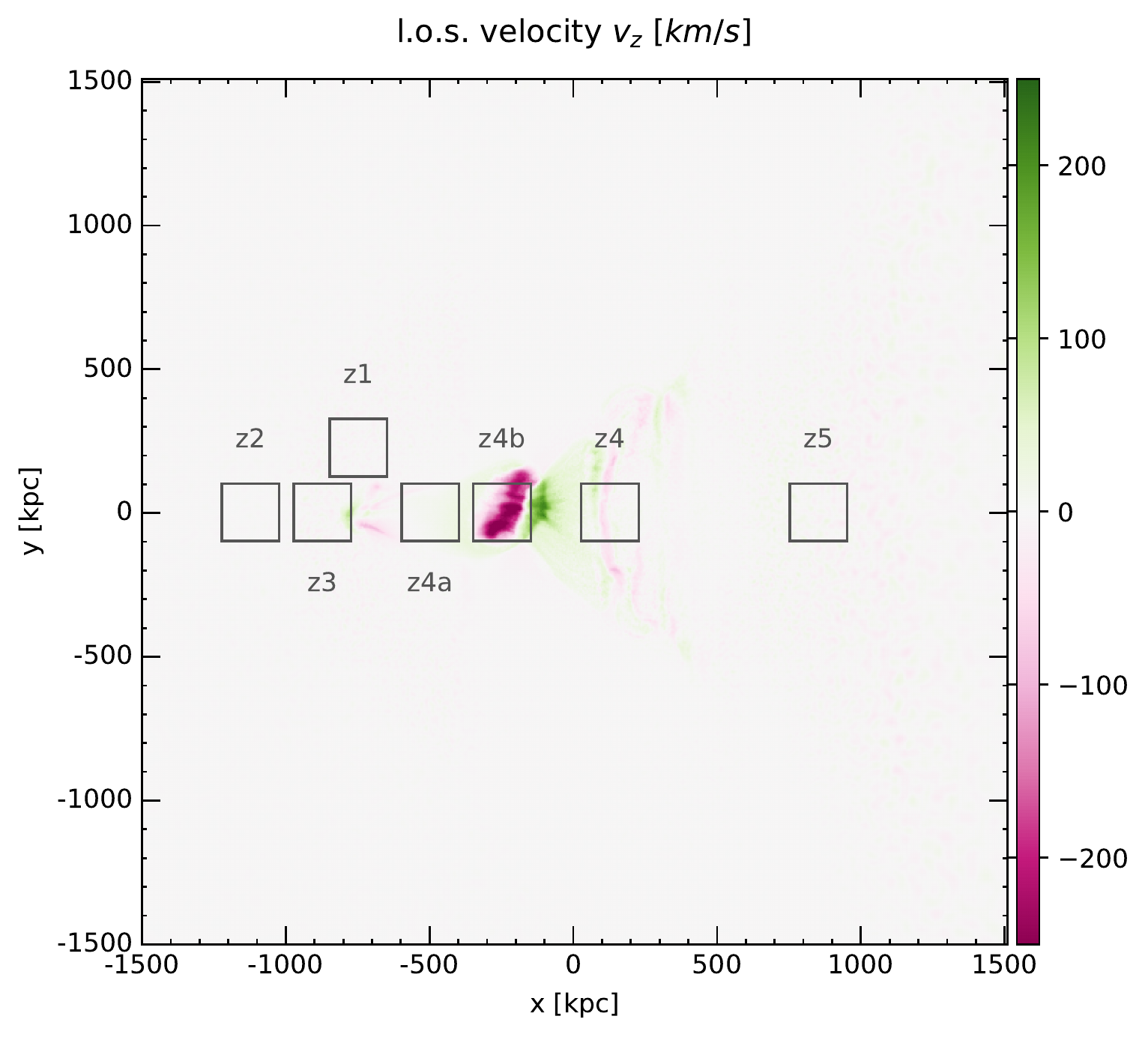}
\qquad
\includegraphics[width=0.43\textwidth]{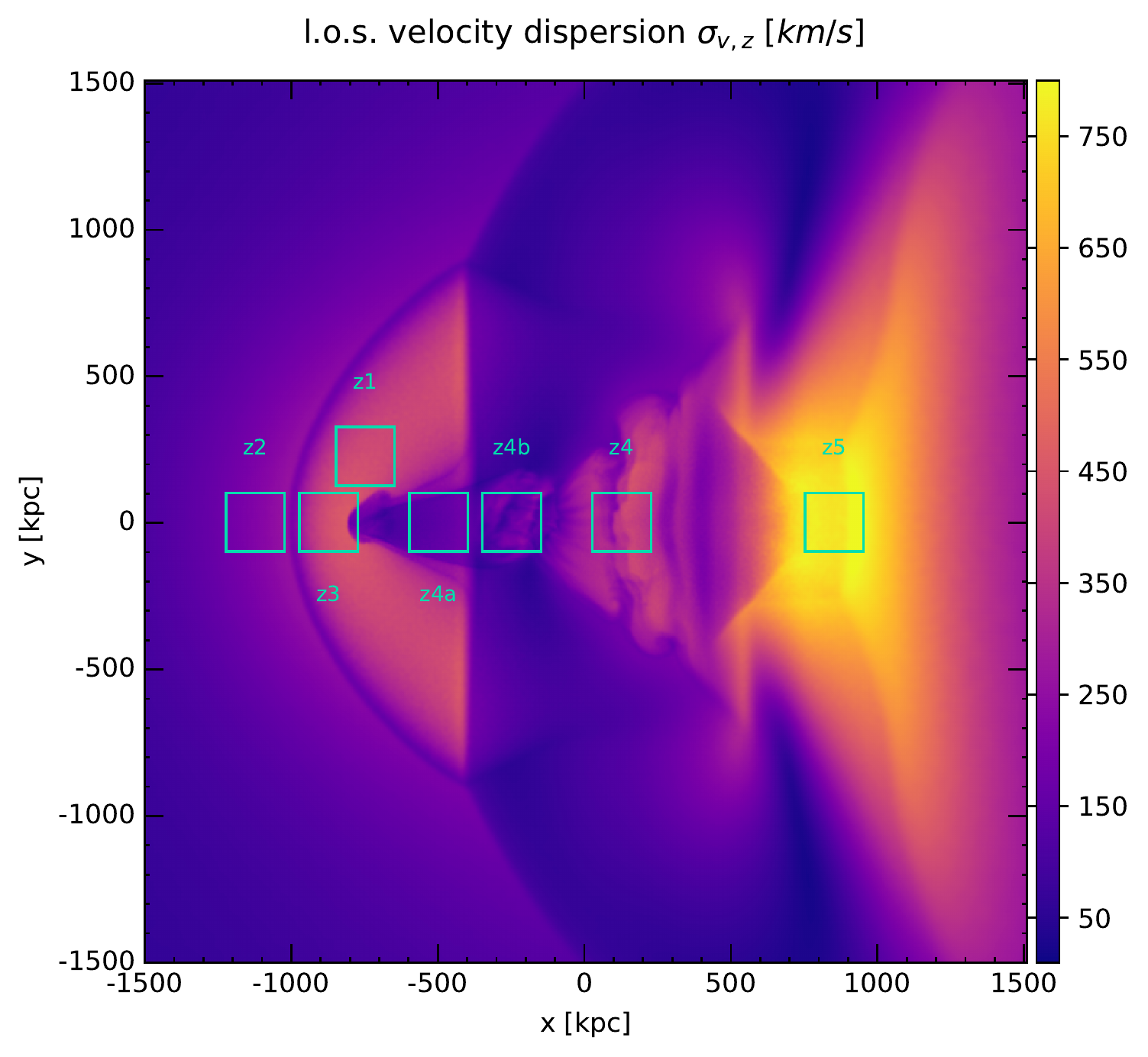}
  \caption{Maps of X-ray flux in the 0.5--7\,keV band (top-left), spectroscopic-like temperature (top-right), EM-weighted l.o.s.\ velocity (bottom-left) and
    EM-weighted l.o.s.\ velocity dispersion (bottom-right), for $z$-axis projection (i.e. for the merger in the plane of the sky). The projection is performed over 10\,Mpc along the l.o.s.\ (z-axis).
    Overplotted, the seven pointings of $200$\,kpc per side (roughly $3'$ at the assumed redshift of $z=0.057$) analysed in Sec.~\ref{sec:z-proj}.
    \label{fig:map}}
\end{figure*}

We first consider the projection aligned with the $z$ axis of the simulation box, perpendicular to the plane of the merger.
In Fig.~\ref{fig:map} we show projected maps, along the $z$ axis, of the gas
X-ray flux (0.5--7\,keV rest-frame band)\footnote{The X-ray flux is calculated from the simulation gas elements, assuming an APEC model with fixed metallicity ($Z=0.3~\rm Z_\odot$).}, spectroscopic-like temperature\footnote{The spectroscopic-like temperature is an estimate of the gas average temperature often used in cluster simulations and defined as 
$T_{\rm sl} \equiv \nicefrac{\int \rho^2 T^a T {\rm d}V}{\int \rho^2 T^a {\rm d}V},$ with $a=-0.75$~\cite[][]{mazzotta2004}.},
l.o.s.\ \emi-weighted velocity and velocity dispersion (from top-left to bottom-right, respectively).
In almost all the X-ray maps, the bright, hot shock front, formed by the secondary bullet halo that passed through the main cluster, is apparent. From the velocity dispersion map, an interesting region with large values of $\sigma_v \gtrsim 600\,$km/s develops on the opposite side of the merger, close to the core of the primary cluster.
Given the symmetry of the merger, that occurs in this projection on the plane of the sky, the l.o.s.\ bulk velocity is instead typically zero (bottom-left panel in Fig.~\ref{fig:map}), except for a small region in the brightest central part of the system, where a turbulent eddy has developed after the core passage and residual net negative and positive velocities remain.

On the maps, we mark seven square regions, centered on interesting regions of the merging system, in the vicinity of the shock (z1, z2 and z3), in the highest velocity dispersion region (z5) and in the brightest part of the system (z4, z4a and z4b).
Each region, 200\,kpc per side, approximates in size the $2.9'\times2.9'$
FoV of the \xrism Resolve instrument at the fiducial redshift of $z=0.057$,
where the cluster is positioned for the synthetic X-ray
observations.

\subsubsection{ICM thermo-dynamical structure along l.o.s.}\label{sec:z-proj2}

\begin{figure*}
  \centering
  \includegraphics[width=0.34\textwidth,trim=0 10 20 25,clip]{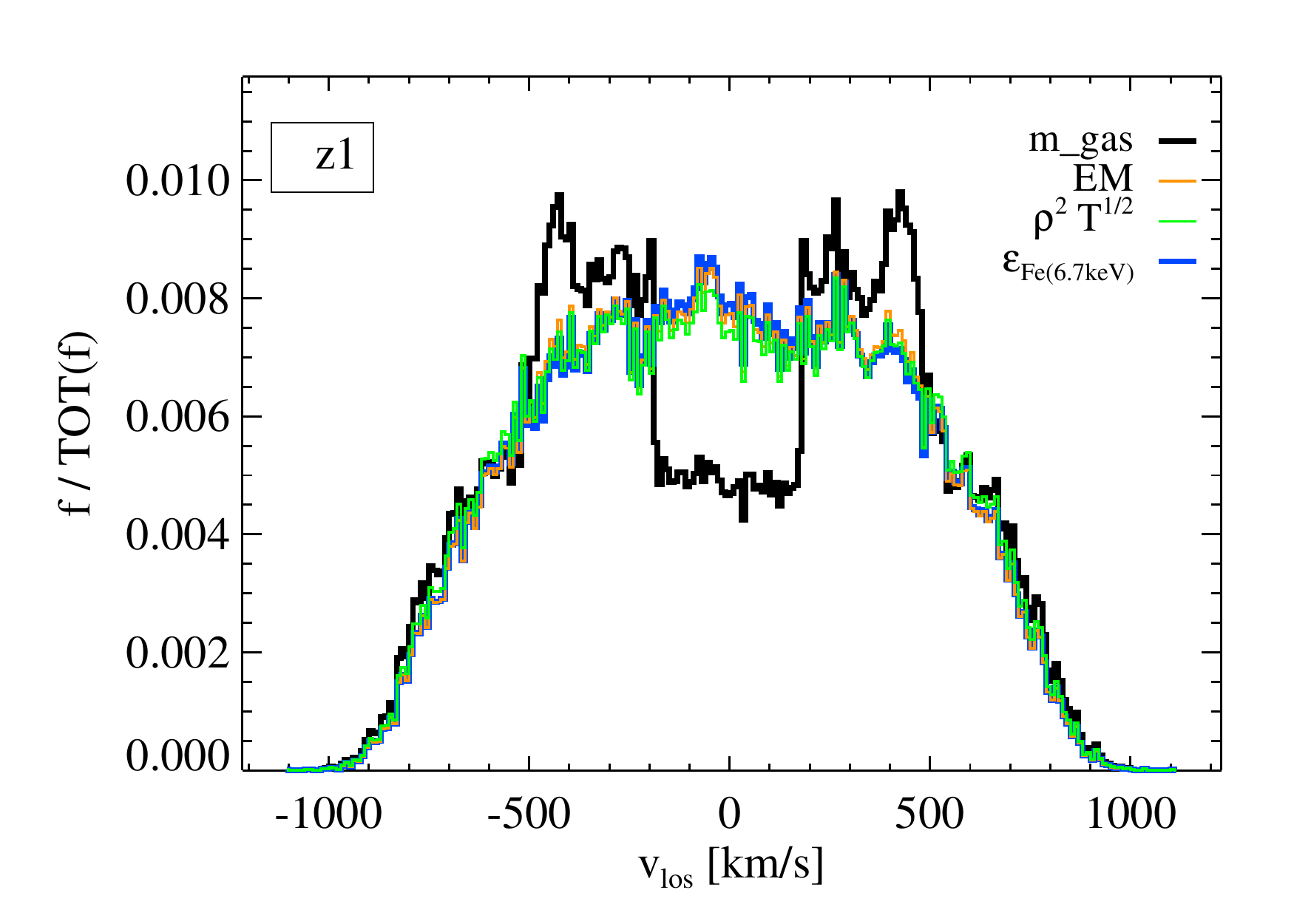}
  \includegraphics[width=0.32\textwidth,trim=100 0 20 10,clip]{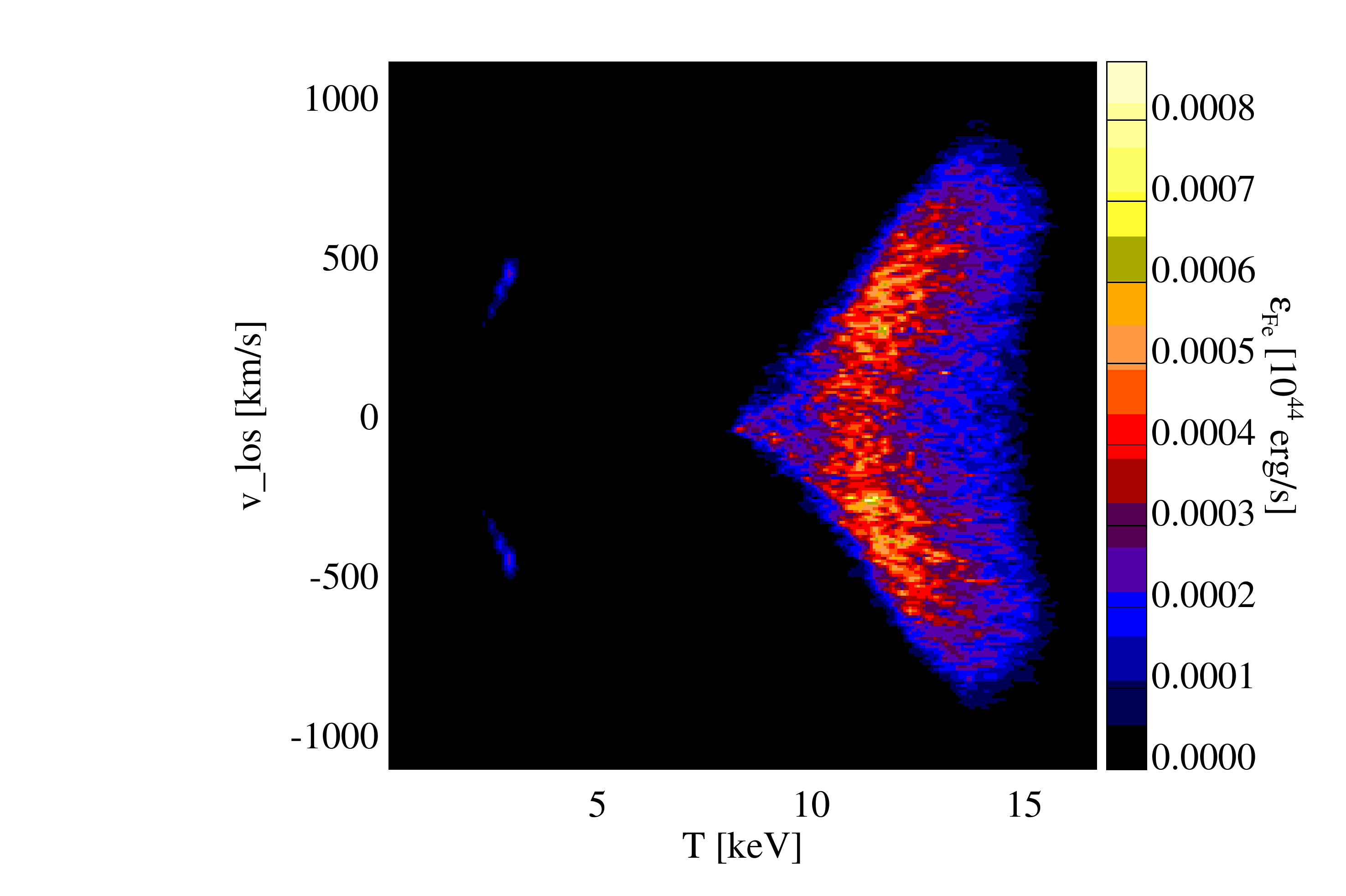}
  \includegraphics[width=0.33\textwidth,trim=20 10 10 20,clip]{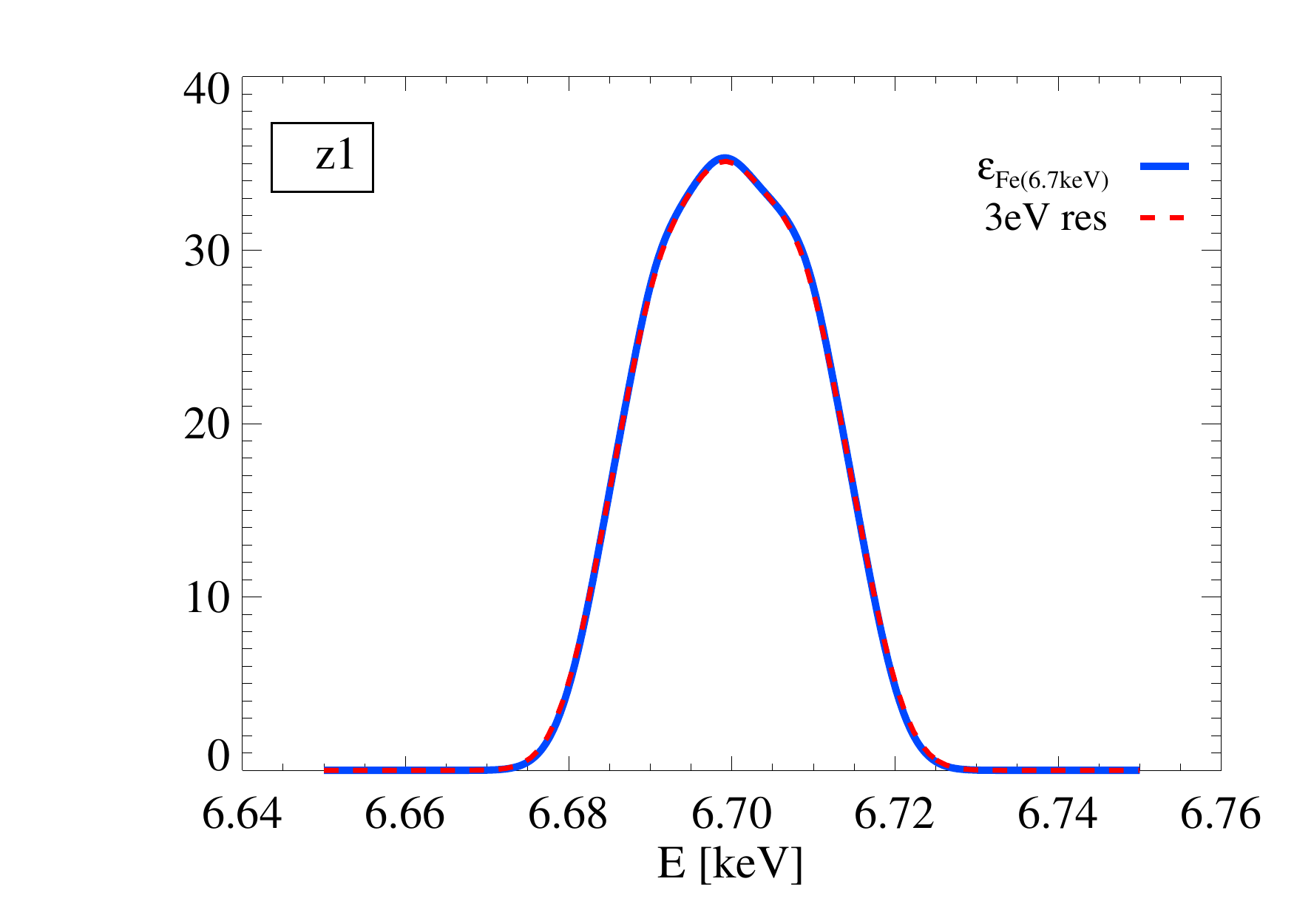}
  \\
  \includegraphics[width=0.34\textwidth,trim=0 10 20 25,clip]{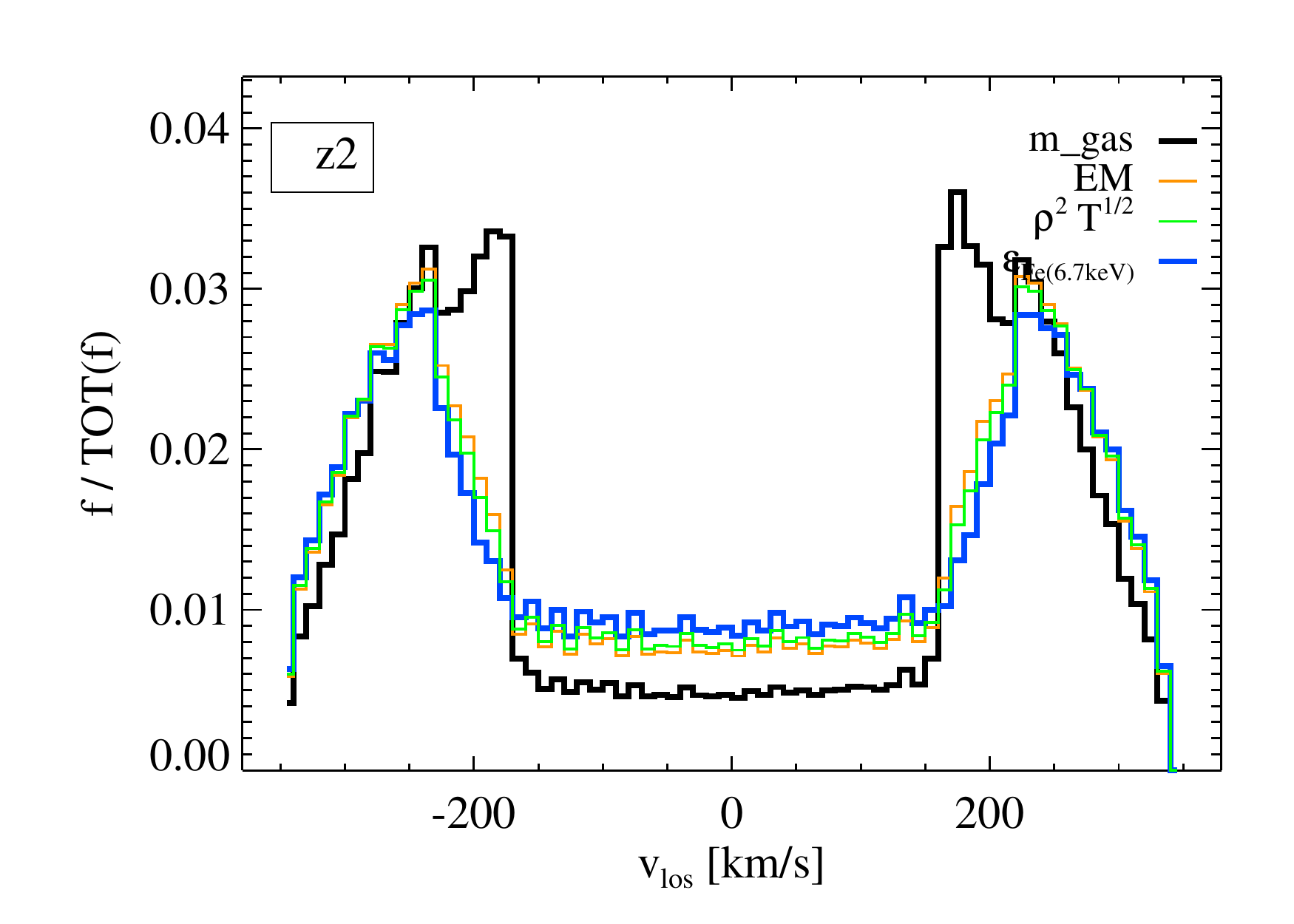}
  \includegraphics[width=0.32\textwidth,trim=100 0 20 10,clip]{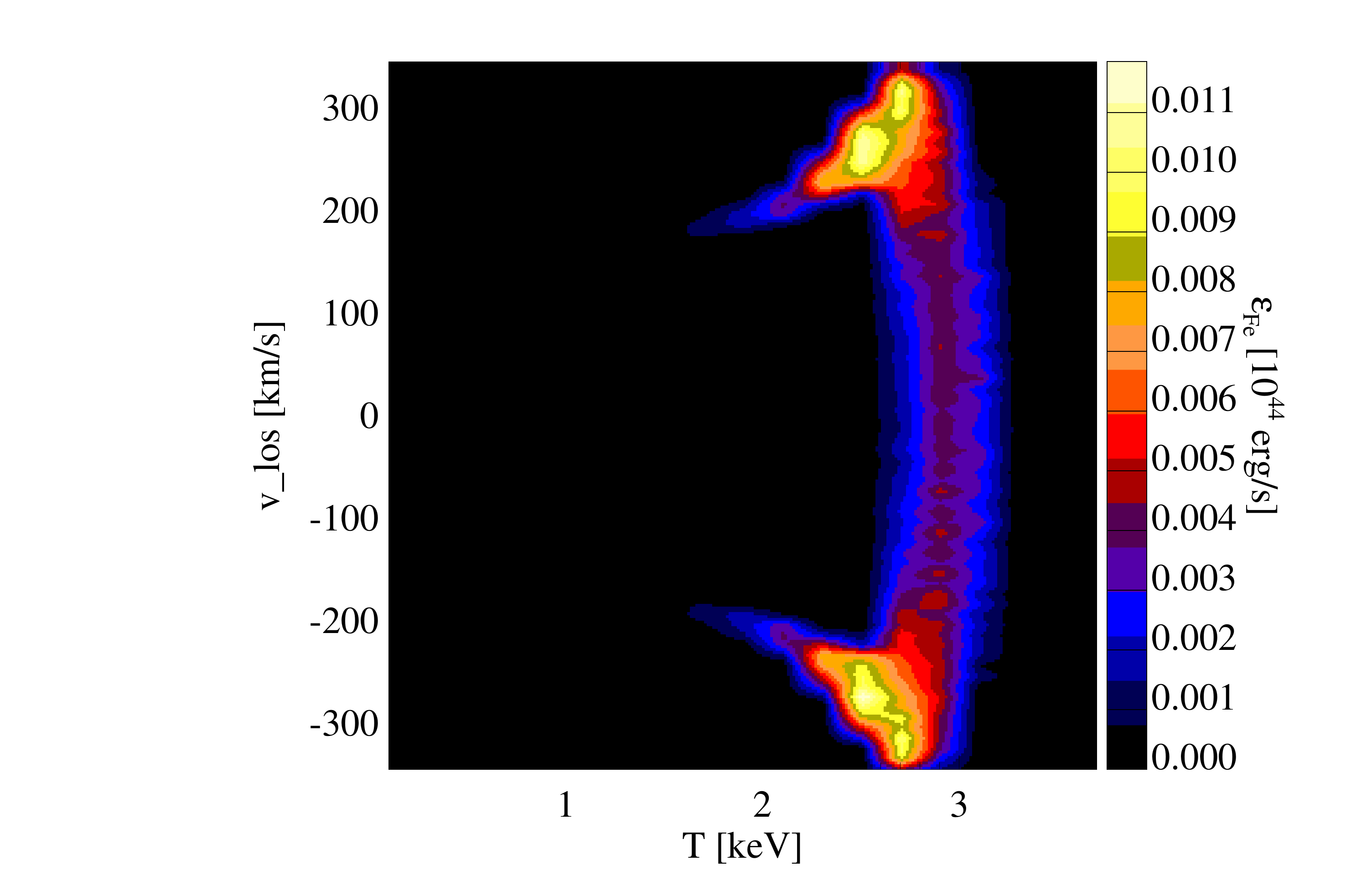}
  \includegraphics[width=0.33\textwidth,trim=20 10 10 20,clip]{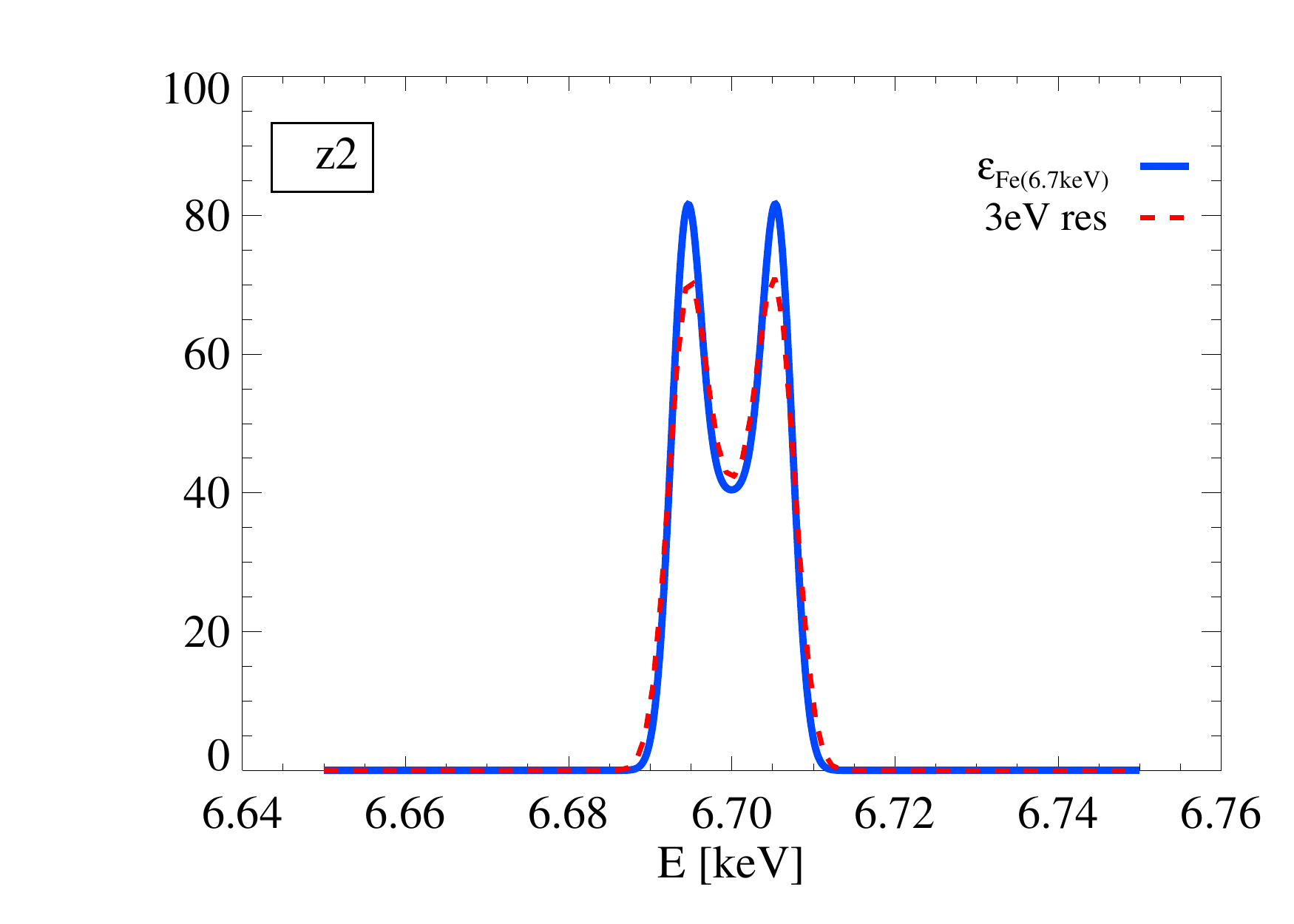}
  \\
  \includegraphics[width=0.34\textwidth,trim=0 10 20 25,clip]{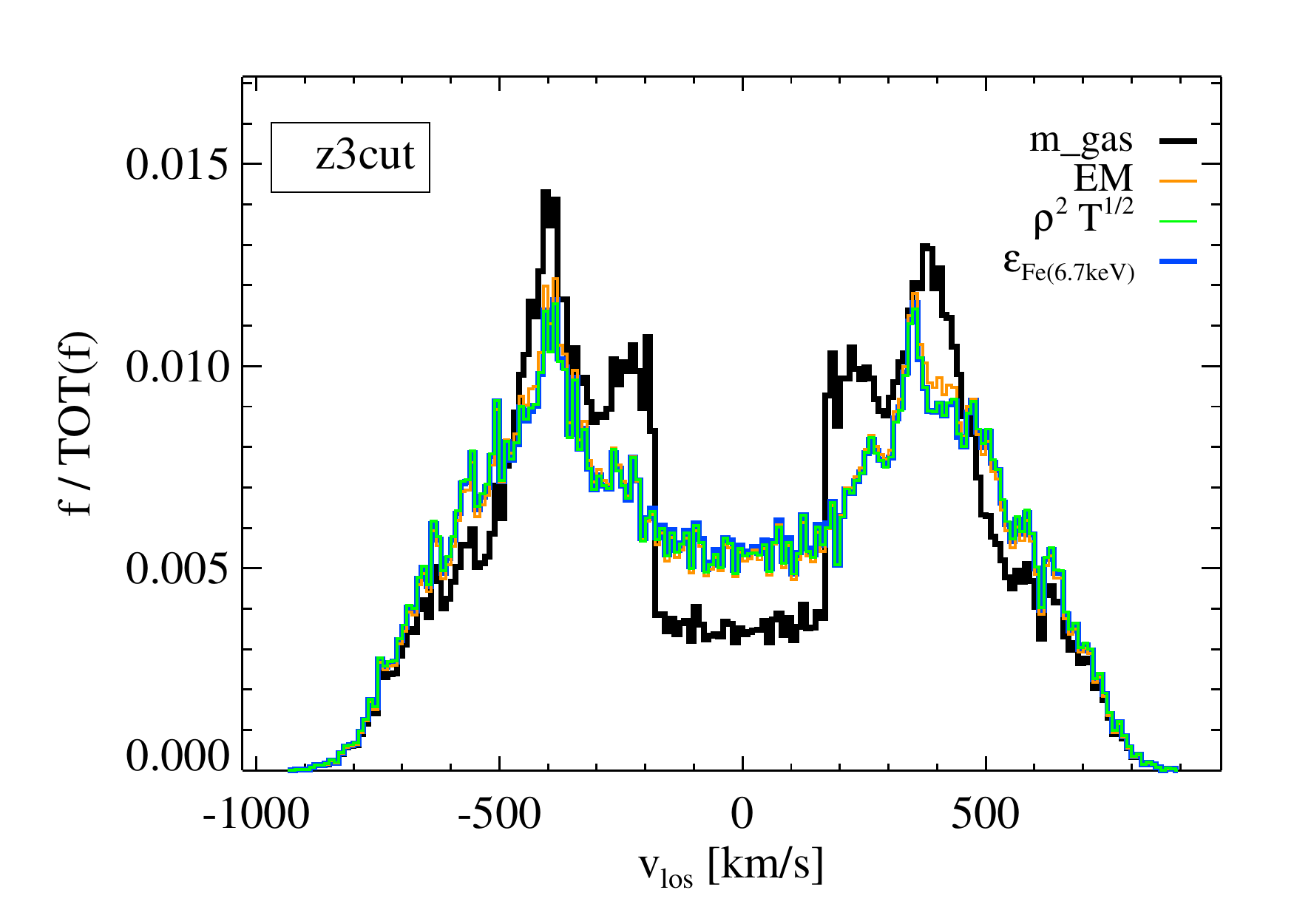}
  \includegraphics[width=0.32\textwidth,trim=100 0 20 10,clip]{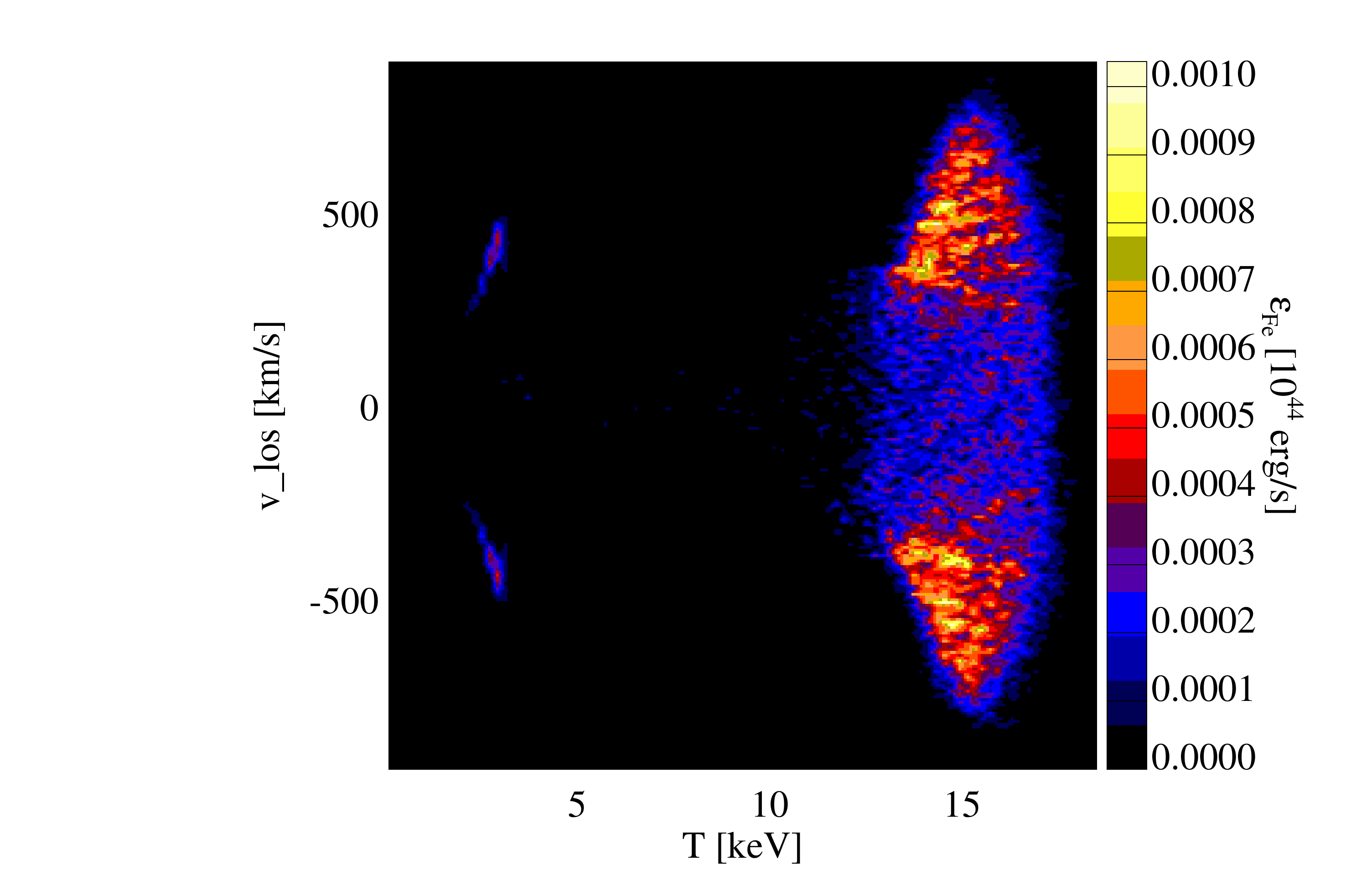}
  \includegraphics[width=0.33\textwidth,trim=20 10 10 20,clip]{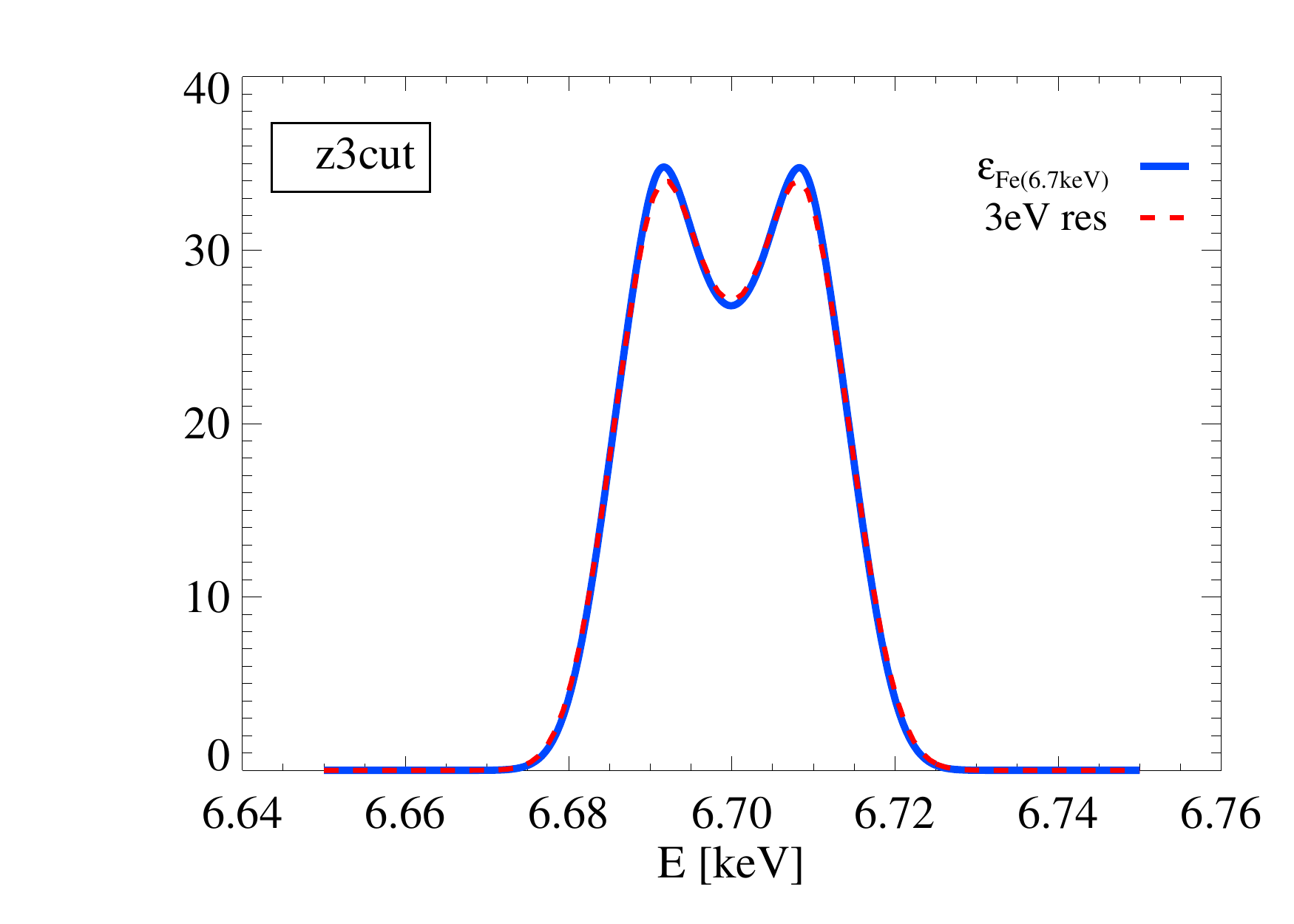}
  \\
  \includegraphics[width=0.34\textwidth,trim=0 10 20 25,clip]{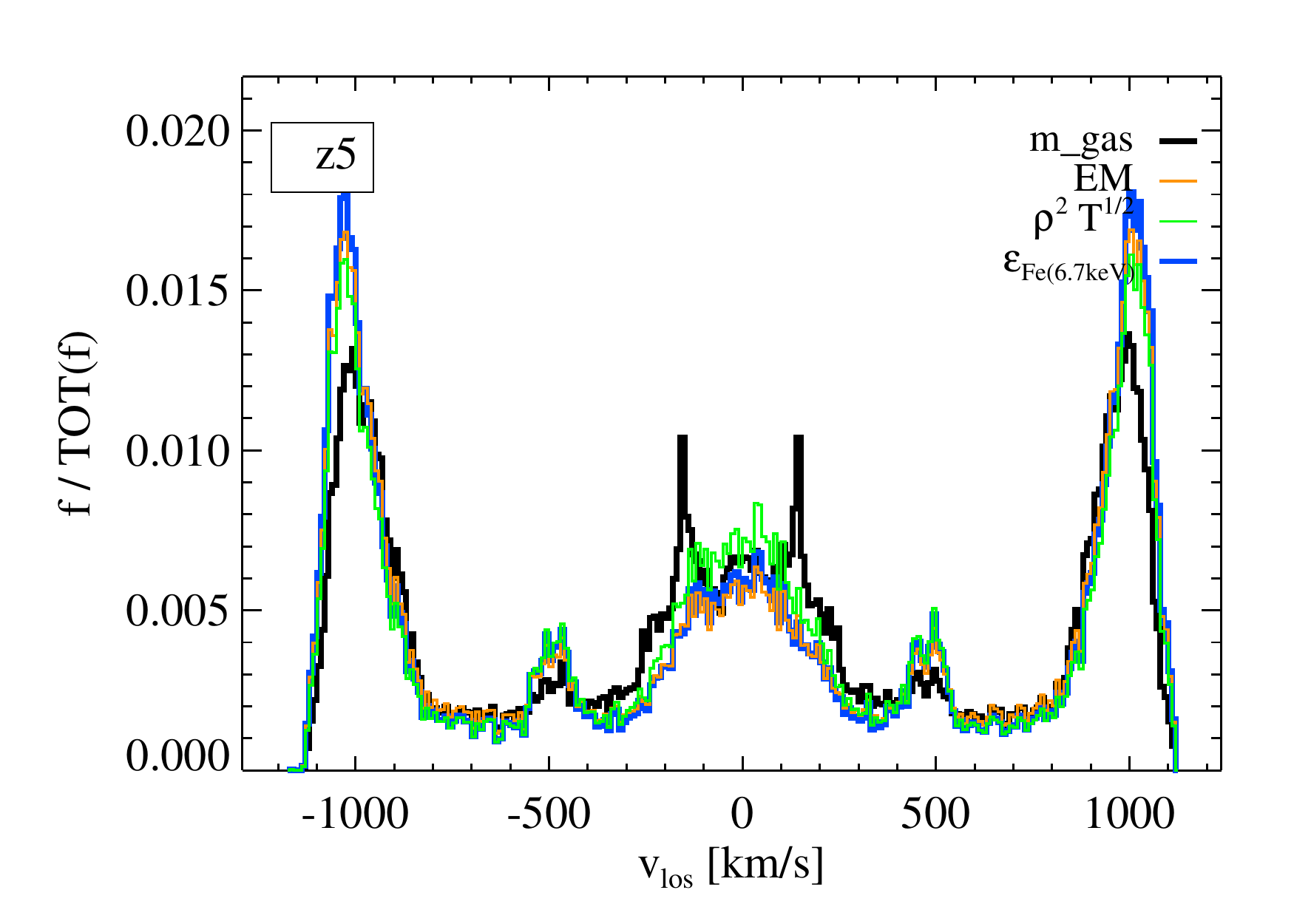}
  \includegraphics[width=0.32\textwidth,trim=100 0 20 10,clip]{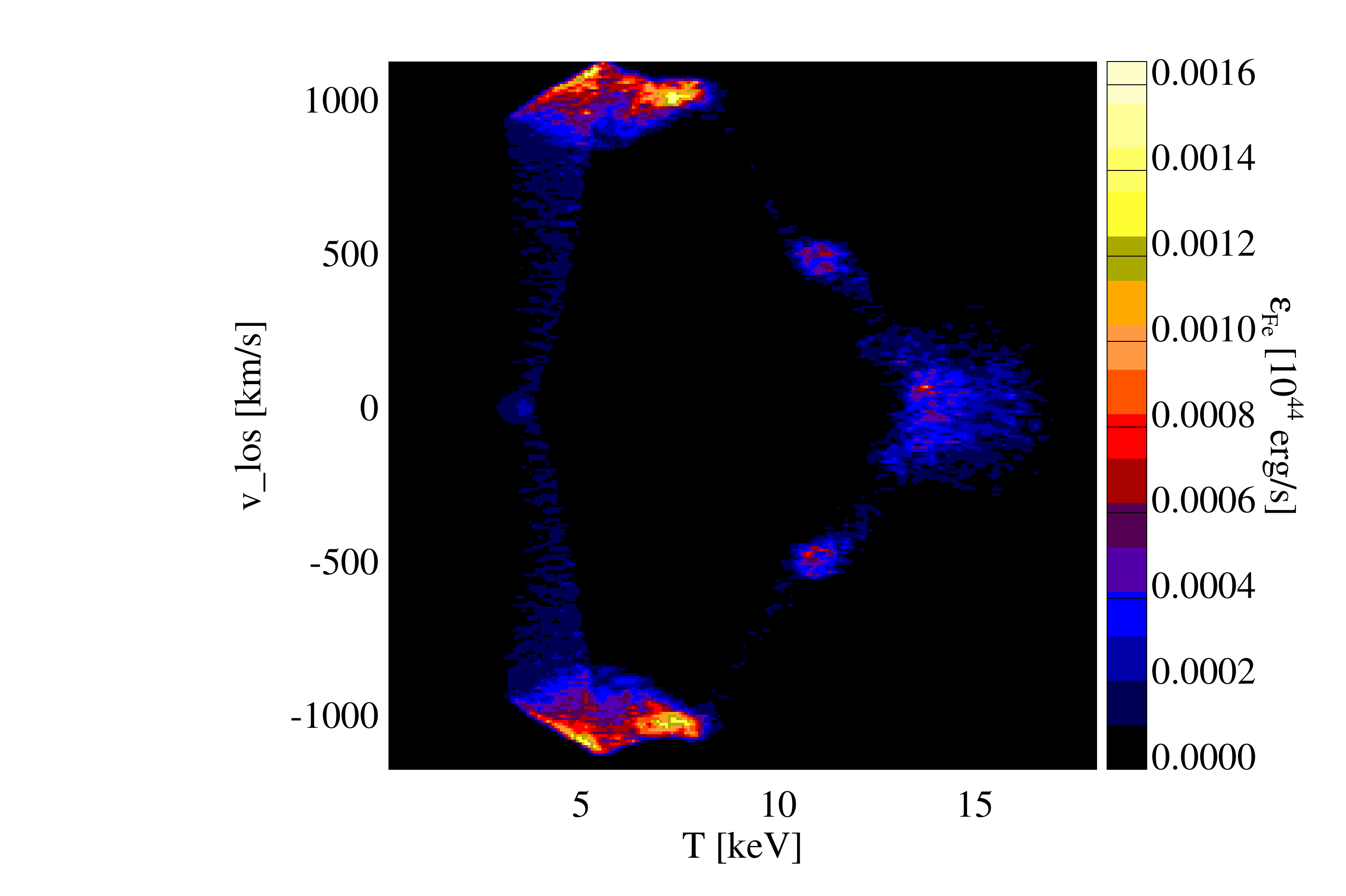}
  \includegraphics[width=0.33\textwidth,trim=20 10 10 20,clip]{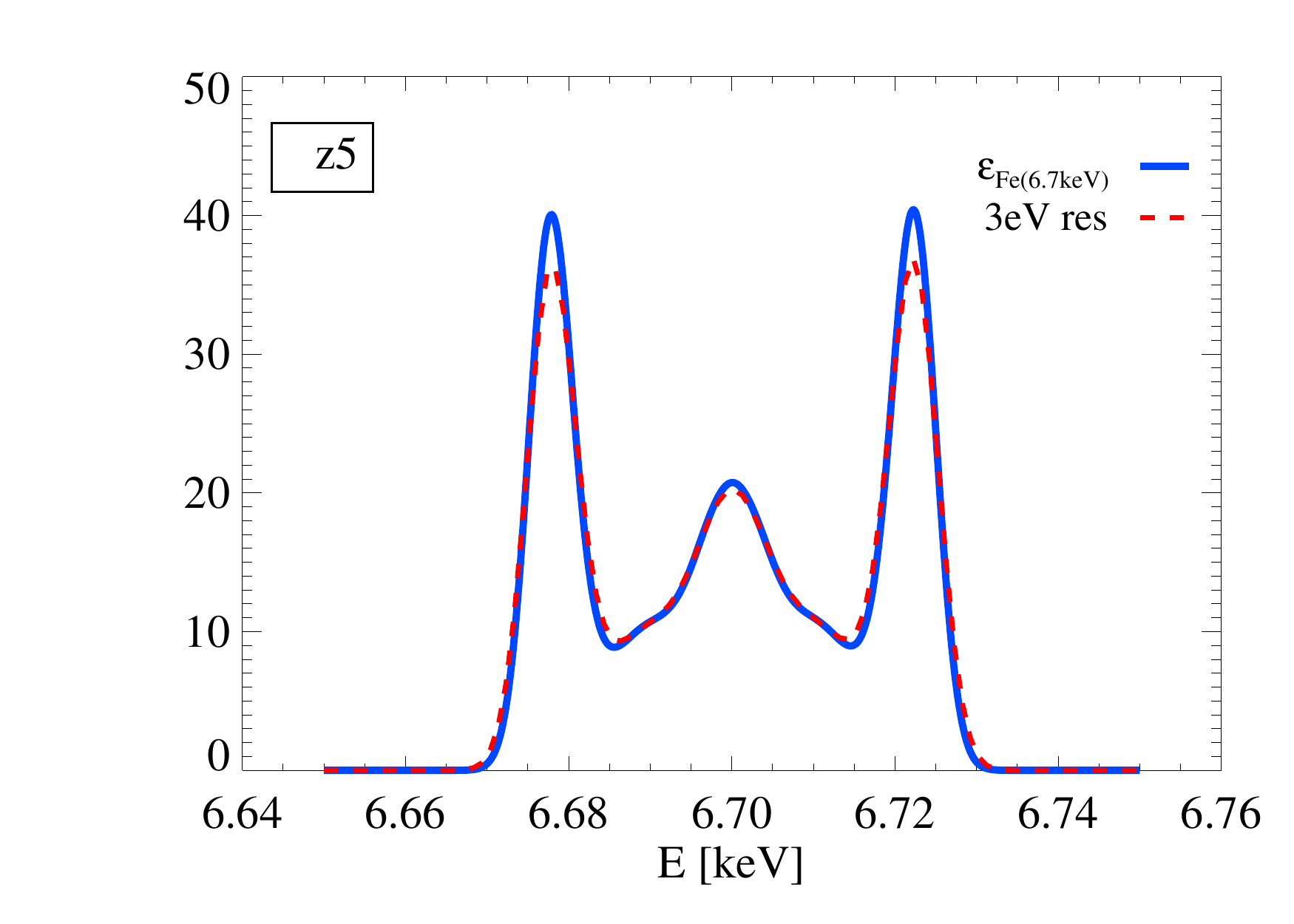}
  \caption{Velocity features of the ICM selected within the z1, z2, z3cut and z5 regions.
  Left panels: velocity distributions, in terms of fractional contribution of the gas in each velocity bin to the total in the selected region, considering different gas properties $f$: gas mass (black), \emi (orange), Bremsstrahlung-like emissivity (green) or Fe emissivity (blue).
  Middle panels: velocity-temperature diagrams, color coded by Fe emissivity. 
  Here, the iron emissivity is computed assuming an APEC model for the 6.5--7.1\,keV rest-frame energy band comprising the Fe He-like K$_{\alpha}$ complex.
  Right panels: theoretical shape of an idealized Fe He-like line centered at $6.7$\,keV, after the convolution with the gas velocity distribution (blue solid lines), and with an additional smoothing for a representative 3\,eV energy resolution (red dashed lines). The flux reported on the y-axis is in arbitrary units.
  \label{fig:z-pdfs1}}
\end{figure*}

\begin{figure*}
  \centering
  \includegraphics[width=0.34\textwidth,trim=0 10 20 25,clip]{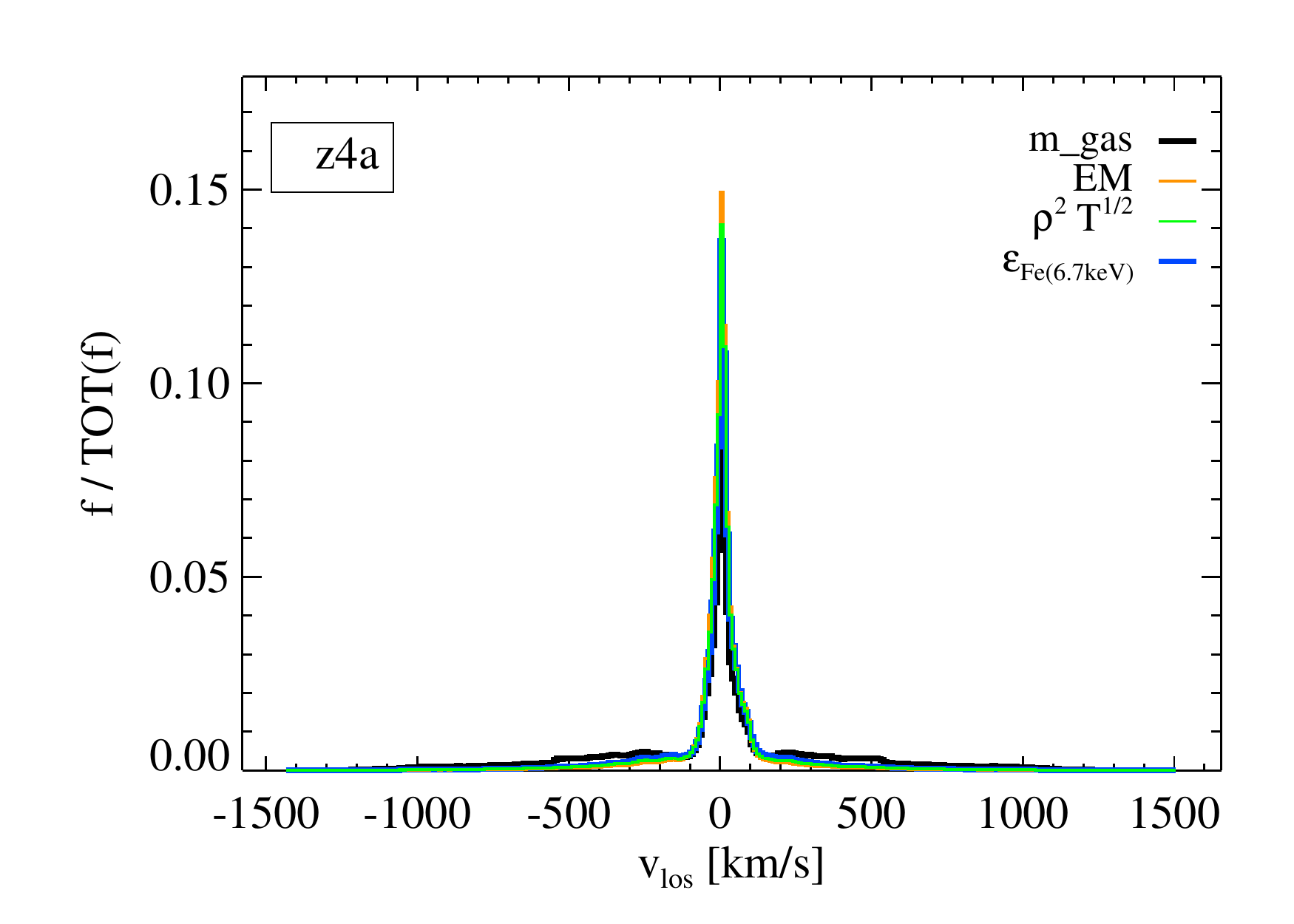}
  \includegraphics[width=0.32\textwidth,trim=100 0 25 10,clip]{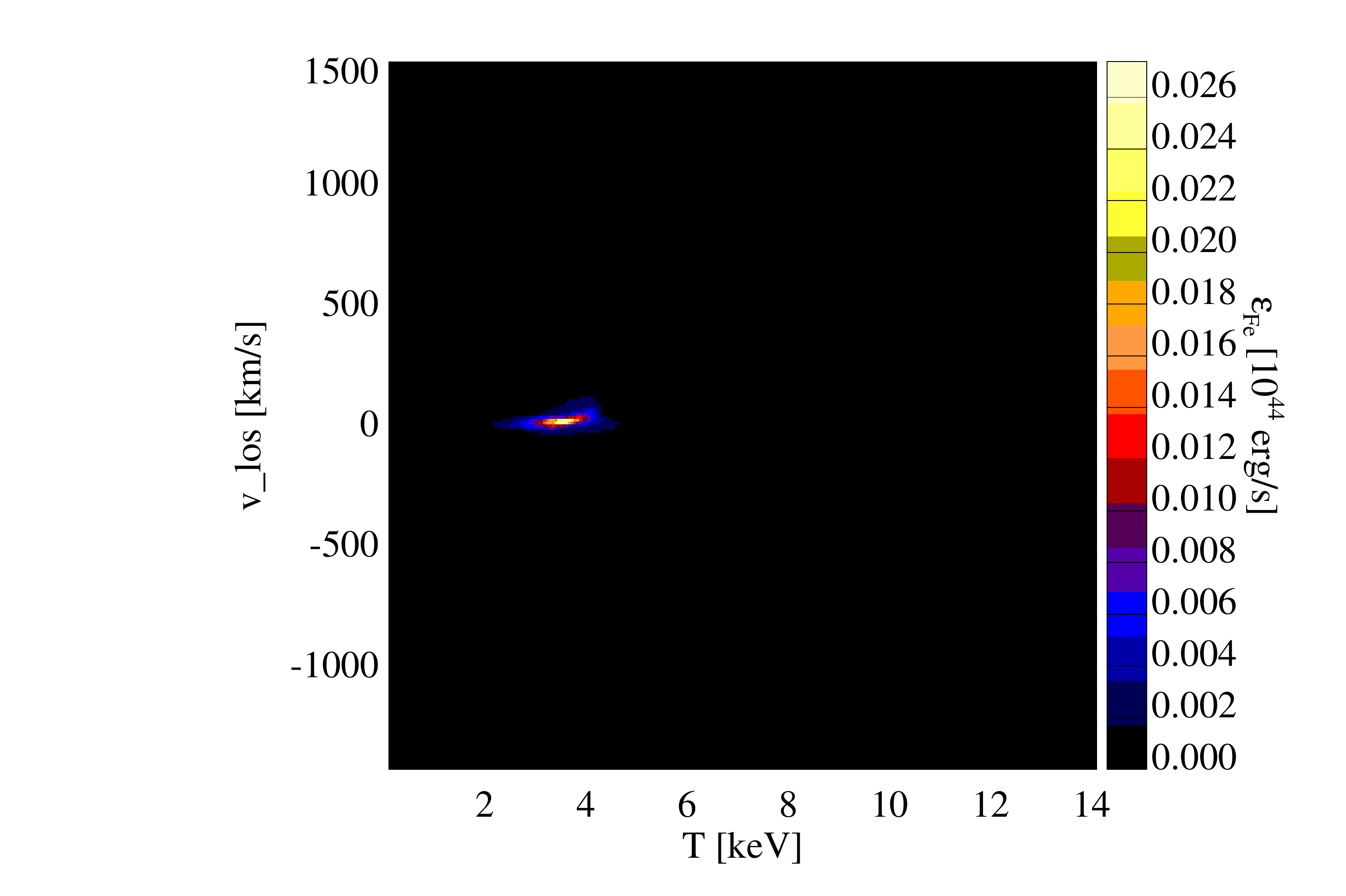}
  \includegraphics[width=0.33\textwidth,trim=20 10 10 20,clip]{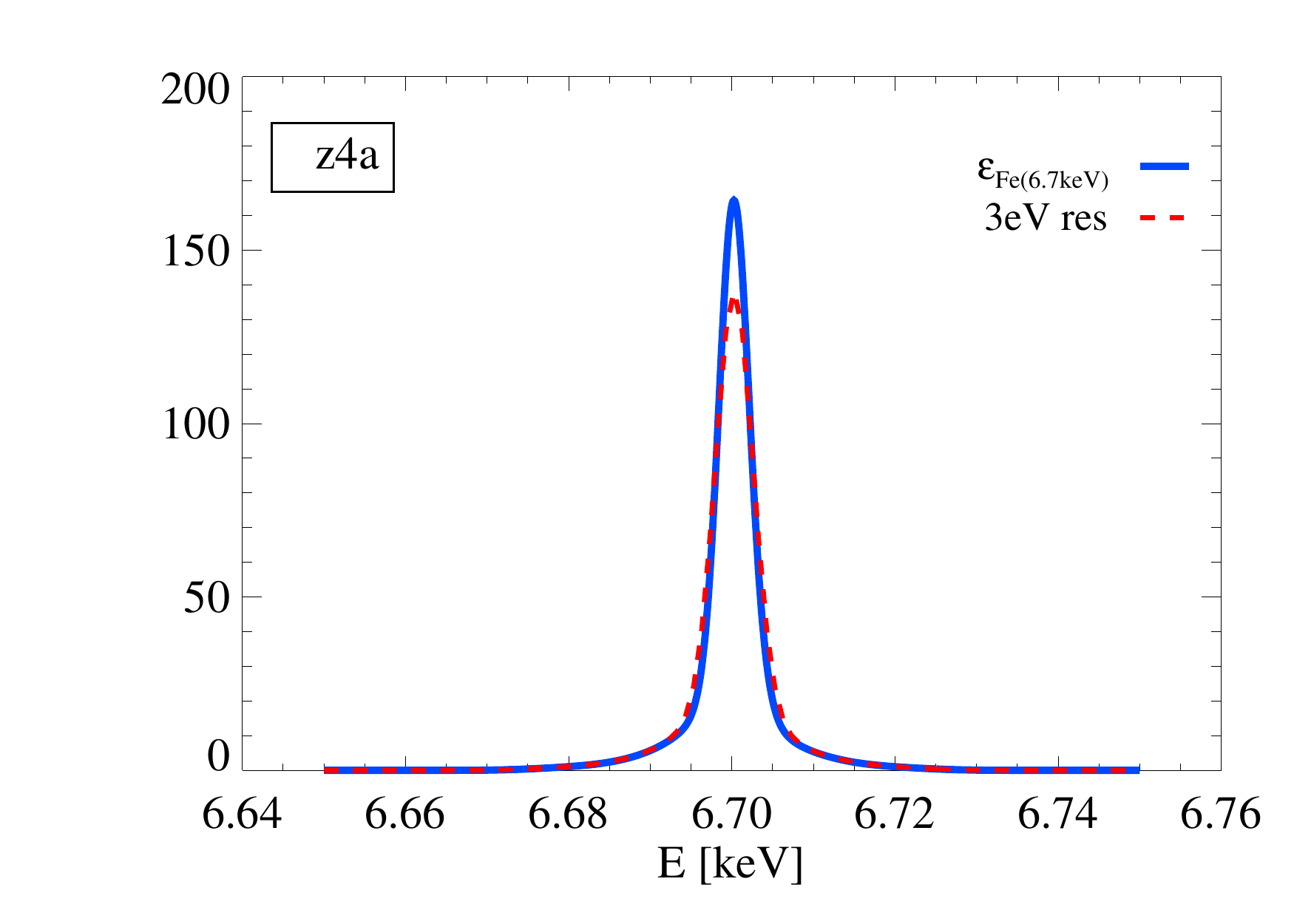}
  \\
  \includegraphics[width=0.34\textwidth,trim=0 10 20 25,clip]{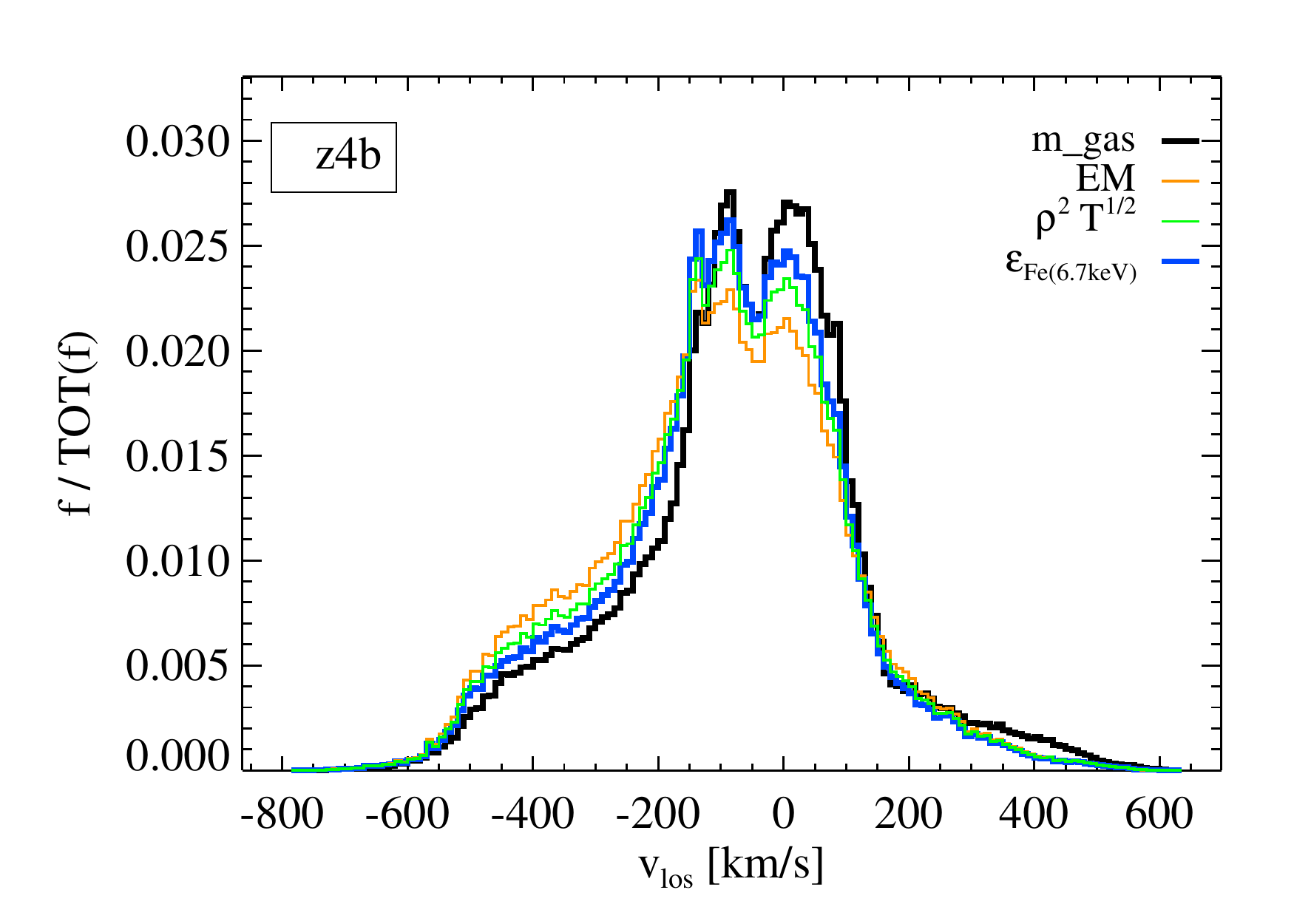}
  \includegraphics[width=0.32\textwidth,trim=100 0 25 10,clip]{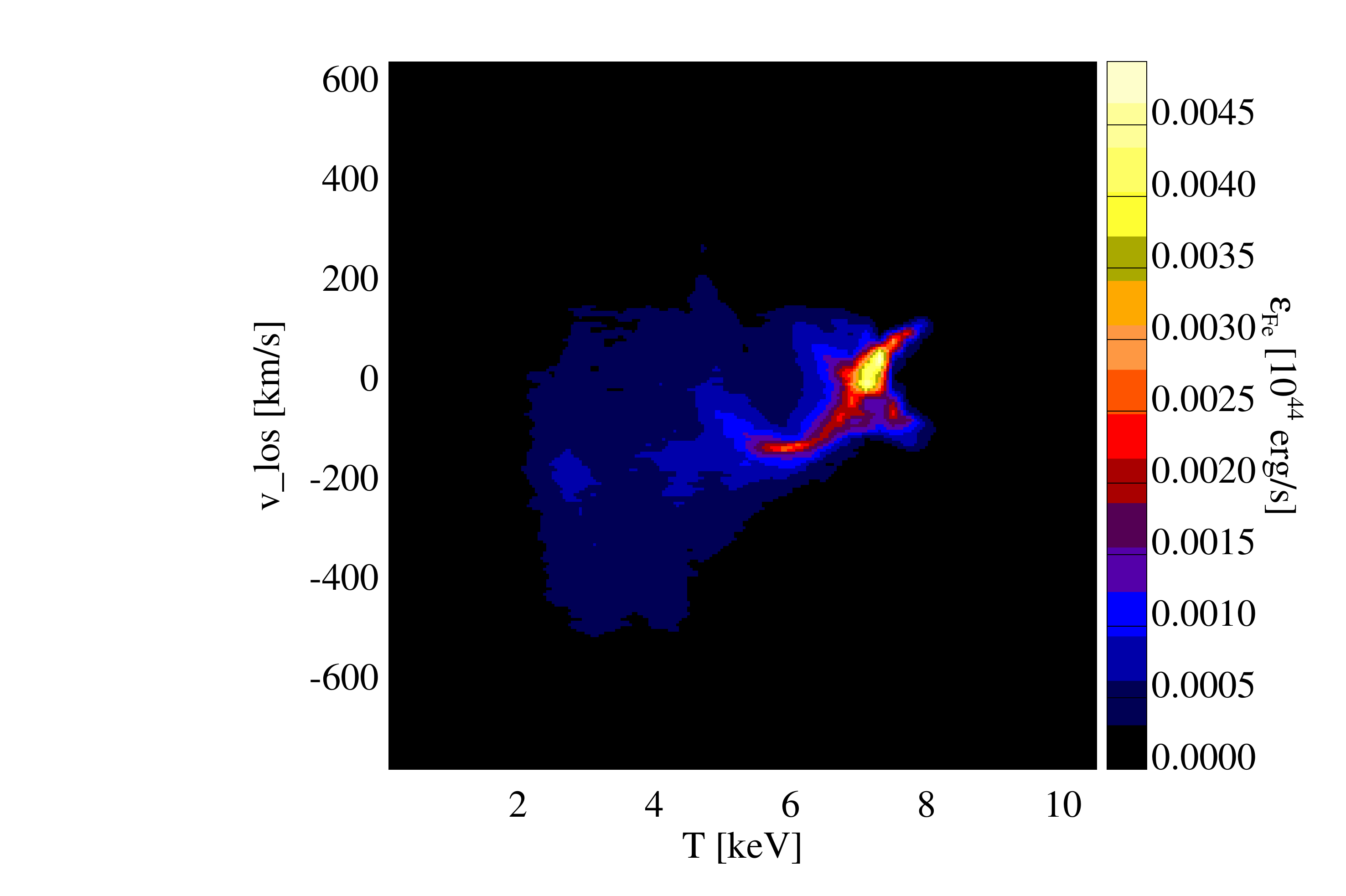}
  \includegraphics[width=0.33\textwidth,trim=20 10 10 20,clip]{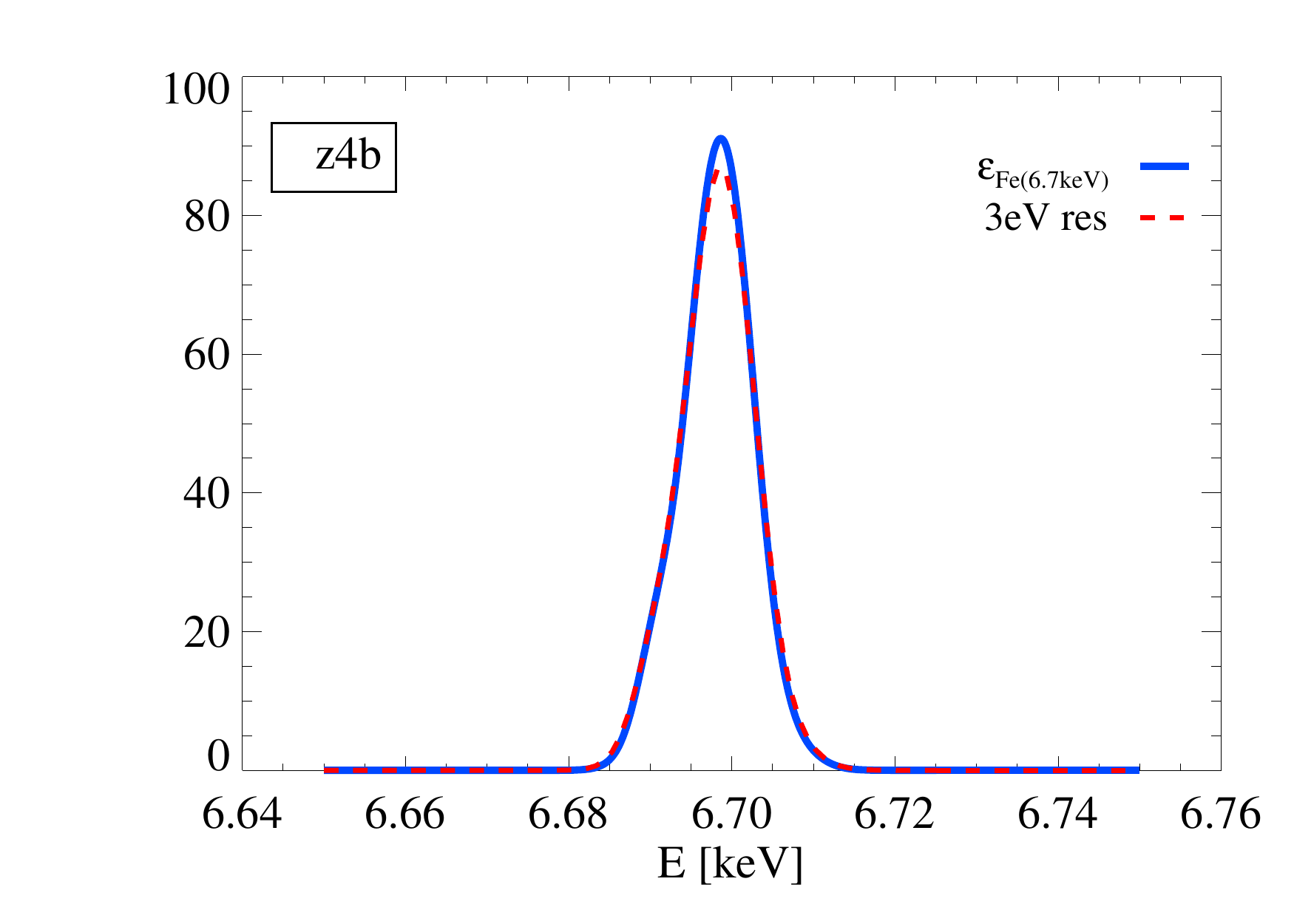}
  \\
  \includegraphics[width=0.34\textwidth,trim=0 10 20 25,clip]{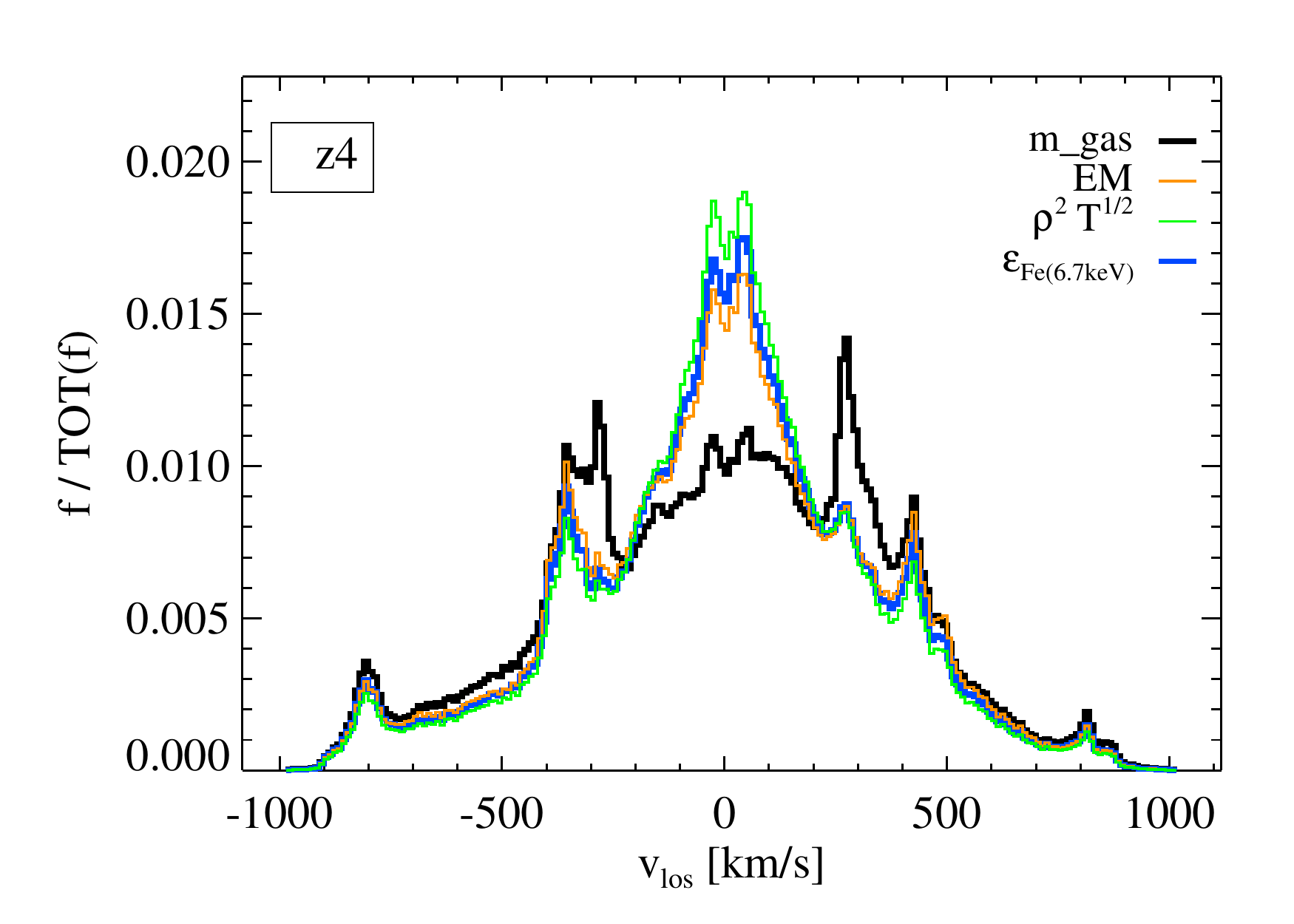}
  \includegraphics[width=0.32\textwidth,trim=100 0 25 10,clip]{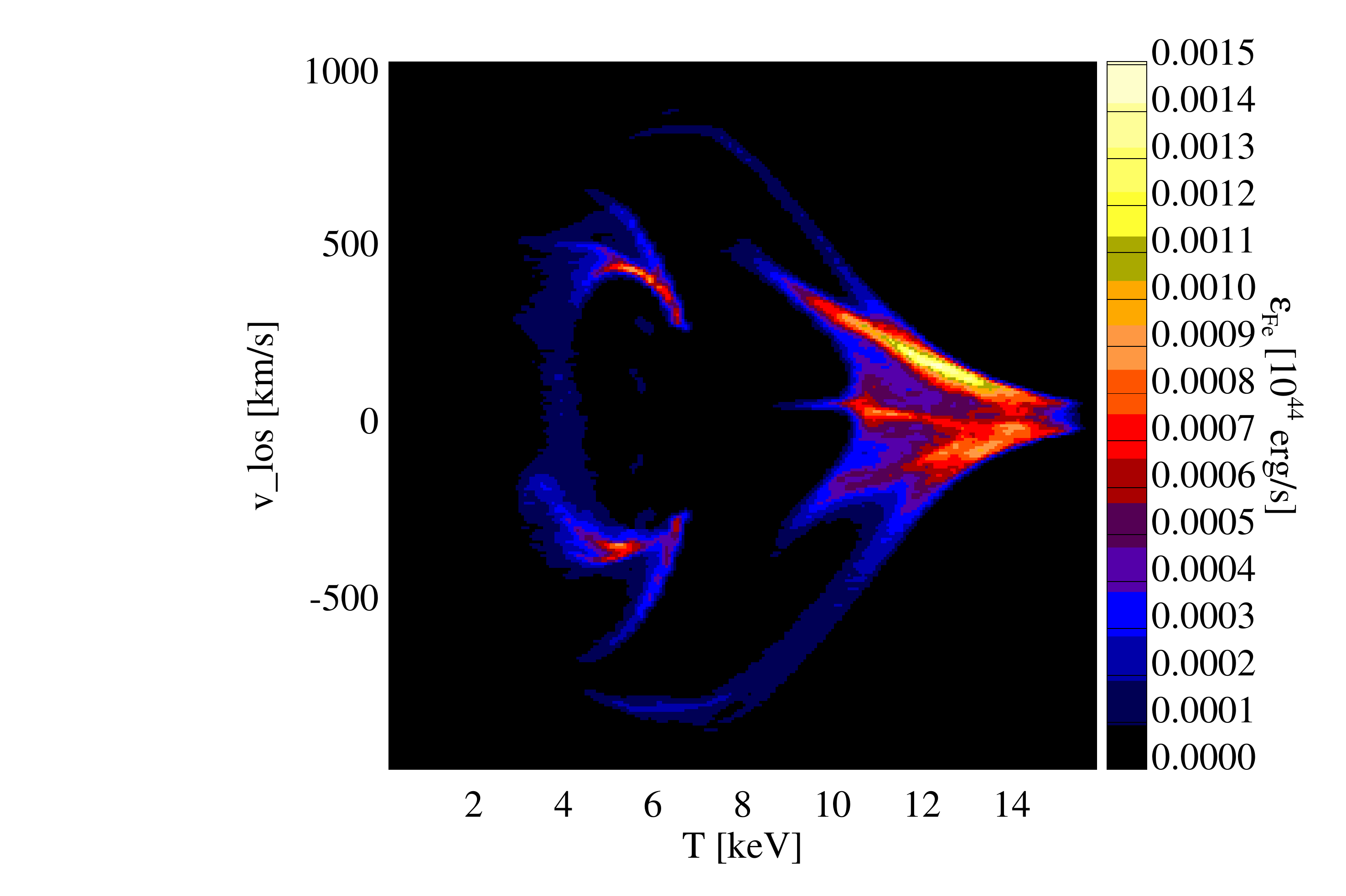}
  \includegraphics[width=0.33\textwidth,trim=20 10 10 20,clip]{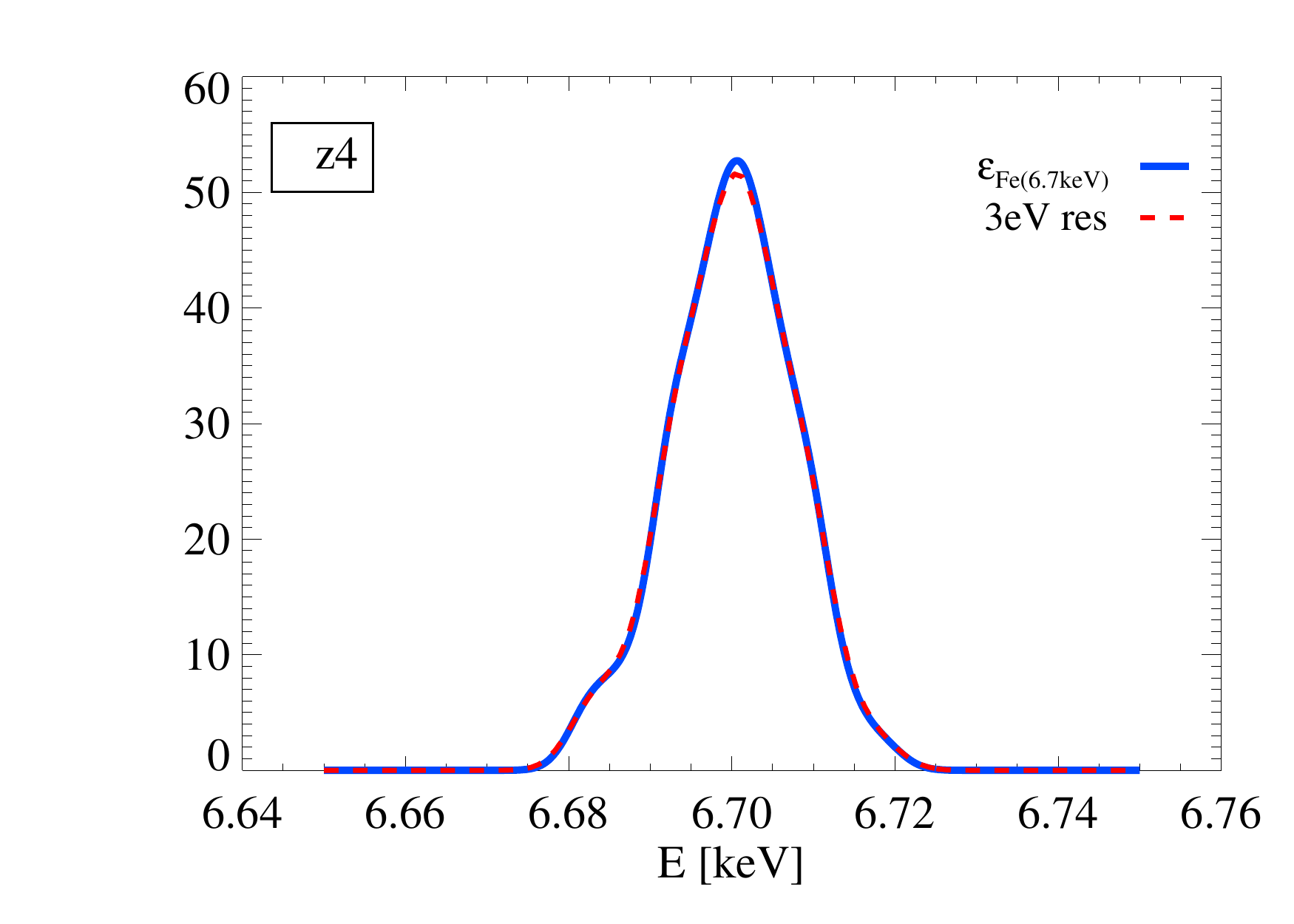}
  \caption{Same as Fig.~\ref{fig:z-pdfs1}, but for regions z4a, z4b and z4.
  \label{fig:z-pdfs2}}
\end{figure*}

For each region marked in Fig.~\ref{fig:map}, we investigate the l.o.s.\ velocity distribution of the gas directly from the simulation and relate it to the gas thermal and emission properties.
Results of this analysis are reported in Fig.~\ref{fig:z-pdfs1}, for the low-luminosity peripheral regions of the merger (i.e.\ z1, z2, z3, z5), and in Fig.~\ref{fig:z-pdfs2}, for the brighter parts of the system (i.e.\ z4, z4a, z4b).
In the left-hand-side panels of the figures, we show the distributions of the l.o.s.\ velocity component, computed directly from the gas elements in the simulation and considering different weights --- namely gas mass (black histograms), emission measure (\emi; orange), Bremsstrahlung-like emissivity (green) and emissivity of the Fe He-like iron at 6.7\,keV (blue).

Phase diagrams mapping the gas velocity and temperature distributions along the l.o.s.\ are shown in the central column. 
Here, the color coding marks the total expected emissivity of the Fe K-$\alpha$ complex (as in Sec.~\ref{sec:num_estimate}) for the gas in each velocity-temperature bin. The Fe K-$\alpha$ line is a complex of multiple lines, the most prominent of which has a rest-frame energy of $6.7$\,keV. This emission feature is an optimal target to unveil peculiar motions of the emitting plasma due to the low level of thermal broadening expected for the heavy Fe$_{\rm XXV}$ ion~\cite[e.g.][]{inogamov2003,rebusco2008}.

Additionally, we construct an idealized model of the helium-like iron complex, by assuming a single Gaussian centered on the rest-frame energy ($6.7$\,keV) of the most prominent line component and apply the thermal broadening expected for iron at the given plasma temperature. 
For each selected region, the right-hand-side plots of Fig.~\ref{fig:z-pdfs1} show the shape of the toy-model line, convolved with the gas l.o.s.\ velocity distribution (blue lines). 
We also apply an ideal energy smoothing assuming a representative energy resolution of $3$\,eV in the 6--7\,keV energy range (red dashed lines), comparable to the expected resolutions of future microcalorimeters (like those that will be on board \xrism, \athena, or \textit{Lynx}). We note that this has no significant impact on the global features of the model lines.

The l.o.s.\ velocity distribution shows the presence and relative contribution of various velocity components along the l.o.s. in each region.
For all the regions investigated, the differences between the various emission-weighted distributions and average values are not significant, so that we will mainly refer to the Fe-line emissivity case, with the ultimate goal of studying the Fe-line properties as velocity diagnostics from X-ray spectral analysis.
On the other hand, some of the interesting features of the intrinsic, mass-weighted, ICM velocity distribution are weakened or even completely absent in the emission-weighted velocity distributions.
This is the case for z1, an off-axis region located behind this shock front, where the average l.o.s.\ gas velocity dispersion is $\sigma_v\approx 400$--$500$\,km/s.
In particular, the l.o.s.\ gas-mass-weighted velocity distribution of z1 (Fig.~\ref{fig:z-pdfs1}, top row) features two separate components corresponding to the two symmetrical sides of the shock front expanding outwards in the $z$ direction, which are moving towards and away from the observer along the l.o.s.
In the case of a velocity distribution weighted by any of the emission-related weights, the two velocity components cannot be distinguished and a single broad distribution is instead observed.
This broad distribution, due to the superposition of l.o.s.\ distinct bulk motions rather than to turbulent motions, is in fact reflected in the mildly non-Gaussian shape of the ideal Fe-like emission line (top-right panel in Fig.~\ref{fig:z-pdfs1}).

Unlike z1, the other regions in Fig.~\ref{fig:z-pdfs1} are all located along the merger axis, respectively in front of (z2) and behind (z3) of the shock-front edge\footnote{In z3, we exclude a rectangular region of $33\,$kpc~$\times~200$\,kpc on the right-most end, to study the gas within the shock front and exclude the small fraction of highly-emitting gas associated with the core of the merging secondary cluster --- the cut sub-region considered and shown is referred to as z3cut.}, in regions with average velocity dispersion of about 100--200\,km/s and 400--500\,km/s respectively, and in a faint region on the opposite side with extremely high velocity dispersion (z5, $\sigma_v\approx 700$\,km/s).

Similarly to z1, the other regions in Fig.~\ref{fig:z-pdfs1}, all relatively faint, show very symmetric double-peaked velocity distributions, due to the symmetry of the merger geometry. 
In the case of z2, z3 and z5, however, these features are also present in the mass-weighted velocity distribution and reflected in the velocity-convolved Fe-line shape. Thus, for these regions, the velocity distribution cannot be approximated by a single Gaussian, whereas a Gaussian fit can be attempted for z1.

In Table~\ref{tab:xrism-z}, we report the numerical estimates of l.o.s.\ velocity and velocity dispersion computed from Eq.~\ref{eq:shift-sigma}, and compare them to the results from the Gaussian fit of the simulated velocity distributions, when available.

\begin{table*}
  \caption{Comparison between direct simulation estimates and spectral
  best-fit values of velocity shift and dispersion, 
  for the \xrism Resolve mock observations of the seven regions 
  marked in Fig.~\ref{fig:map}.
  In the last column we report the exposure time of the mock \xrism observation.
  \label{tab:xrism-z}}\vspace{3pt}
  \centering \renewcommand{\arraystretch}{1.2}
    \small \begin{tabular}{c|cc|cc|cc|c}
    \hline
    & \multicolumn{2}{c|}{Theoretical estimate} & \multicolumn{2}{c|}{Gaussian fit} & \multicolumn{3}{c}{\xrism mock --- XSPEC fit}\\[3pt]
    \hline
     & $\mu_{wFe}$ & $\sigma_{wFe}$  
             & $\mu_{1}$ & $\sigma_{1}$  
             & $\mu_{1}$ & $\sigma_{1}$ & exp
             \\
             & [km/s] & [km/s] 
             & [km/s] & [km/s] 
             & [km/s] & [km/s] & [Ms]\\[2pt]
    \hline
    z1    & -1.7  & 410.3  & -7.3   & 485.1 & $-21.5\pm 200.3$ & $580.3\pm 213.6$ & 0.5\\
    z2    & -0.1  & 220.8  & -- & -- & -- & --  & 0.5\\
    z3cut & -2.2  & 416.3  & --   & -- & -- & -- & 0.5 \\
    z4a   & 8.3   & 200.4  & 7.8    & 23.7 &  $6.4 \pm 29.8$ & $106.9 \pm 46.3$ & 0.1\\
    z4b   & -85.4 & 182.5  & -62.0  & 153.1 & $-77.9 \pm 32.0$ & $189.7 \pm 31.2$ & 0.1\\
    z4    & 3.9   & 327.0  & 20.9   & 288.3 & $-3.0 \pm 62.5$ & $380.6 \pm 60.1$ & 0.1\\
    z5    & 3.8   & 779.7  & -- & -- & -- & --  & 0.5\\
    \hline
  \end{tabular}
\end{table*}

The l.o.s.\ temperature structure of the gas can further impact the ICM X-ray emission and therefore its spectral modelling.
The idealized symmetrical origin of the velocity distributions in Fig.~\ref{fig:z-pdfs1} (left) is accompanied by a relatively simple thermal
structure, shown in the $v_{l.o.s.}-kT$ phase diagrams (central column panels), where the majority of the Fe-line emission in all cases is dominated by a single
temperature component.
This also contributes to the negligible differences among the various
emission-weighted distributions.
As we will discuss in the following section, the low surface
brightness of these regions poses a serious challenge for
deriving constraints on gas velocities from X-ray spectra.

The brighter regions z4, z4a and z4b, reported in Fig.~\ref{fig:z-pdfs2}, exhibit a single-peaked gas velocity distribution along the l.o.s. 
For the z4 and z4b regions, in particular, the velocity distributions can be reasonably approximated by a single Gaussian curve, despite the complex velocity-temperature structure of the selected gas (central column). The best-fit Gaussian $\sigma_{wFe}$ is in fairly good agreement with the velocity dispersion value computed directly from the simulation (Eq.~\ref{eq:shift-sigma}; see Table~\ref{tab:xrism-z}).  
Differently, the left-most z4a pointing shows a very narrow velocity distribution, for which the Gaussian approximation presents large residuals from the distribution wings, which extend to very large velocity values.  These extended faint wings are most likely related to the presence of the shock envelope surrounding the brighter central region, which is characterized by larger gas velocities but is too faint to contribute significantly to the region X-ray emission (see the corresponding velocity-temperature diagram in Fig.~\ref{fig:z-pdfs2}). We find indeed a significant difference between the best-fit Gaussian $\sigma_{wFe}$ ($\sim 24$\,km/s) and the value of the velocity dispersion computed directly from the simulations ($\sim 200$\,km/s), with the latter capturing the broader and fainter velocity component as well. Similar conclusions can be derived for the bulk l.o.s.\ velocity estimates (Table~\ref{tab:xrism-z}). The main features of the ICM velocity distribution in the regions in Fig.~\ref{fig:z-pdfs2} are captured by the toy-model Fe line.

The possibility to constrain these l.o.s.\ velocity distributions via high-resolution X-ray spectroscopy
will be explored in the following via synthetic X-ray observations, assuming the characteristics of next-generation X-ray microcalorimeters, such as \xrism Resolve or \athena X-IFU.

\subsubsection{\xrism Resolve mock observations}\label{sec:z-xrism}

Our SOXS-based model for the \xrism Resolve soft X-ray spectrometer has a FoV of about $2.9' \times 2.9'$ ($6\times 6$ pixels of $30''$ each), a Gaussian PSF with a FWHM of $1.3'$, and a RMF with $5$\,eV spectral resolution. The requirement for the effective area is of $\geq 160(210)\,{\rm cm}^2$ at $1(6)\,$keV.
In this section, we present results on ICM velocity diagnostics
from mock \xrism Resolve observations of the pointings marked in Fig.~\ref{fig:map}, whose size ($200$\,kpc per side) approximates the \xrism FoV at the fiducial redshift of $z=0.057$.
In all cases no background component is included. 
For bright sources, like the z4, z4a and z4b regions, the background (both physical and instrumental) is in fact subdominant at the energies where the spectral model is constrained~\cite[i.e. in the Fe-K complex band; see][]{hitomi2018,xrism2020}.
For z1, z2, z3, and z5, given the very faint emission, the background level would be definitely important to consider. Since the signal is so low even in the ideal case with no background, it is not relevant to proceed further and study the complete mock including the background for these particular regions.

\begin{figure*}
  \centering
  \includegraphics[width=0.43\textwidth,trim=0 0 0 10,clip]{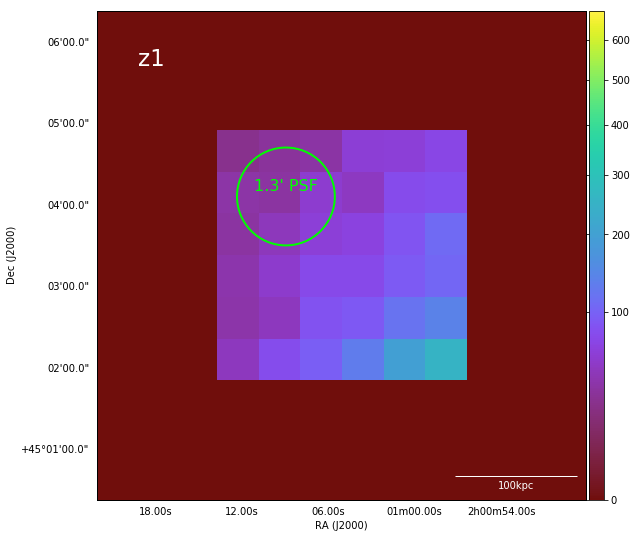}\qquad
  \includegraphics[width=0.43\textwidth,trim=0 0 0 10,clip]{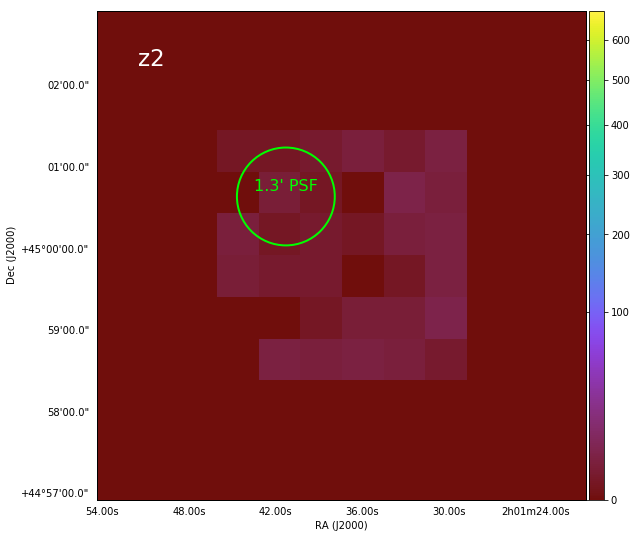}\\[5pt]
  \includegraphics[width=0.43\textwidth,trim=0 0 0 10,clip]{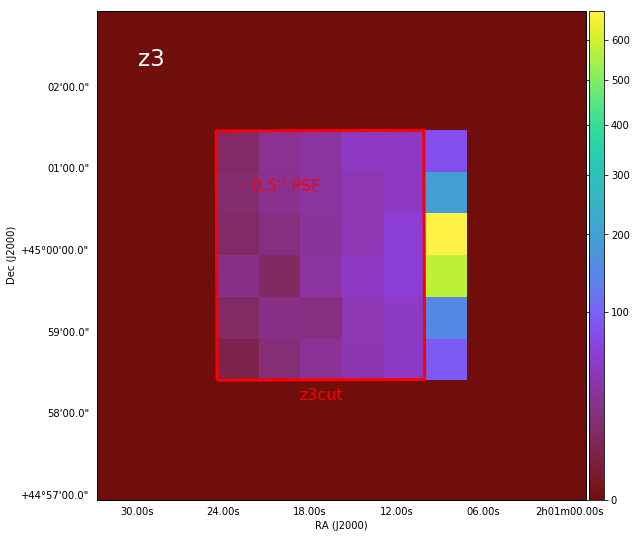}\qquad
  \includegraphics[width=0.43\textwidth,trim=0 0 0 10,clip]{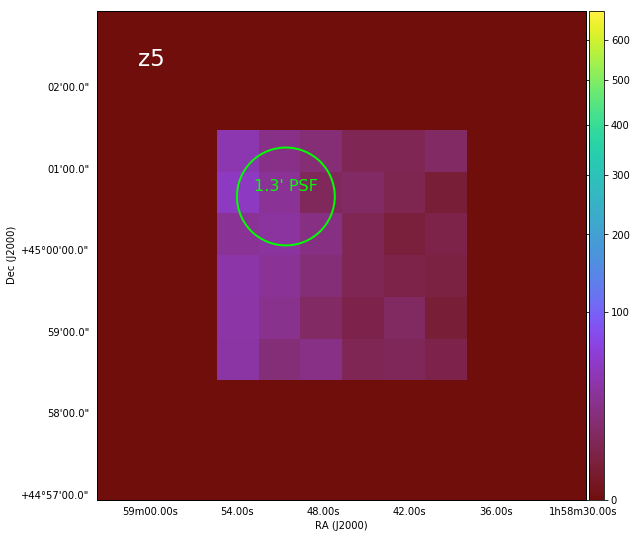}
  \caption{\xrism Resolve images ($0.5$--$2$\,keV band) for the faint pointings shown in Fig.~\ref{fig:z-pdfs1}: z1, z2, z3 and z5, from top-left to bottom-right. The images refer to a $500$\,ks esposure observation. 
  The physical size of each pointing is $200$\,kpc per side ($\sim 3'$ at the assumed redshift $z=0.057$) and the \xrism Resolve PSF is $1.3'$. A physical size of $100$\,kpc is marked for comparison on the z1 image.
  The z3 image has been derived for an artificially high spatial resolution (adopting an ideal \textit{Chandra}-like PSF of $0.5"$) and we mark in red the sub-region used to extract the spectrum. \label{fig:xrism_faint}}
\end{figure*}

\begin{figure*}
  \centering
  \includegraphics[width=0.37\textwidth,trim=0 0 0 10,clip]{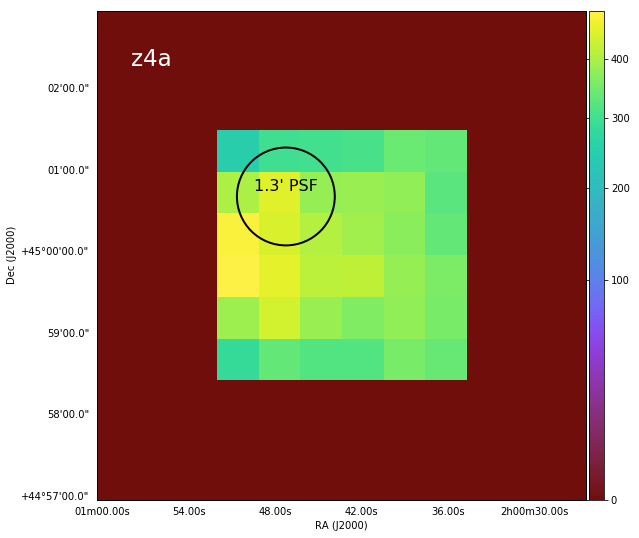}
  \includegraphics[width=0.45\textwidth,trim=0 40 40 50,clip]{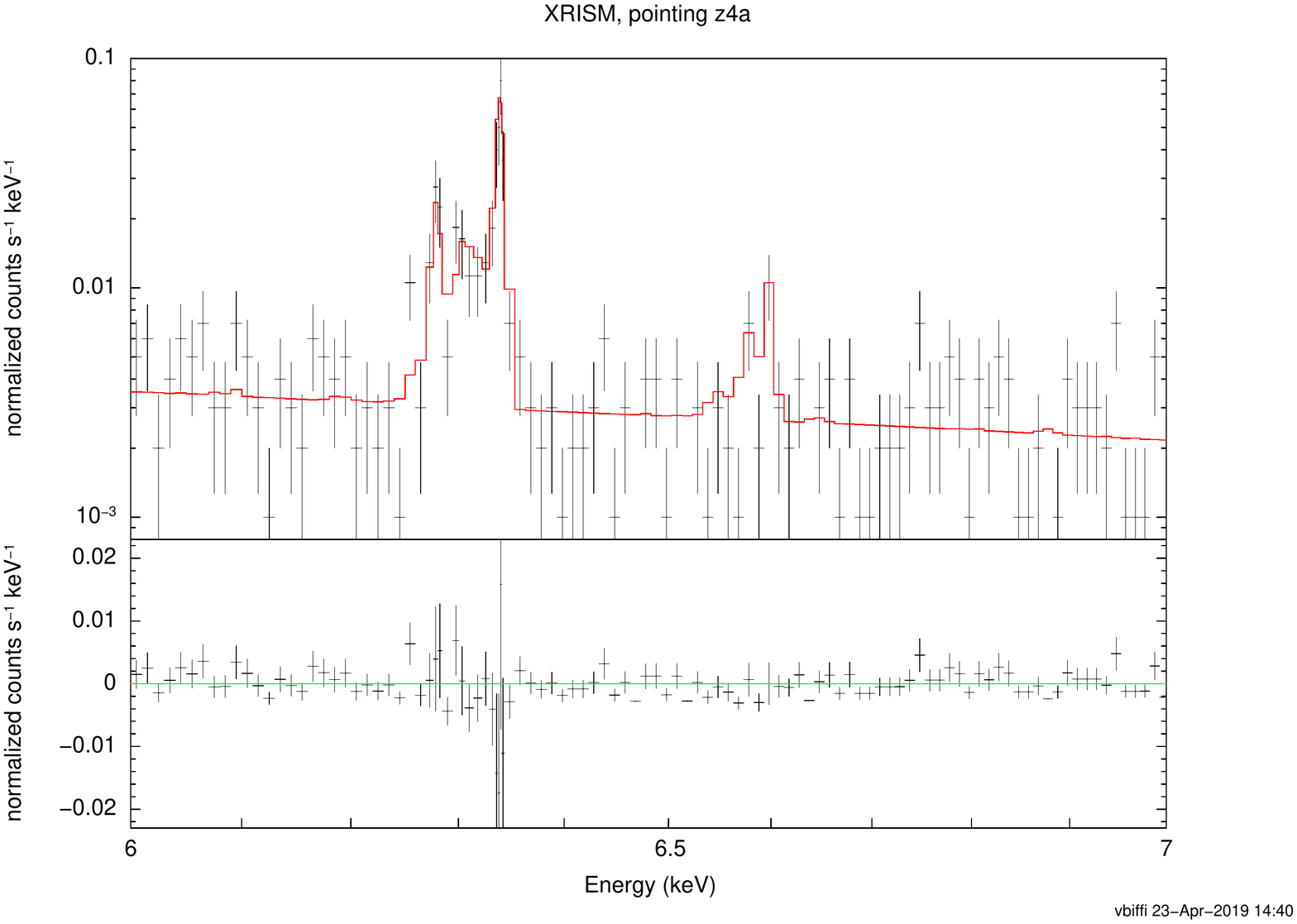}
  \\[3pt]
  \includegraphics[width=0.37\textwidth,trim=0 0 0 10,clip]{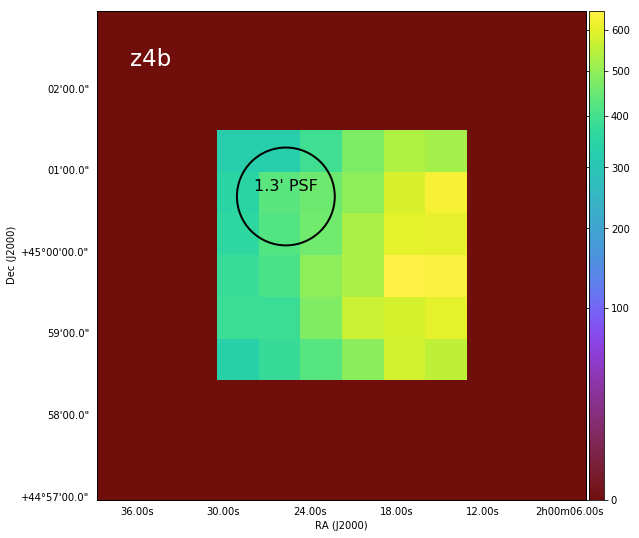}
  \includegraphics[width=0.45\textwidth,trim=0 40 40 50,clip]{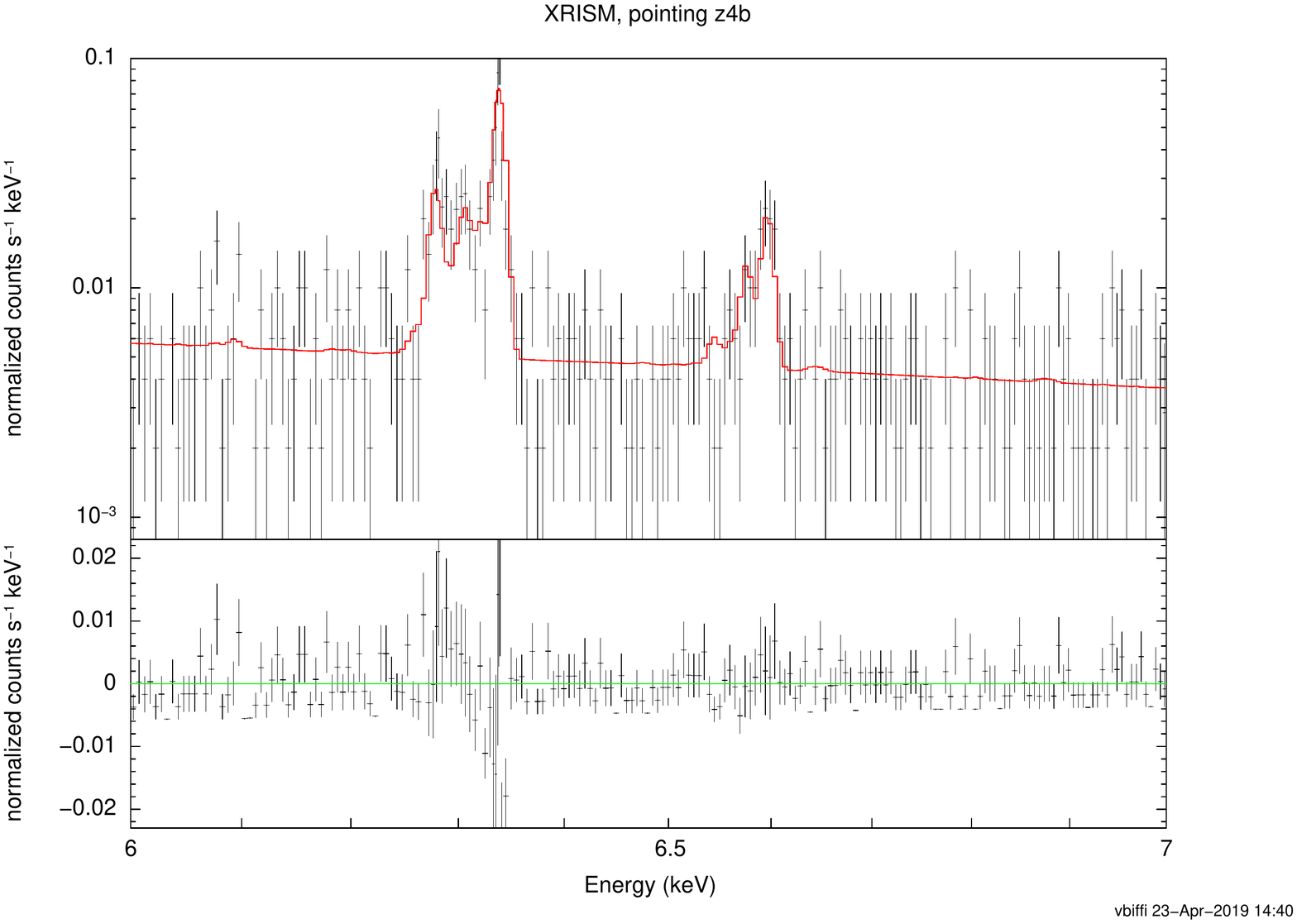}
  \\[3pt]
  \includegraphics[width=0.37\textwidth,trim=0 0 0 10,clip]{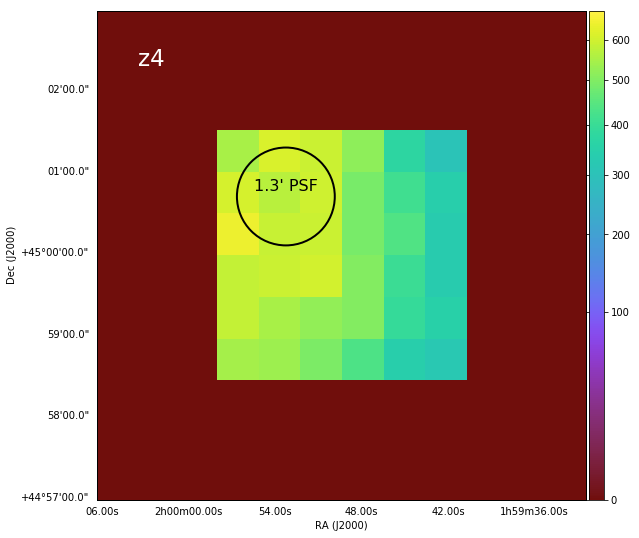}
  \includegraphics[width=0.45\textwidth,trim=0 40 40 50,clip]{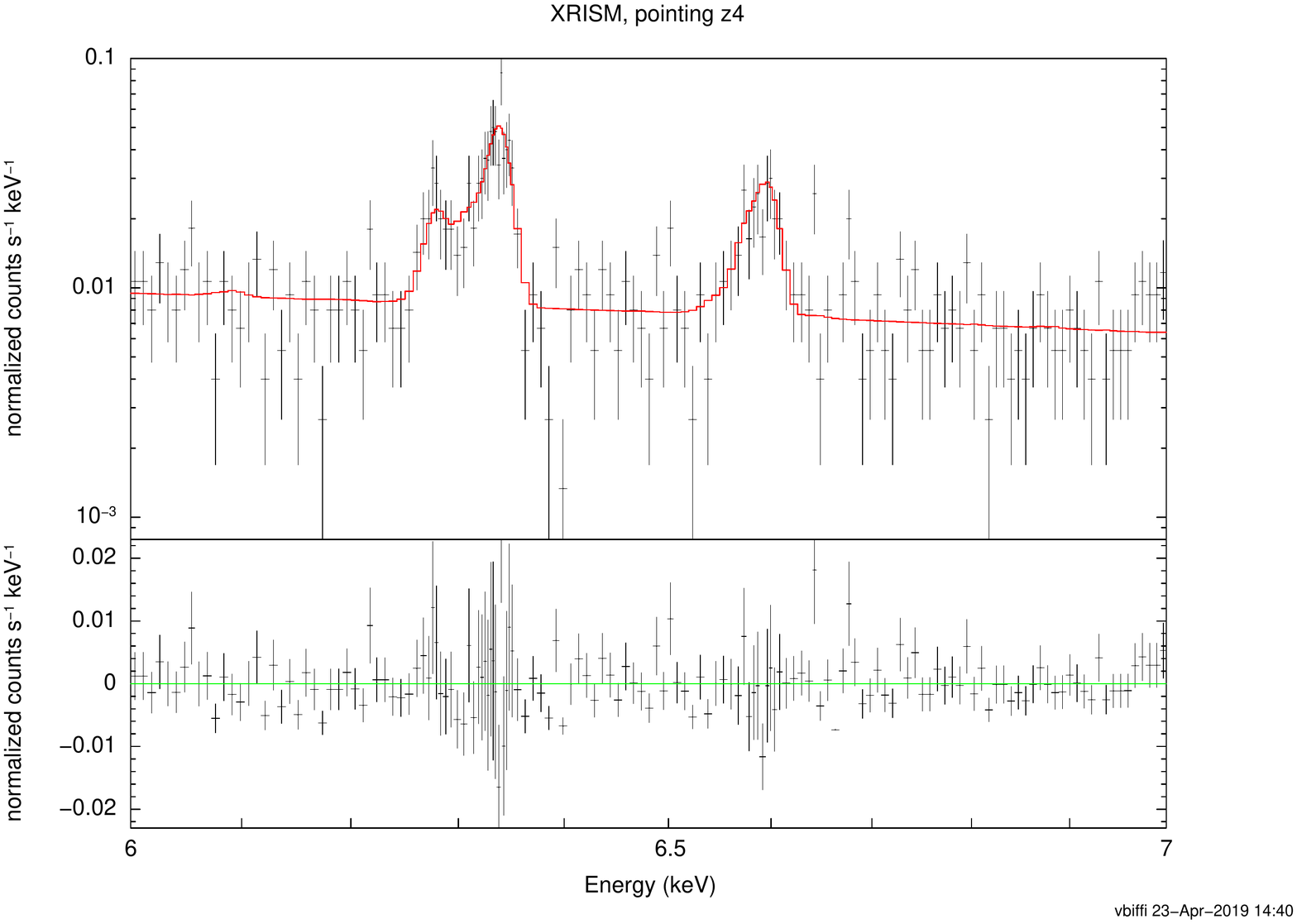}
  \caption{{\it Left:} \xrism Resolve images ($0.5$--$2$\,keV band) for the bright z4a, z4b and z4 pointings shown in Fig.~\ref{fig:z-pdfs2}. 
  The images refer to a 100\,ks exposure observation. 
  The physical size of each pointing is $200$\,kpc per side (as marked in the z1 image, Fig.~\ref{fig:xrism_faint}) and the \xrism Resolve PSF is $1.3'$.
  {\rm Right:} corresponding spectra, best-fit models and residuals around the Fe-K lines (6--7\,keV energy band) for the z4a, z4b and z4 pointings. \label{fig:xrism_bright}}
\end{figure*}

\paragraph{Constraints from spectral analysis.}

The mock \xrism Resolve images for regions 
z1, z2, z3 and z5 are shown in
Fig.~\ref{fig:xrism_faint}. 
As already noticeable from the idealized flux map 
(top-left panel in Fig.~\ref{fig:map}), these
pointings are located in faint regions of
the system, and we adopt therefore an exposure of
500\,ks.
Even so, only z1 has about 200 counts in the 6--7\,keV band, 
whereas the others have much lower number counts 
(fewer than 30 photons in the
case of z2 and z5). This is not sufficient to perform a spectral fit of the Fe-K complex and reliably constrain the velocity broadening, especially given the complex line structures shown in Fig.~\ref{fig:z-pdfs1}.  
Also, the emission of z3 is contaminated by the bright core falling in the rightmost part of the FoV, especially since the $\sim 1'$ PSF scatters photons originating from parts of the core which do not even fall on the Resolve FoV. 

For this reason, we treat this pointing
differently: we generate an ideal mock observation where the PSF is
artificially set to an optimally low value (with a FWHM of $0.5''$) and we mask the
core area (rightmost pixel column), while we extract 
the spectrum from the ``z3cut'' region (marked by the red box, in the lower-left panel of Fig.~\ref{fig:xrism_faint}). 
By adopting the \textit{Chandra}-like optimal spatial resolution we minimize the contamination (about 25\%) due to
photons scattered from surrounding regions and especially from the
bright core due to the wide \xrism PSF. Instead, by extracting the spectrum from the z3cut sub-region and therefore
excluding the core emission
(which contributes 30--40\% of the total photon
counts), we specifically sample the region on the shock front. 
The spectrum obtained in this way has nevertheless
only $\approx 80(160)$ counts in the 6--7\,keV region for the
500\,ks(1\,Ms) exposure.  Since we have found that $>200$ counts are required to place stringent constraints on
the velocity shift and broadening of the Fe-K line, much longer exposures ($>1$\,Ms) would be required.

The challenge of constraining the velocity broadening from the iron K$\alpha$ complex at $\sim$6.7~keV for these faint pointings, due to the poor count statistics, is consistent with previous studies. 
Predictions for \textit{Astro-H}~(\textit{Hitomi}) indicate in fact that higher number of photons, of order $10^3$--$10^4$, are required to accurately model the iron complex shape with multiple velocity components~\cite[e.g.][]{shang2012,zuhone2016b}. In Table~\ref{tab:xrism-z} (three rightmost columns) we therefore only report the best-fit results for the z1 mock spectrum, which are consistent with the numerical simulation estimates within the uncertainties.

A different picture is given by the bright pointings located on the central axis of the merging system, z4a, z4b and z4, for which \xrism Resolve images and spectra 
(along with best-fit models and residuals) are
shown in Fig.~\ref{fig:xrism_bright}.  
With 100\,ks exposure, we obtain significantly larger number counts in the \xrism Resolve spectra, ranging between $\approx400$ and $\approx1000$ photons in the 6--7\,keV band around the Fe-K complex.
This allows us to fit an APEC model with line broadening (BAPEC)
to the line, in the 6--7\,keV band, for which best-fit
results for velocity shift and broadening are reported in Table~\ref{tab:xrism-z}.

The spectral best-fit values reported in Table~\ref{tab:xrism-z} are consistent with the numerical estimates computed directly from the simulation, within the 1-$\sigma$ uncertainties. The difference in the velocity shifts, in particular,
are limited to a factor of few-to-ten km/s.
The only exception is the case of z4a, for which the spectral fit provides a lower velocity dispersion value compared to the numerical estimate, despite still being consistent within $\sim 2\,\sigma$. This was also noted for the Gaussian-fit estimate of the velocity broadening, which is consistently much lower, and related to the shape of the velocity distribution for z4a (Fig.~\ref{fig:z-pdfs2}, upper-left panel), composed by the superposition of a very narrow central component and fainter broader wings.

Since the spectral fit of the iron complex also assumes a Gaussian distribution of the velocities, any discrepancy between the spectral best-fit values and those estimated directly from the simulation indicate that the Gaussian approximation cannot capture correctly the complexity of the velocity distribution and the presence of multiple l.o.s.\ velocity components.  

\paragraph{1000 mock realizations.}

We note that the spectral best-fit values reported in Table~\ref{tab:xrism-z}, for all the \xrism Resolve pointings, correspond to average values computed over 1000 realizations for each pointing of the mock observations.
This is done to minimize the impact of low-number statistics,
related to the random sampling of model emission spectra with low number of counts
while generating the mock observations.
From the analysis of the 1000 synthetic spectra, we then compute the velocity shift and broadening estimates, and their uncertainties, as the median and robust dispersion of the distributions of XSPEC best-fit values.
Median and robust dispersion provide reliable estimates against outliers, therefore mitigating the impact of poor spectra for which the fit does not converge.

\subsubsection{\athena X-IFU mock observations}\label{sec:z-athena}

In this section we further investigate the possibility to constrain the ICM velocity field through high-resolution X-ray spectroscopy, in regions
characterised by low surface brightness.
To this scope, we employ the characteristics of the X-IFU spectrometer, planned to be on-board the ESA \athena satellite and designed to grant a much larger collecting area ($A_{\rm eff}\sim 15000(1600)\,{\rm cm}^2$ at $1(7)$\,keV), while still securing a very fine spectral resolution ($\sim 2.5$\,eV at $E < 7$\,keV) and an angular resolution of $\sim 5''$~\cite[][]{xifu2018}.

\begin{figure}
  \centering
\includegraphics[width=0.45\textwidth]{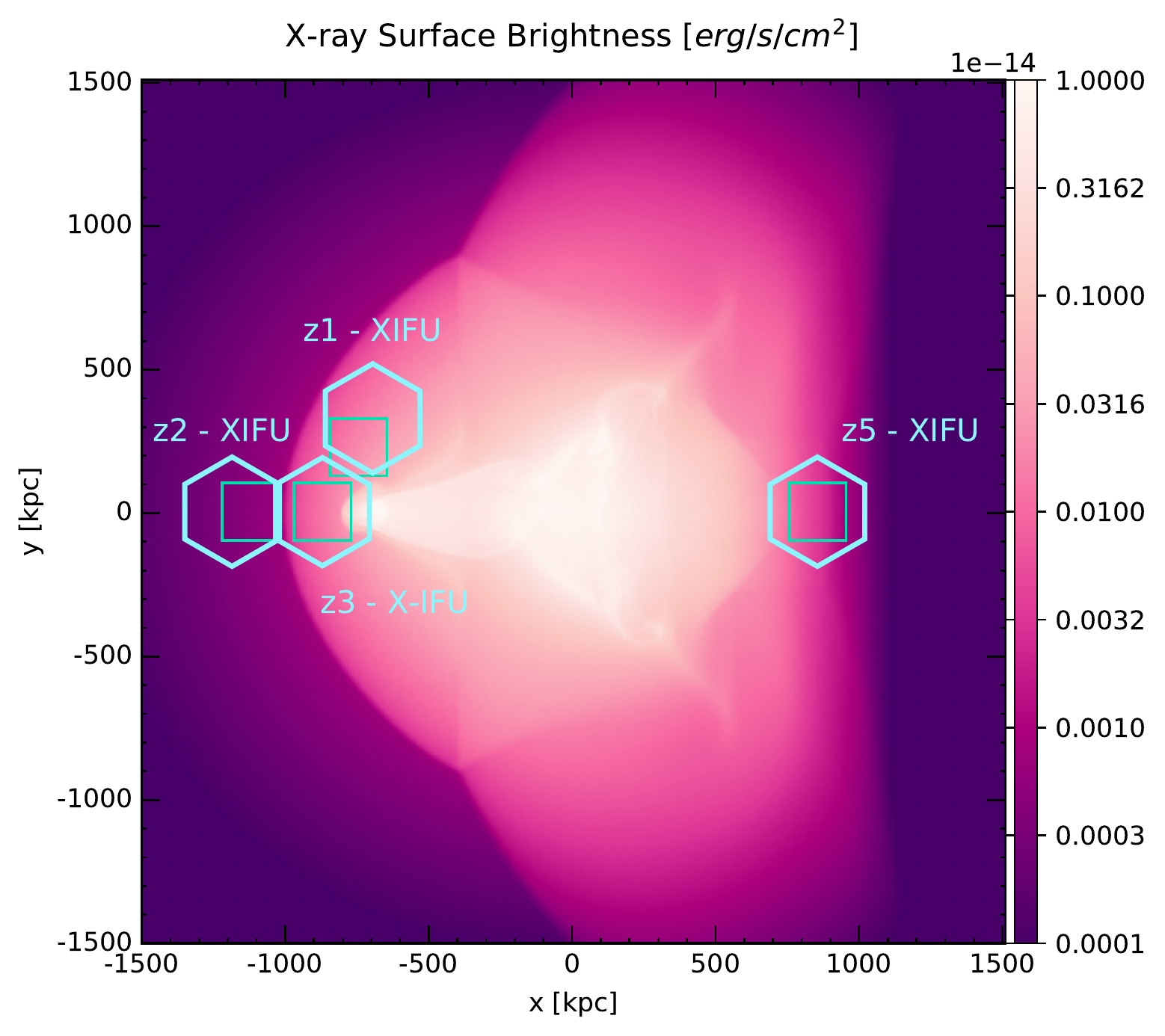}
  \caption{Simulation map of the X-ray flux in the $0.5$--$7$\,keV band, same as in Fig.~\protect\ref{fig:map}. Overplotted, the 4 hexagonal X-IFU-like pointings, centered on the faintest regions discussed in Sec.~\protect\ref{sec:z-xrism} (z1, z2, z3 and z5) -- marked here as thin green squares for comparison. The X-IFU hexagonal pointing covers a region of $5'$ equivalent diameter (corresponding to $\sim 330$\,kpc at the fiducial redshift of $z=0.057$).
    \label{fig:xifu-sims}}
\end{figure}

We generate X-IFU synthetic observations of the faint regions centered on z1, z2, z3, and z5, for which the gas velocity field could not be well constrained with \xrism.
The location of these pointings and the corresponding areas covered by the X-IFU hexagonal FoV of $\sim 5'$ (equivalent diameter) are shown in Fig.~\ref{fig:xifu-sims} (cyan hexagons), where the \xrism pointings are also reported for comparison (green thin squares).

We take advantage of the X-IFU large FoV and collecting area and assume here the reasonable exposure of $100$\,ks. This allows us to obtain a sufficiently large number of photon counts  from the ICM and to perform a reliable spectral fit, in the full $0.2$--$10$\,keV energy band (with $\sim 160000,~5000$ and $50000$ ICM counts for z1, z2 and z5, respectively). 
In the $6$--$7$\,keV range surrounding the Fe-K lines, 
the net number of counts is about five times larger than in the \xrism spectra, reaching $\sim 1000$ photons for z1. For z2 and z5, however, it is still quite small, namely $\sim 200$ and about 10 counts, respectively).

Including the soft part of the spectrum and extending the fit to the full band,
for the fainter z2 and z5 regions
it is thus important to evaluate the impact of the total background emission on the ICM modelling.
We thus generate background \athena X-IFU spectra, including both the instrumental Non-X-ray background (NXB) and the cosmic X-ray background (CXB)~\cite[e.g.][]{cucchetti2018_xifu}. 
We assume the NXB level equal to the expected mission requirement, namely a constant value of $5\times10^{-3}\,{\rm cts\,cm}^{-2}\,{\rm s}^{-1}\,{\rm keV}^{-1}$ across the X-IFU FoV, to be renormalized for the full detector area.
The physical background comprises both the diffuse unresolved CXB and the astrophysical Galactic foreground. This is modelled as the sum of a soft thermal emission model at low energies ($<1$\,keV), associated with the Galactic plane (unabsorbed) and halo (absorbed) hot plasma, and an absorbed power-law component at higher energies, due to the integrated emission of unresolved extragalactic AGNs. The model and parameters, for an extraction area of $1\,{\rm arcmin}^2$, are those reported by~\cite{lotti2014} (as in their Table~2, in which the normalization of the extragalactic component reported in their Table~2 assumes that in real observations of diffuse sources with the \athena X-IFU, given its spatial resolution, AGNs can be mostly resolved to up to $80\%$).

For the z2 region, given the low gas temperature, the emission from the cluster ICM is in general fainter (lower intrinsic normalization) than the emission from the background. While we can still attempt an overall fit of the entire $0.2$--$10$\,keV energy band, it is not possible to aim for a measurement of the l.o.s.\ velocity dispersion via the spectral analysis of the Fe-K complexes in the $6$--$7$\,keV range alone.
Similarly, in z5, the emission in the $6$--$7$\,keV range is also dominated by the (instrumental) background, so that a reliable measurement of the line broadening cannot be obtained.

\begin{figure*}
  \centering
  \includegraphics[width=0.28\textwidth]{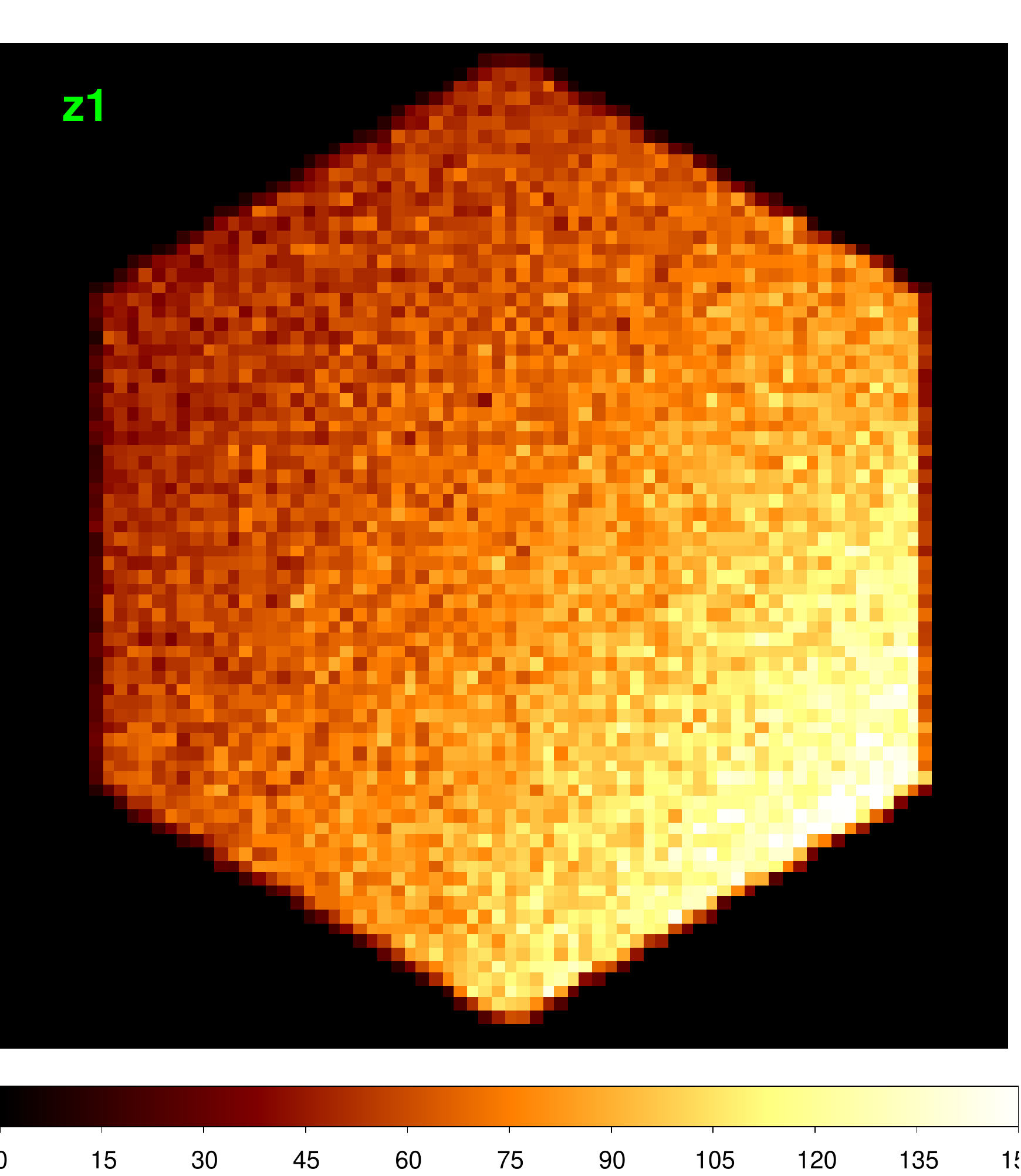}\hspace{7pt}
  \includegraphics[width=0.5\textwidth,trim=0 20 55 35,clip]{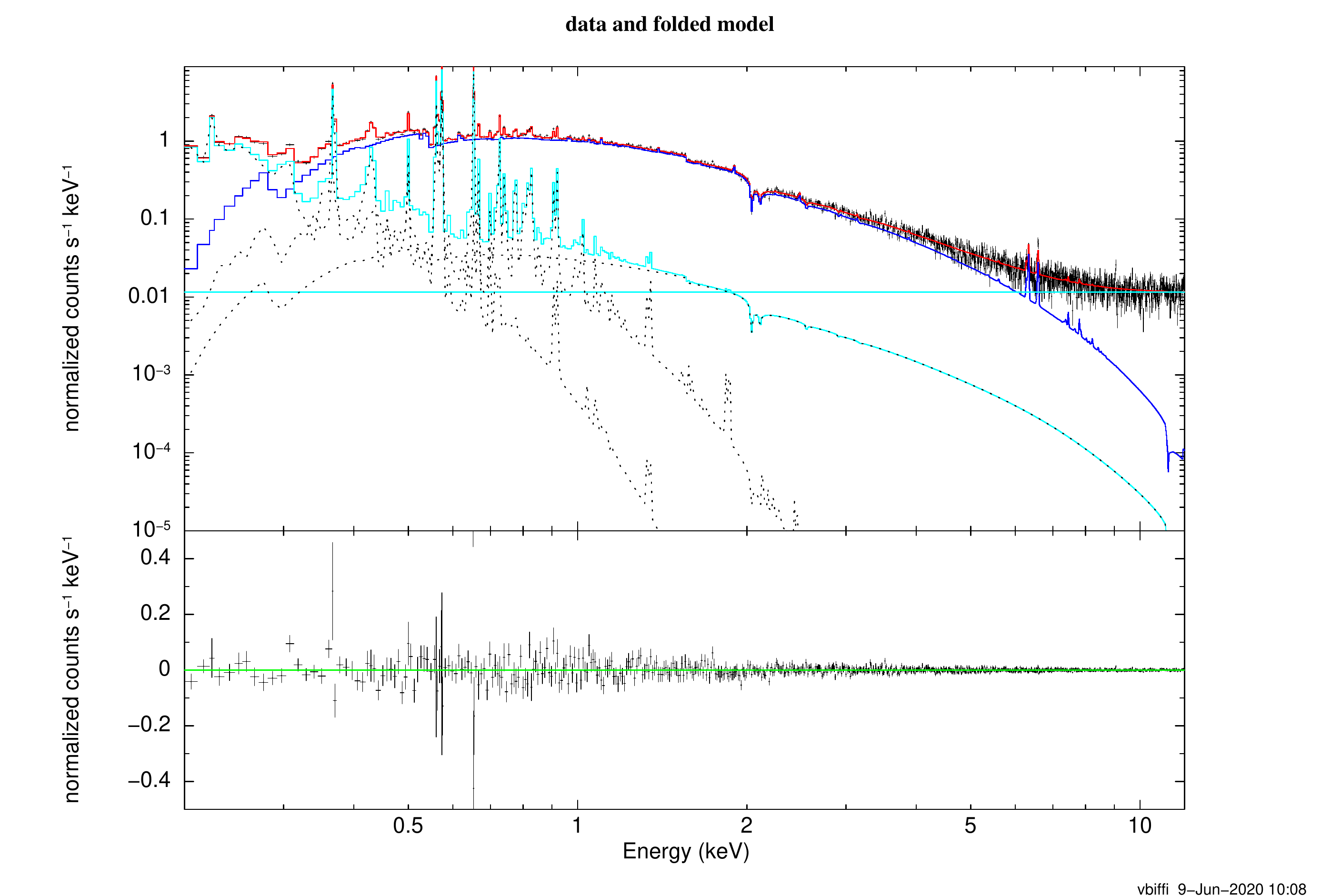}\\[3pt]
  \includegraphics[width=0.28\textwidth]{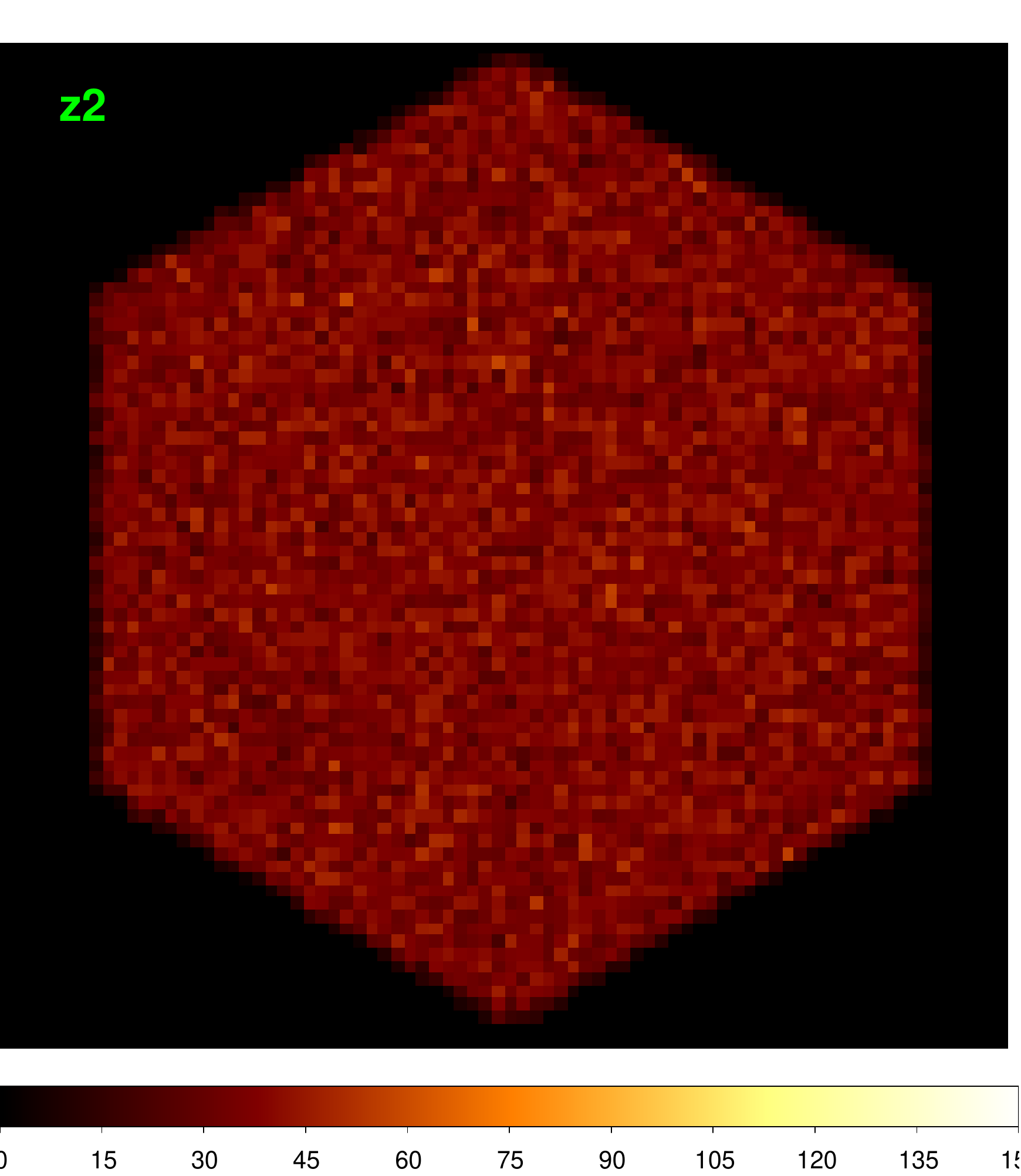}\hspace{7pt}
  \includegraphics[width=0.5\textwidth,trim=0 20 55 35,clip]{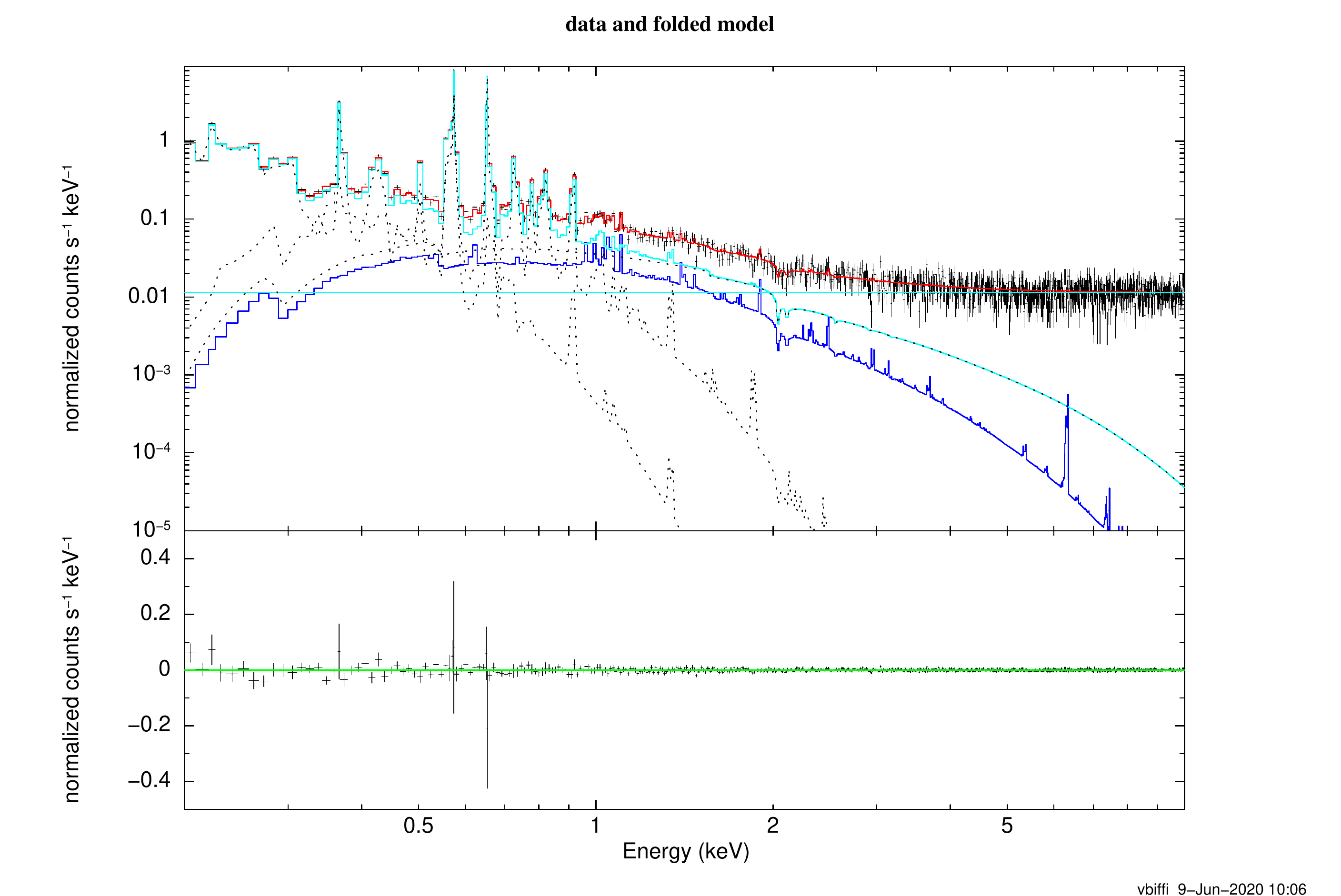}\\[3pt]
  \includegraphics[width=0.28\textwidth]{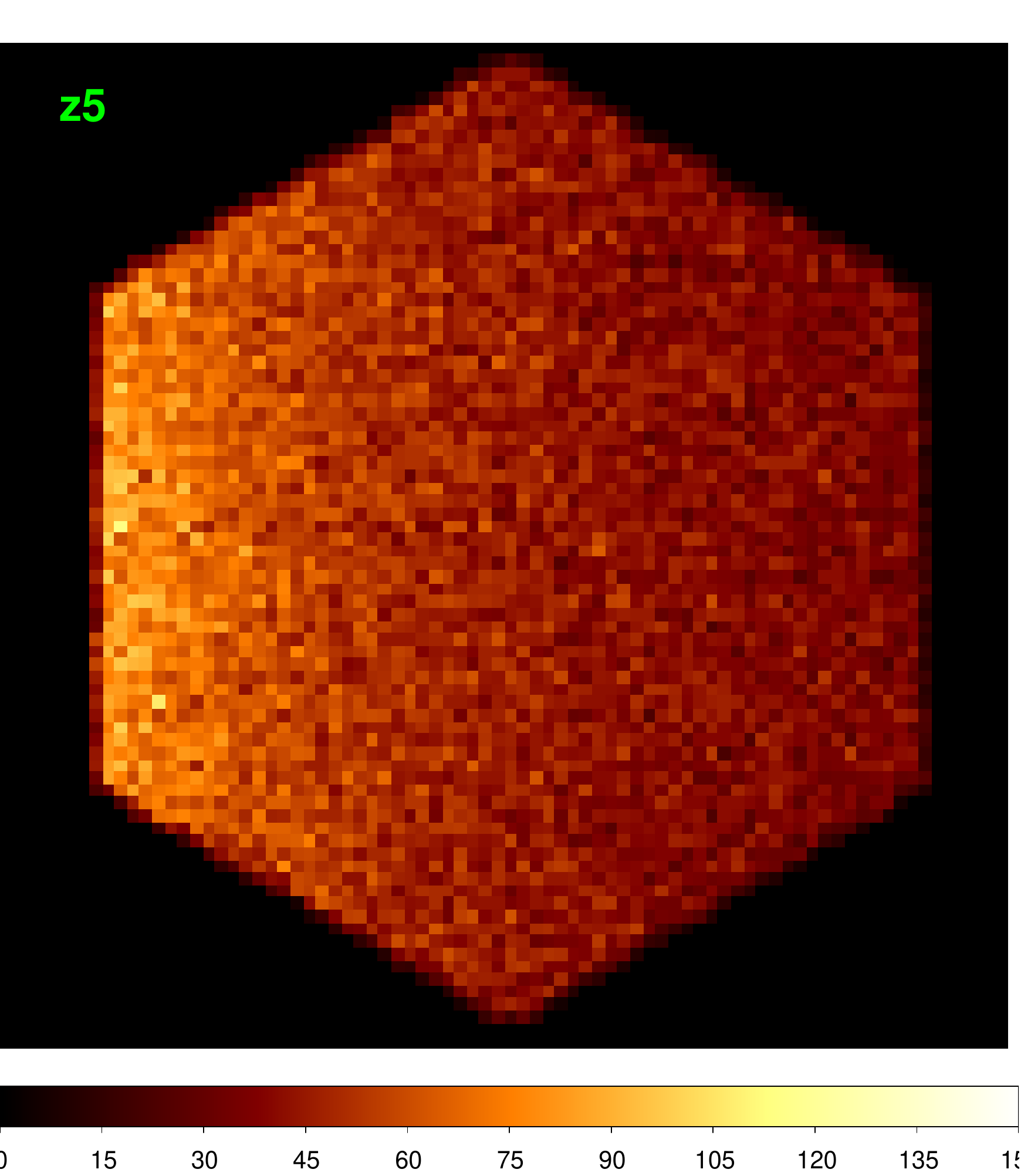}\hspace{7pt}
  \includegraphics[width=0.5\textwidth,trim=0 20 55 35,clip]{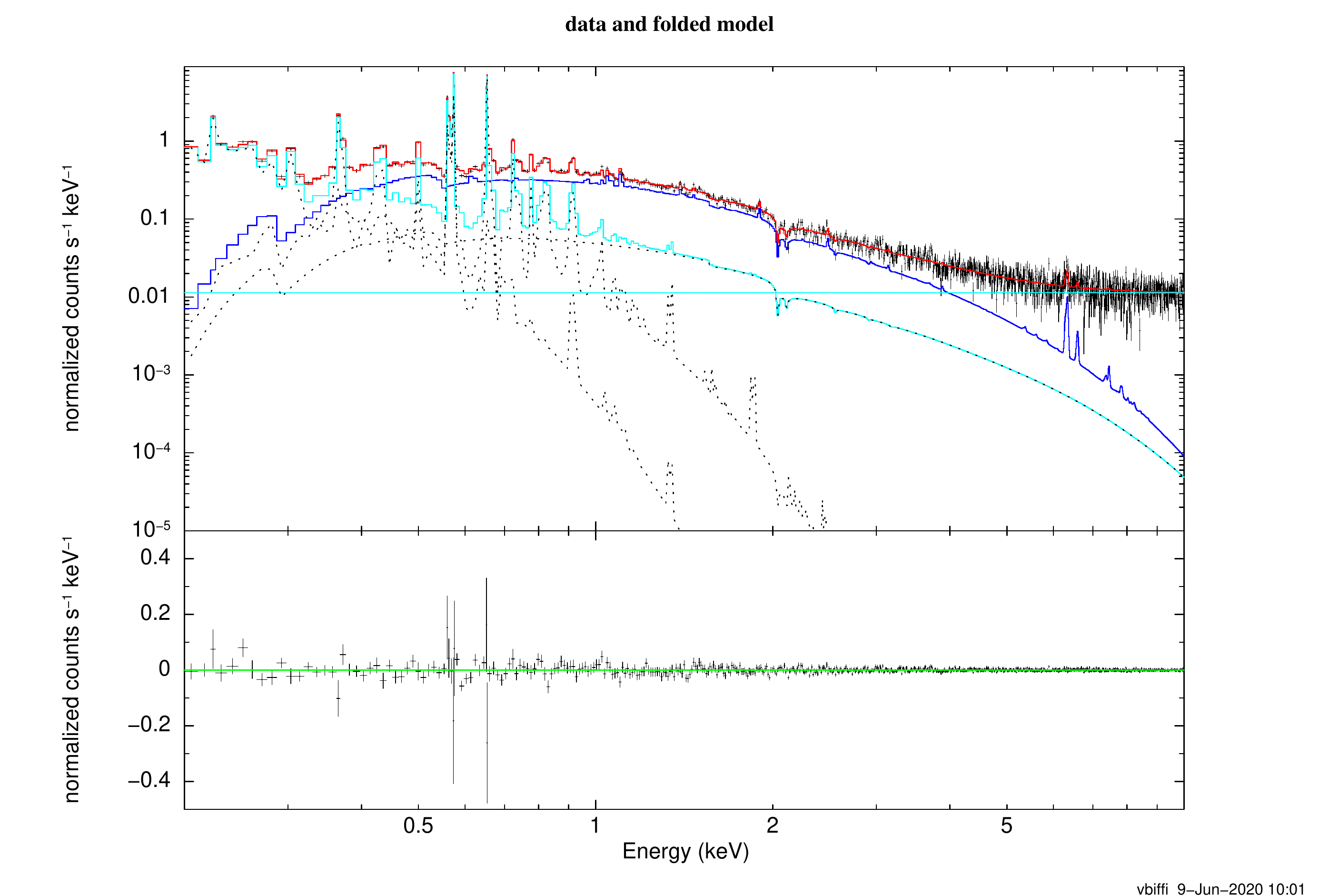}
  \caption{\athena X-IFU $100$\,ks mock observations of the faint z1,
    z2 and z5 pointings: full-band images (l.h.s. panels) and spectra
    (r.h.s. panels). The best-fit models (red curves) are overplotted
    on the spectra, with separate components marking the background
    model (cyan) and cluster emission (blue).\label{fig:xifu-mocks}}
\end{figure*}

For these reasons, we use the full $0.2$--$10$\,keV energy band to fit the z1, z2, and z5 \athena X-IFU spectra, assuming for the ICM a single thermally broadened APEC model (BAPEC in XSPEC), both with and without background.
In all the fits, we fix the abundance and Galactic absorption values to those initially adopted to generate the synthetic observations.
The $100$\,ks X-IFU images of z1, z2 and z5 are shown in Fig.~\ref{fig:xifu-mocks} (l.h.s. panels).  The spectra, with best-fit models (red curves) and residuals, are shown in the right-hand-side panels of the Figure, where the ICM and background components are marked in blue and cyan, respectively. Best-fit results on the l.o.s.\ velocity and velocity dispersion are reported in
Table~\ref{tab-xifu}, both for the full mock observations (background included) and for the ICM-only ones.

\begin{table*}
  \caption{Comparison between theoretical values computed directly from the simulations and 
  spectral best-fit values of velocity shift and dispersion, 
  for the X-IFU pointings centered on z1, z2 and z5. We report the best-fit results for the spectra of the ICM emission only (columns 3 and 4) and for the complete observations including both instrumental and physical background (columns 5 and 6). The exposure time of the mock X-IFU observations is 100\,ks for all pointings (last column).\label{tab-xifu}}
  \centering
  \renewcommand{\arraystretch}{1.2} \small
  \begin{tabular}{c|cc|cccc|c}
    \hline
    & \multicolumn{2}{c|}{Theoretical estimate}
    & \multicolumn{5}{c}{\athena mock -- XSPEC fit}
    \\[3pt]
\hline
    & $\mu_{wFe}$ & $\sigma_{wFe}$  
    & $\mu_{1}$ (ICM only) & $\sigma_{1}$ (ICM only)  
    & $\mu_{1}$ & $\sigma_{1}$
    & exp
    \\
    & [km/s] & [km/s] 
    & [km/s] & [km/s] 
    & [km/s] & [km/s]
    & [Ms]
    \\[2pt] 
    \hline
    z1    & -1.9  & 395.3
    & $-37.9\pm 57.8$ & $430.3 \pm 54.2$
    & $-26.0\pm 67.5$ & $497.9\pm 67.9$ 
    & 0.1
    \\
    z2    & -0.3  & 201.3
    & $-65.5\pm 49.2 $ & $ 356.0\pm 54.7$
    & $50.6\pm 48.1$ & $125.2\pm 91.4$
    & 0.1
    \\
    z5    & 3.6   & 766.9
    & $-120.1\pm 108.1$ & $908.3\pm 98.6$
    & $63.1\pm 108.4$ & $803.6\pm 105.5$
    & 0.1
    \\
    \hline
  \end{tabular}
\end{table*}

We also re-compute the values of velocity shift and dispersion directly from the simulation for the new regions with X-IFU FoV sizes. Compared to these numerical estimates, all the best-fit values are roughly within the 1- or 2-$\sigma$ uncertainties, even when the full instrumental and physical background are included in the mock observation.
Although the numerical average values from the simulation are weighted by iron emissivity while the fits are performed over the full $0.2$--$10$\,keV band, we expect the comparison to be robust, given that the shapes of the velocity distributions show negligible differences when different emission-like weights are employed (see Fig.~\ref{fig:z-pdfs1}).

For both z2 and z5, we find that the best-fit spectral results differ from the values computed directly from the simulations, even when the spectrum of the ICM alone without background emission is considered.
These discrepancies are related to the intrinsic complexity of the thermodynamical and kinematic properties of the gas in these regions.  
From the inspection of the l.o.s.\ velocity distribution of the ICM in these regions --- not shown here, but already shown in the smaller corresponding \xrism pointings (see Fig.~\ref{fig:z-pdfs1}) --- we note that poor photon statistics compound the difficulties in fitting a one-component model due to the multi-component velocity structure of the gas. 
In particular, this is especially the case for the z2 pointing, where the low number of counts and the complex shape of the 2-component velocity distribution results in a poor constraint on the velocity broadening. 
To a lesser extent, also in the z5 region, the velocity-broadened line shape is very complicated and two temperature components challenge the fitting, although the velocity measurement is nevertheless closer to the theoretical value.
In neither of these two pointings, however, is the Gaussian approximation expected to be a fair representation of the velocity distribution (and consequently of the emission line shape). Discrepancies between the XSPEC best-fit
parameters and the theoretical values are therefore expected.

For the z1 pointing, the larger number of photons in the spectrum and the simpler shape of the intrinsic velocity distribution allow for a better reconstruction of the velocity parameters.
Line shift and broadening best-fit values extracted from the ICM-only spectra agree within 1-$\sigma$ with those computed directly from the simulation.
When we include the background components in the mocks, the fit provides poorer constraints, with the best-fit velocity dispersion departing further from the simulation value, although still consistent 
within 2-$\sigma$.
In this case, the brighter ICM emission also allows us to perform a fit of the Fe-complex spectral range separately, finding $\mu_{v}\,[6\textrm{--}7\,{\rm keV}] = -61.6\pm 99.6$\,km/s and $\sigma_{v,z}\,[6\textrm{--}7\,{\rm keV}] = 426.5\pm 89.8$\,km/s, when both instrumental and physical background are included. 
This is still roughly consistent with the full-band fit and is actually in better agreement with the simulation values.

\subsubsection{Velocity gradient across the shock front}\label{sec:z3-shock}

One of the interesting regions of the merging system studied here is the shock front developing after the first core passage of the two haloes. 
The z3 pointing, marked in Fig.~\ref{fig:map}, covers this region.
Mock observations with \xrism pointed out the difficulty to constrain robustly the complex velocity structure of the gas. 
Indeed, the gas velocity distribution along the l.o.s.\ is characterized by two components and a very complex, double-peaked, non-Gaussian shape (see Fig.~\ref{fig:z-pdfs1}, third row from the top).
This region is moreover one of the fainter ones explored, and suffers also from the photon contamination from the bright cluster core nearby. We thus generate mock observations with the \athena X-IFU instrument, to take advantage of its much larger collecting area and superior angular resolution, while preserving the high spectral resolution. 

\begin{figure}
  \centering
  \includegraphics[width=0.4\textwidth]{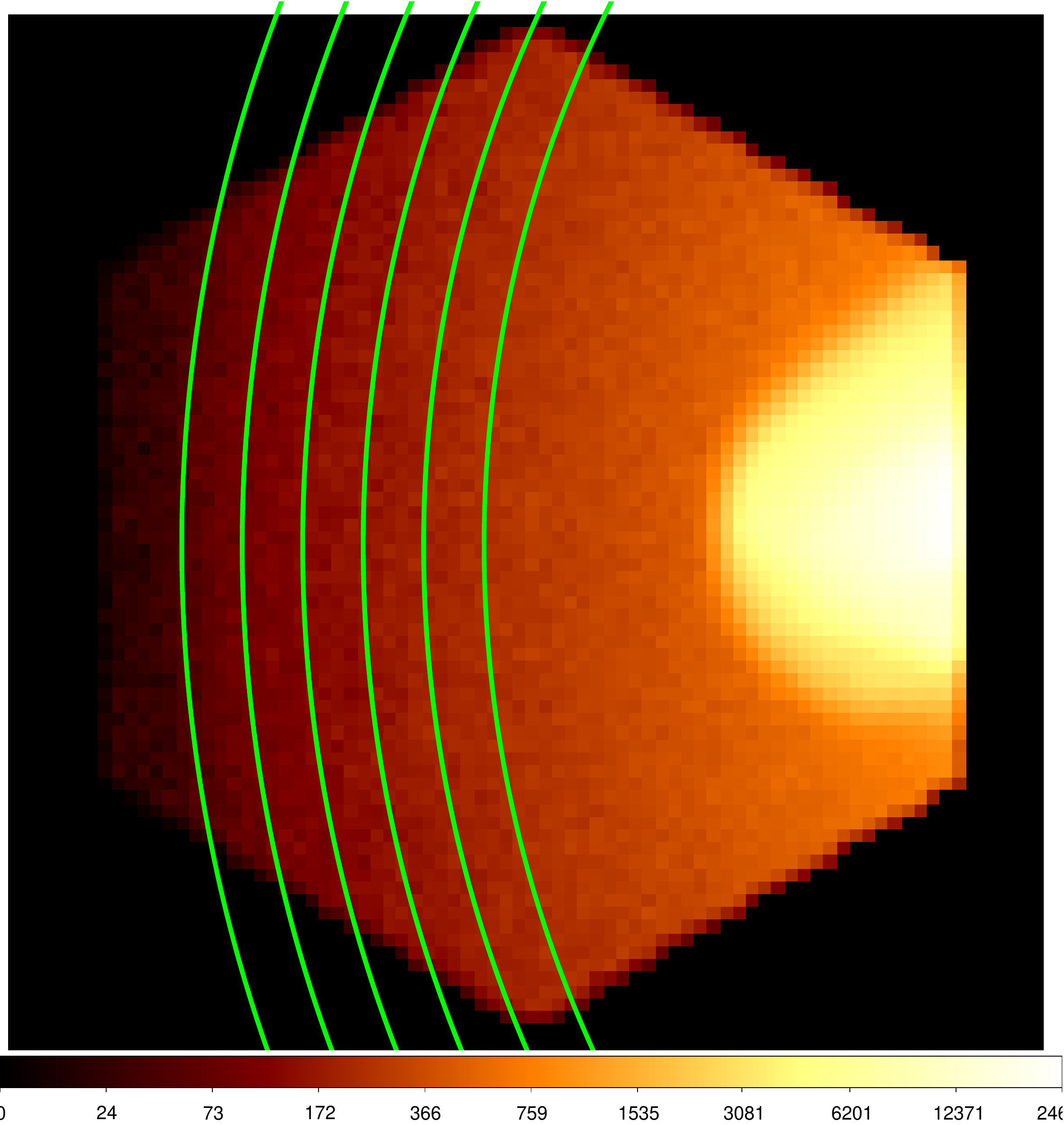}
    \caption{$1$\,Ms X-IFU mock observation (full band) of the z3 region.
    We mark in green the five $20''$-wide annuli from which we extract the spectra.\label{fig:z3_xifu}}
\end{figure}

The symmetric geometry of this merger and the velocity field of the gas in the shock indicate the presence of a velocity gradient across the front. 
To measure the gradient, spectra from several few-kpc-wide annuli must be obtained, requiring a sufficiently large number of photons in each annulus to constrain the gas velocity. 
To this end, we assume an exposure of 1\,Ms for the mock observation with the X-IFU.
The 1\,Ms image for z3 is shown in Fig.~\ref{fig:z3_xifu}, with five annuli defined to map the shock front marked in green. 
The annuli are separated by $20''$, roughly corresponding to $20$\,kpc.
We obtain between $\sim 76000$ and $\sim 17000$ photons per annulus, in the $0.2$--$10$\,keV band. The full band is adopted for the spectral fit due to the much lower number of photons in the $6$--$7$\,keV band, dropping below 200 in the fainter annuli.

We fit each spectrum with a BAPEC model. As in previous Sections, we fix the metal abundance and Galactic absorption to the known input parameters, and vary temperature, redshift, velocity dispersion and normalization. 
We concentrate on the emission of the ICM only, and do not include the background.

\begin{figure}
  \centering
  \includegraphics[width=0.43\textwidth]{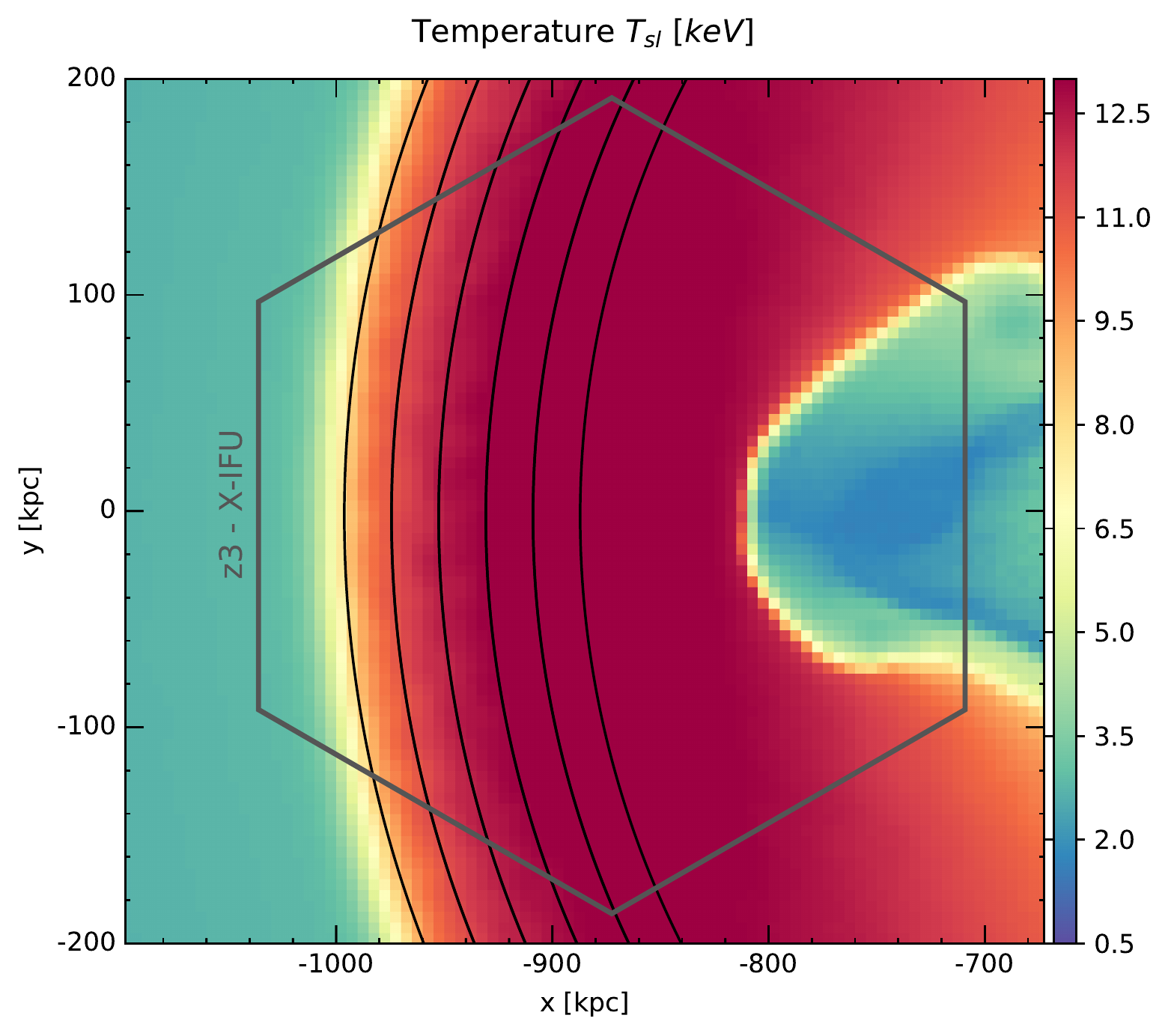}
  \includegraphics[width=0.43\textwidth]{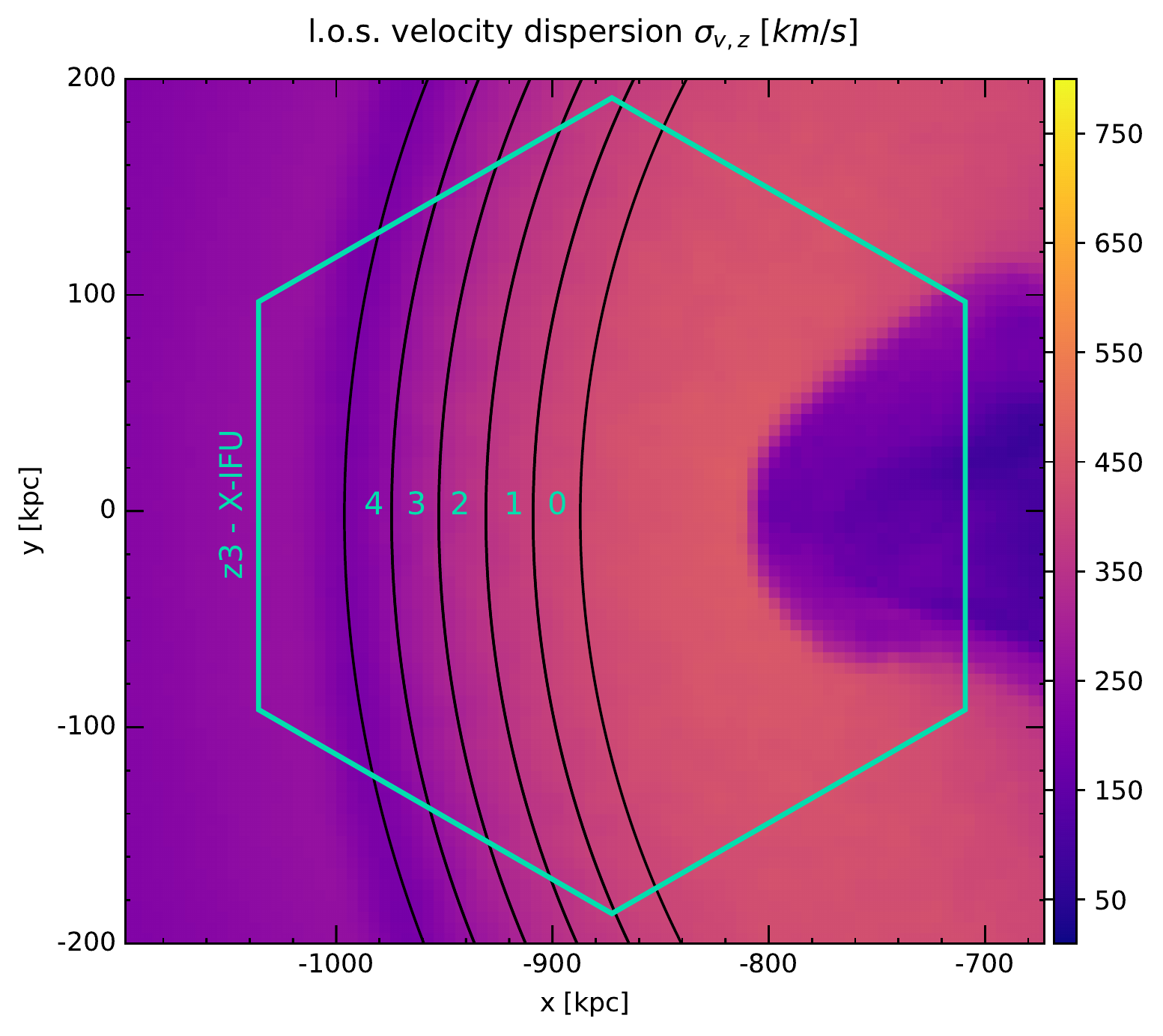}
  \caption{Simulated maps of the gas spectroscopic-like temperature
  and velocity dispersion around the z3 pointing.
    The \athena X-IFU hexagonal FoV ($\sim 5'$ equivalent diameter, corresponding to $\sim 330$\,kpc at $z=0.057$) and the annuli are shown.\label{fig:z3_xifu-sims}}
\end{figure}

\begin{figure}
  \centering
  \includegraphics[width=0.45\textwidth]{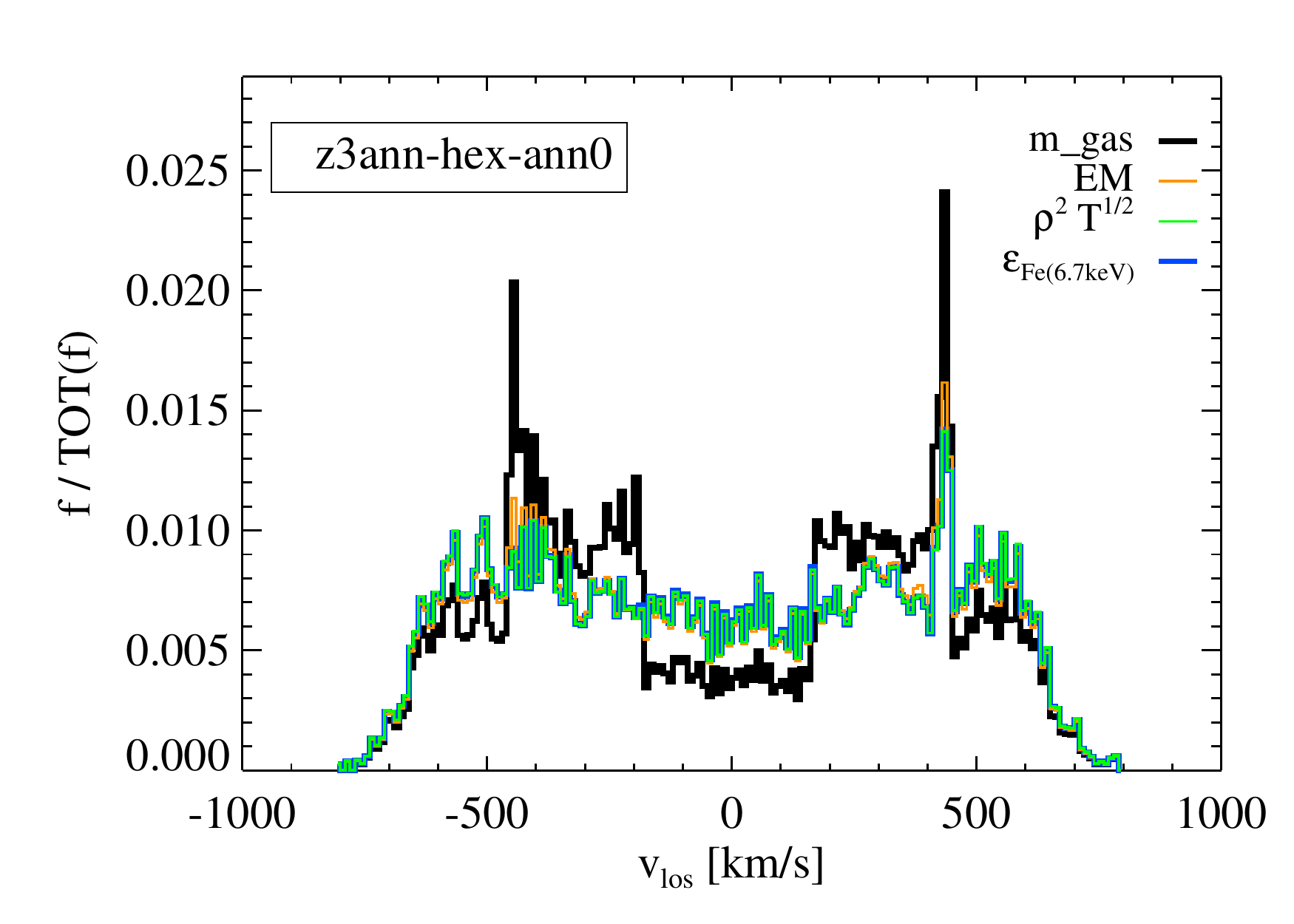}
  \includegraphics[width=0.45\textwidth]{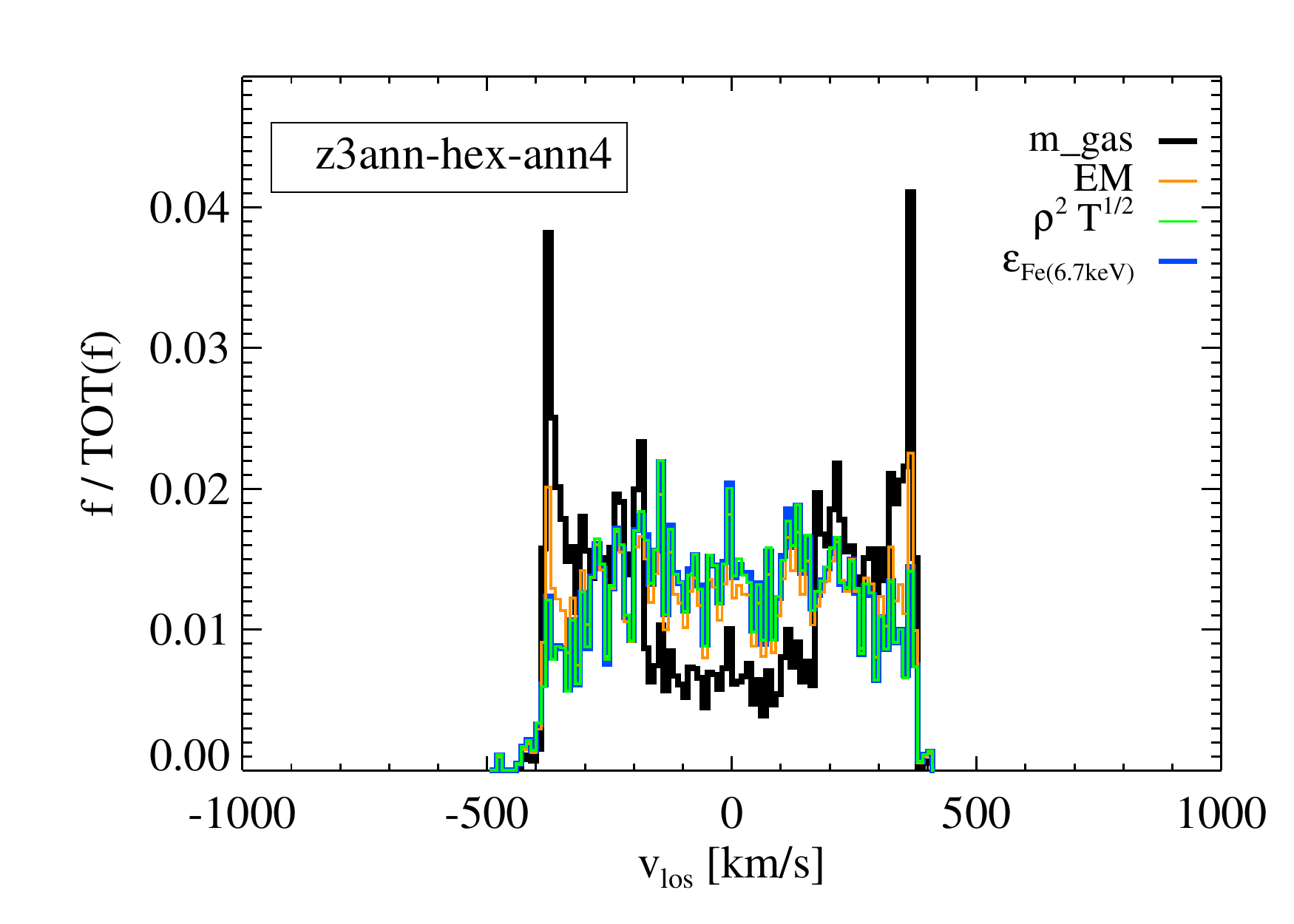}
  \caption{Line-of-sight velocity distributions of gas particles within 
    the innermost and outermost annuli shown in Fig.~\protect\ref{fig:z3_xifu-sims}.
    \label{fig:z3_xifu-vPDF}}
\end{figure}

In Fig.~\ref{fig:z3_xifu-sims}, we show a zoom-in onto the simulated maps of gas spectroscopic-like temperature and l.o.s.\ velocity dispersion, centered on the X-IFU z3 region. In the maps we mark the X-IFU hexagonal FoV and the five annuli on the shock front (numbered from the innermost to the outermost one in the bottom panel).
The velocity distributions for gas elements in the annuli are very similar to the double-peaked one shown in Fig.~\ref{fig:z-pdfs1}, with a spread in velocity that increases with distance from the shock edge. As extreme examples, we only show those for the innermost/brightest (top) and outermost/faintest (bottom) annuli in Fig.~\ref{fig:z3_xifu-vPDF}.

\begin{figure}
  \centering
  \includegraphics[width=0.49\textwidth]{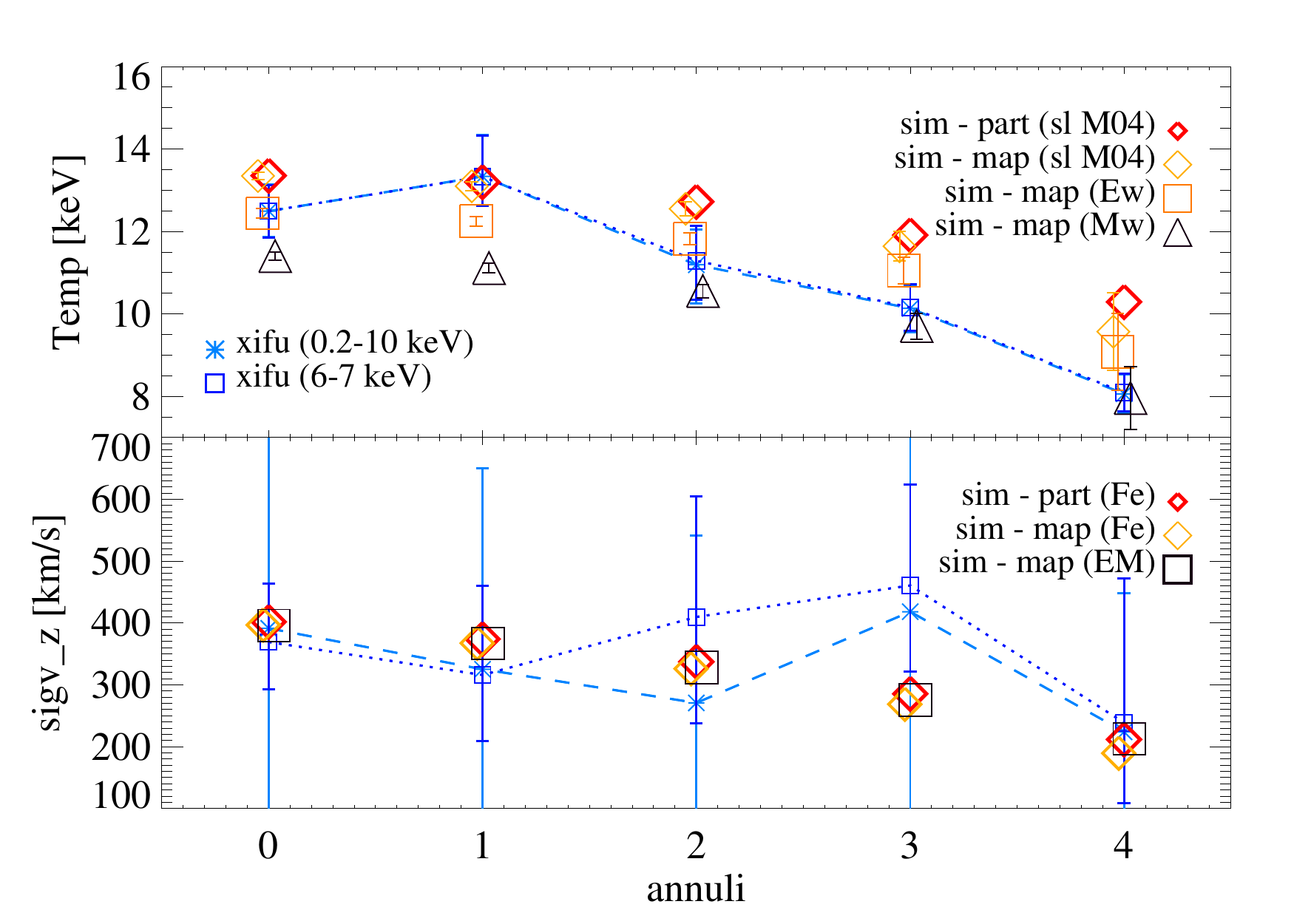}
  \caption{Velocity and temperature gradients across the
    shock (upper and lower panel, respectively). Comparison between simulation predictions and best-fit
    values from mock X-IFU observations for the 5 annuli, from the innermost (0) to the outermost one close to the shock edge (4).
    Various estimates from the simulation and results from the fit of X-IFU spectra in two bands are reported for comparison, as in the legend.
    \label{fig:z3_xifu_grad}}
\end{figure}

We compare best-fit results for temperature and velocity dispersion against the estimates computed directly from the simulation for each annulus (numbered as in Fig.~\ref{fig:z3_xifu-sims}, bottom panel). The temperature and velocity gradients are reported in Fig.~\ref{fig:z3_xifu_grad}.
In the Figure, we report different estimates depending on the band adopted to perform the spectral fit or on the weight used to compute gas velocity and temperature from the simulations, as illustrated in the legend. 
Given the finely spaced annuli considered here, we also compute the average values from the simulations directly from the 2-dimensional regions marked in the projected maps (Fig.\ref{fig:z3_xifu-sims}), to account more properly for the projected gas-cell volume contributing to the emission from such thin annuli.
We note that the temperature best-fit values (blue data points) reproduce the gradient across the annuli, with a decrease of average temperature in the direction towards the shock edge. Compared to the spectroscopic results, however, the emission-weighted values from the simulations tend to overestimate the annulus temperature in the faintest ones. The mass-weighted temperature estimate, on the contrary, tends to underestimate the spectral value in the brightest annuli (i.e. annuli 0 and 1). This difference is induced by the more prominent emission of the hot gas in these annuli, which therefore dominates the spectral temperature.
If we derive the emission-weighted average temperature from the projected maps (orange diamonds and squares), instead of using the gas element distribution directly (red diamonds), the discrepancy is softened.
Overall, the best reconstruction is obtained for the map-derived emission-weighted temperature (orange square), for which the spectral best-fit value is consistent within the 1-$\sigma$ uncertainties, in all annuli.

The velocity broadening is also closely reconstructed by the spectral fit.
For the comparison, we consider the average value in the simulation, weighted either by the emissivity in the Fe-complex (red and orange diamonds), or by emission measure (\emi, black square), which has proved to fairly trace the \athena-like spectroscopic value~\cite[][]{biffi2013,Roncarelli:2018}.
While all of them provide very similar estimates, the best result is obtained by constraining the velocity parameter on the specific 6--7\,keV band, comprising the Fe-complex. 
The stronger discrepancy with respect to the theoretical values is found in the central annuli (ann.\ 2 and 3), where nevertheless the spectral value is still consistent with the theoretical one within 1- and 2-$\sigma$.
When the velocity dispersion is obtained from the fit 
over the whole $0.2$--$10$\,keV band (blue asterisks), the value is typically poorly constrained, with very large uncertainties, although the overall trend is fairly similar.\footnote{We note that the larger $0.2$--$10$\,keV band is in both cases used to constrain the best-fit temperature and normalization.}

Given the large uncertainties on the best-fit values for velocity broadening, the modest velocity gradient across the shock (probing a difference of $\sim 200$\,km/s from innermost annulus to outermost one) cannot be well constrained and the spectral fit values are rather consistent with a flat trend.

\subsection{Merger along the line of sight}\label{sec:x-proj}

Here we explore the extreme case in which the merger axis of the two clusters is aligned with the l.o.s.
Clusters in this configuration, or close to it, might appear as roughly regular clusters in the sky and thus represent a critical challenge for the X-ray mass reconstructions based on hydrostatic equilibrium~\cite[e.g. the merging system Cl 0024+17;][]{ota2004}.
Such systems can also deviate from the expected scaling relations between mass and observable properties, when substantial bulk and turbulent motions in the ICM contribute to the non-thermal pressure support~\cite[][]{biffi2013}.

In such a l.o.s.\ merger, significant gas motions are expected to be measured, in particular in the bright central regions, where the measurement of the double velocity component can thus be more easily obtained.
With net velocities of thousand(s) of km/s, the line shifts can be very easily measured and the two merging components distinguished.

\begin{figure*}
  \centering
    \includegraphics[width=0.33\textwidth]{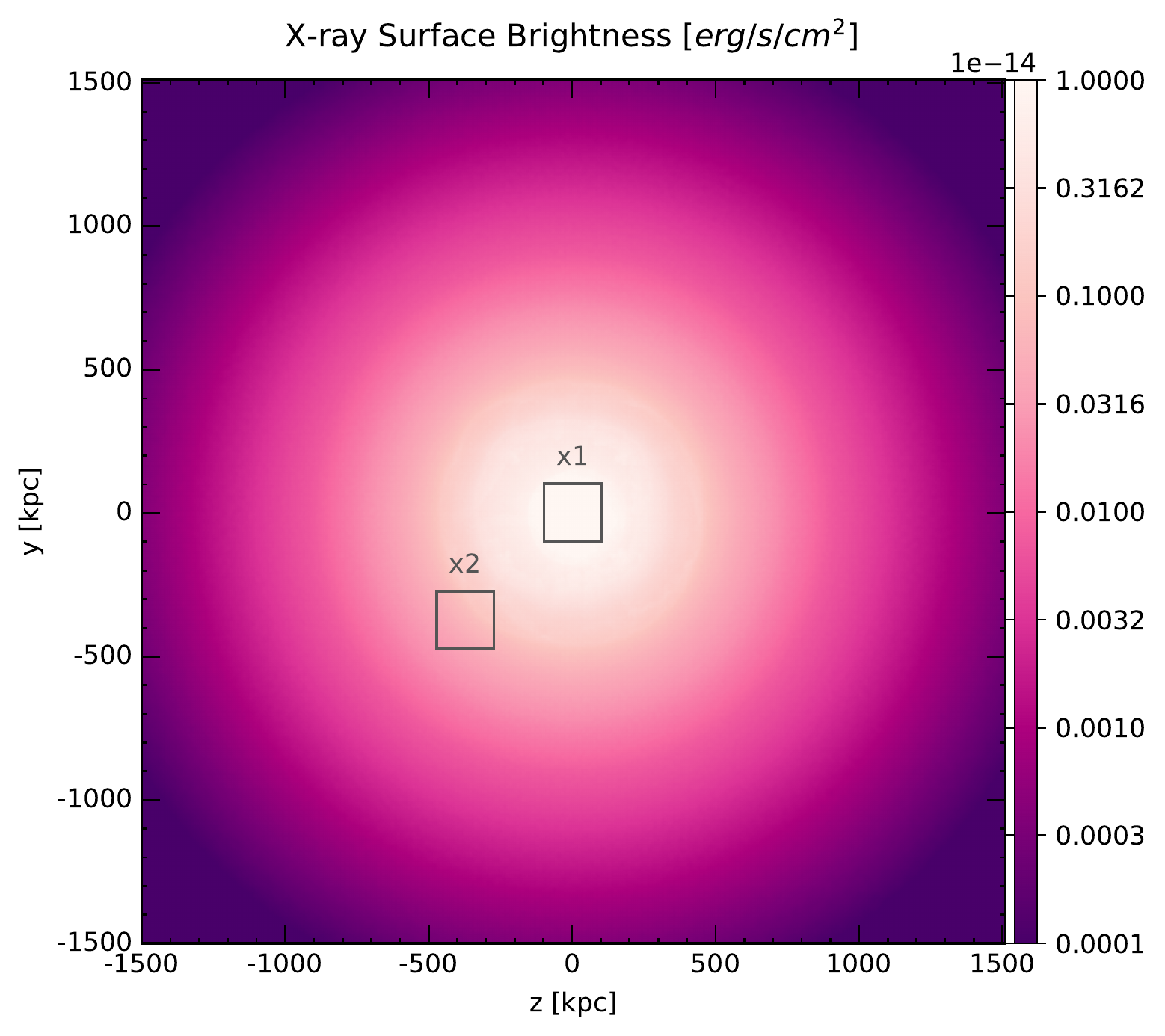} 
    \includegraphics[width=0.33\textwidth]{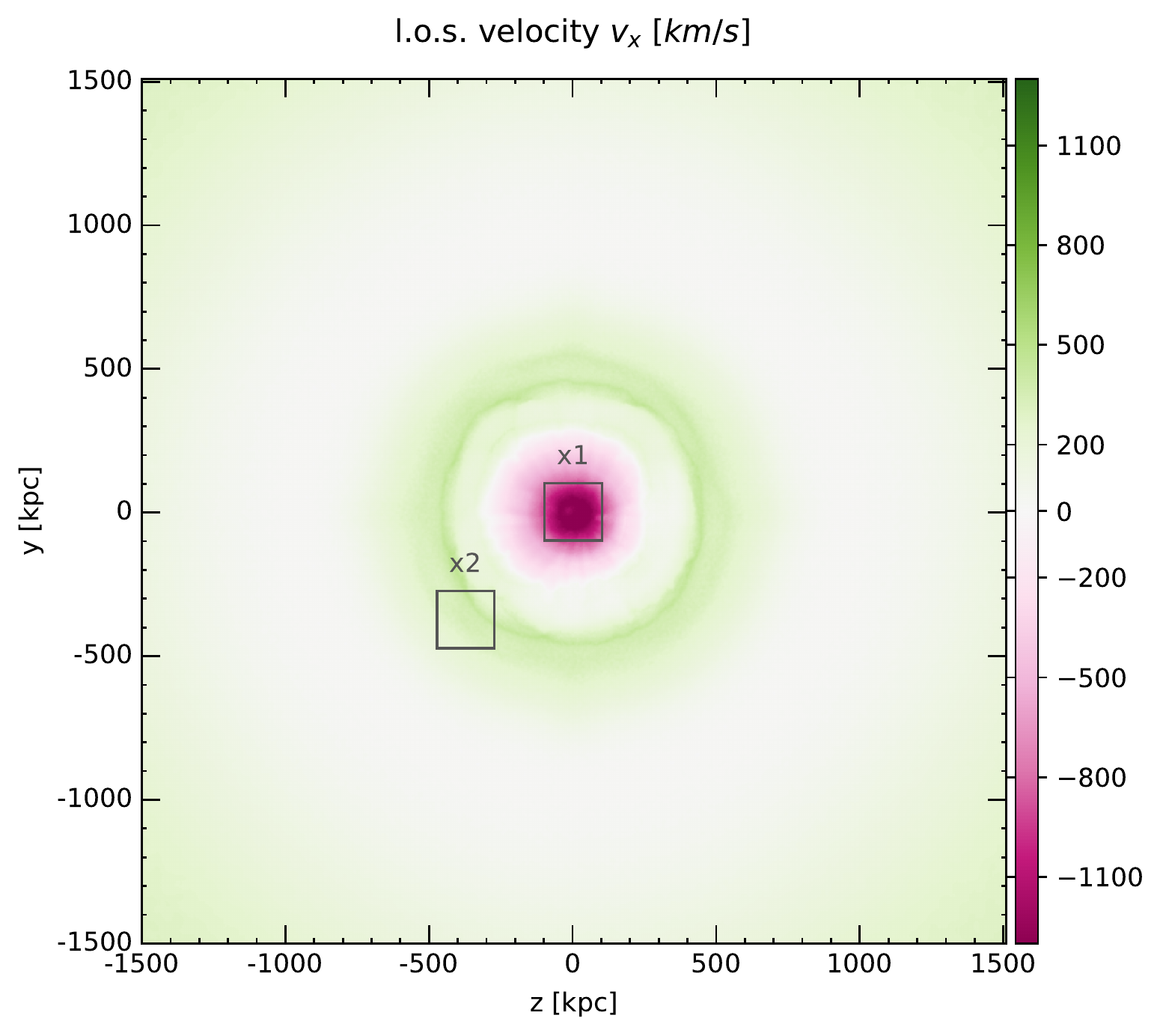} 
    \includegraphics[width=0.32\textwidth]{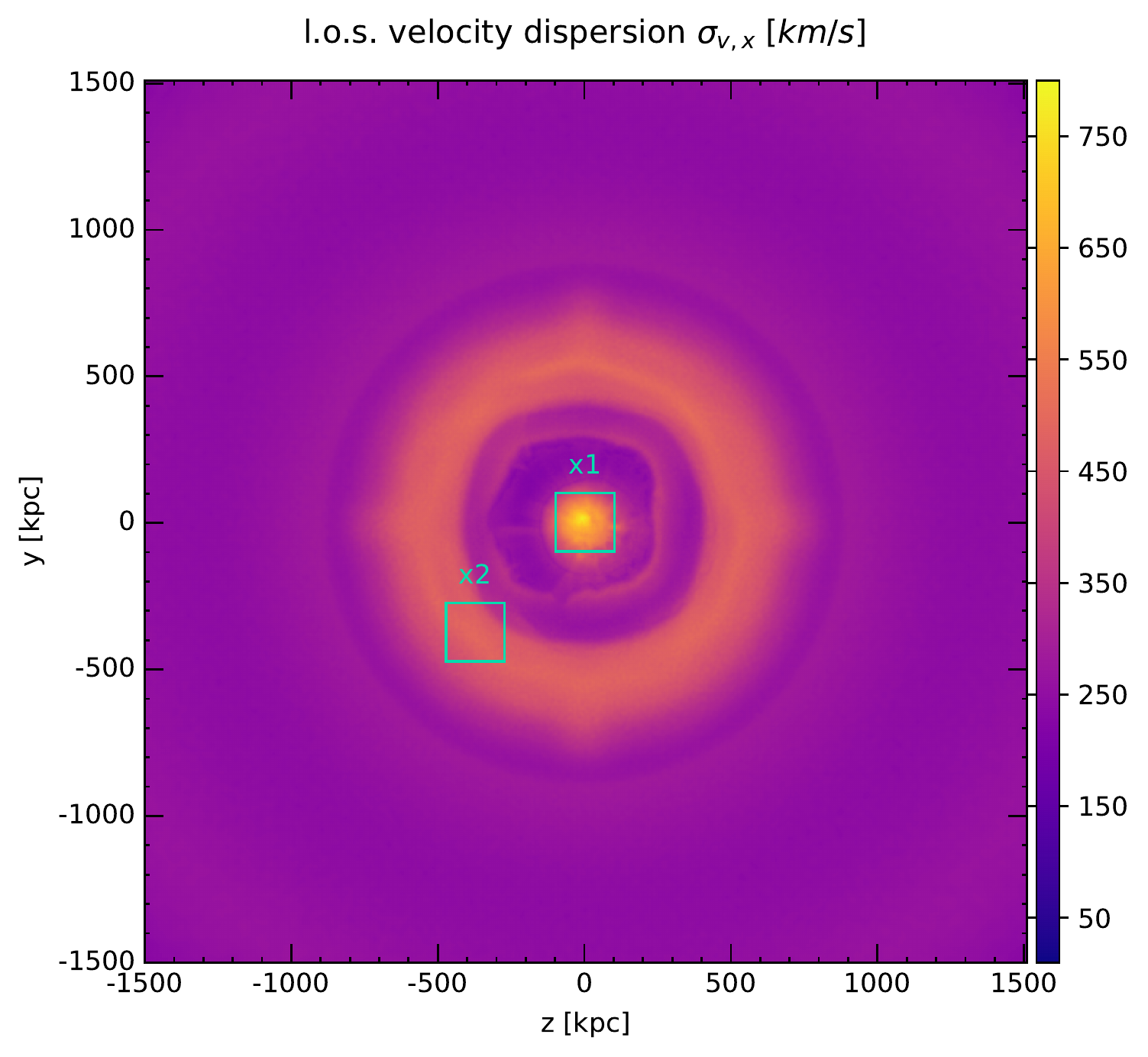} 
  \caption{Projected maps of X-ray flux in the 0.5--7\,keV band (left) and EM-weighted l.o.s.\ velocity (middle) and velocity dispersion (right), for the $x$-axis projection, i.e.\ with the l.o.s.\ aligned with the merger axis. The projection is performed over 10\,Mpc.
  Overplotted, the two square \xrism-like pointings (x1 and x2) of $200$\,kpc per side.
  \label{fig:map.x}}
\end{figure*}

Fig.~\ref{fig:map.x} shows simulated projected maps of the X-ray ($0.5$--$7\,$keV) flux and the gas velocity and velocity dispersion along the $x$-axis l.o.s.\ (from left to right).
The two x1 and x2 regions, approximating in size the \xrism Resolve FoV at the fiducial redshift of $z=0.057$, sample the central brightest region and an offset fainter location, respectively.

\begin{figure*}
  \centering
  \includegraphics[width=0.34\textwidth,trim=0 10 20 25,clip]{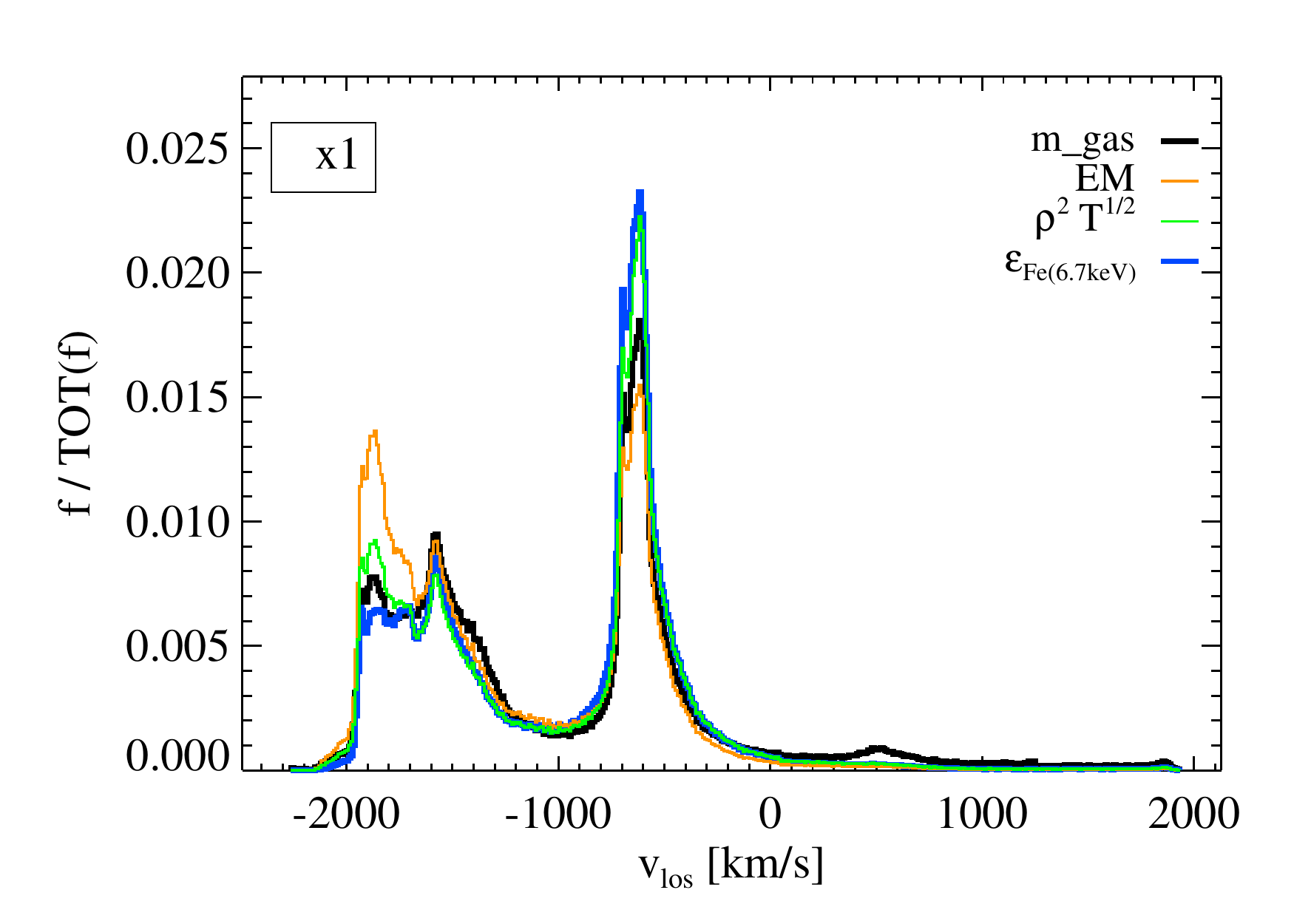}
  \includegraphics[width=0.32\textwidth,trim=100 0 20 10,clip]{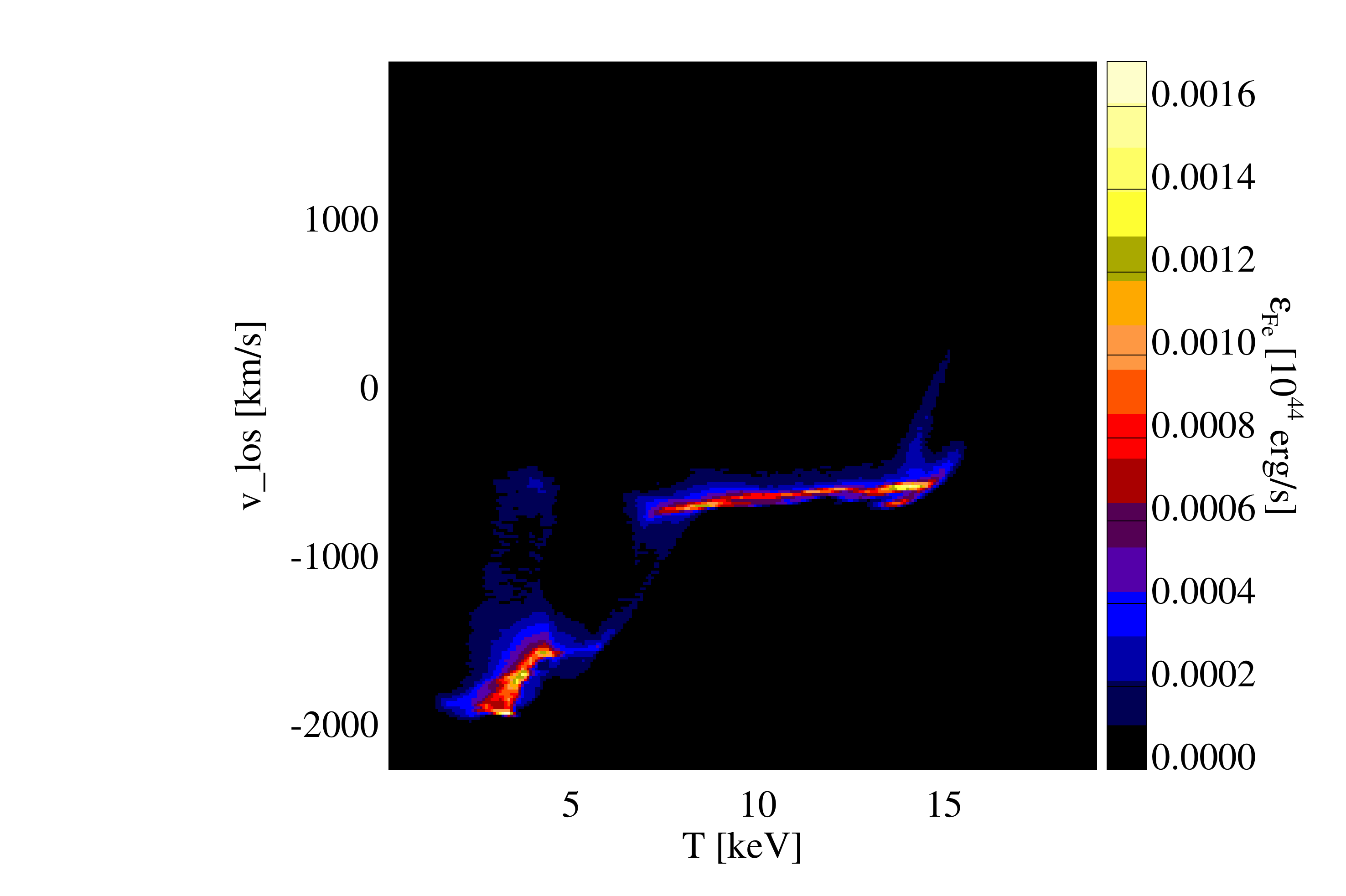}
  \includegraphics[width=0.33\textwidth,trim=20 10 10 20,clip]{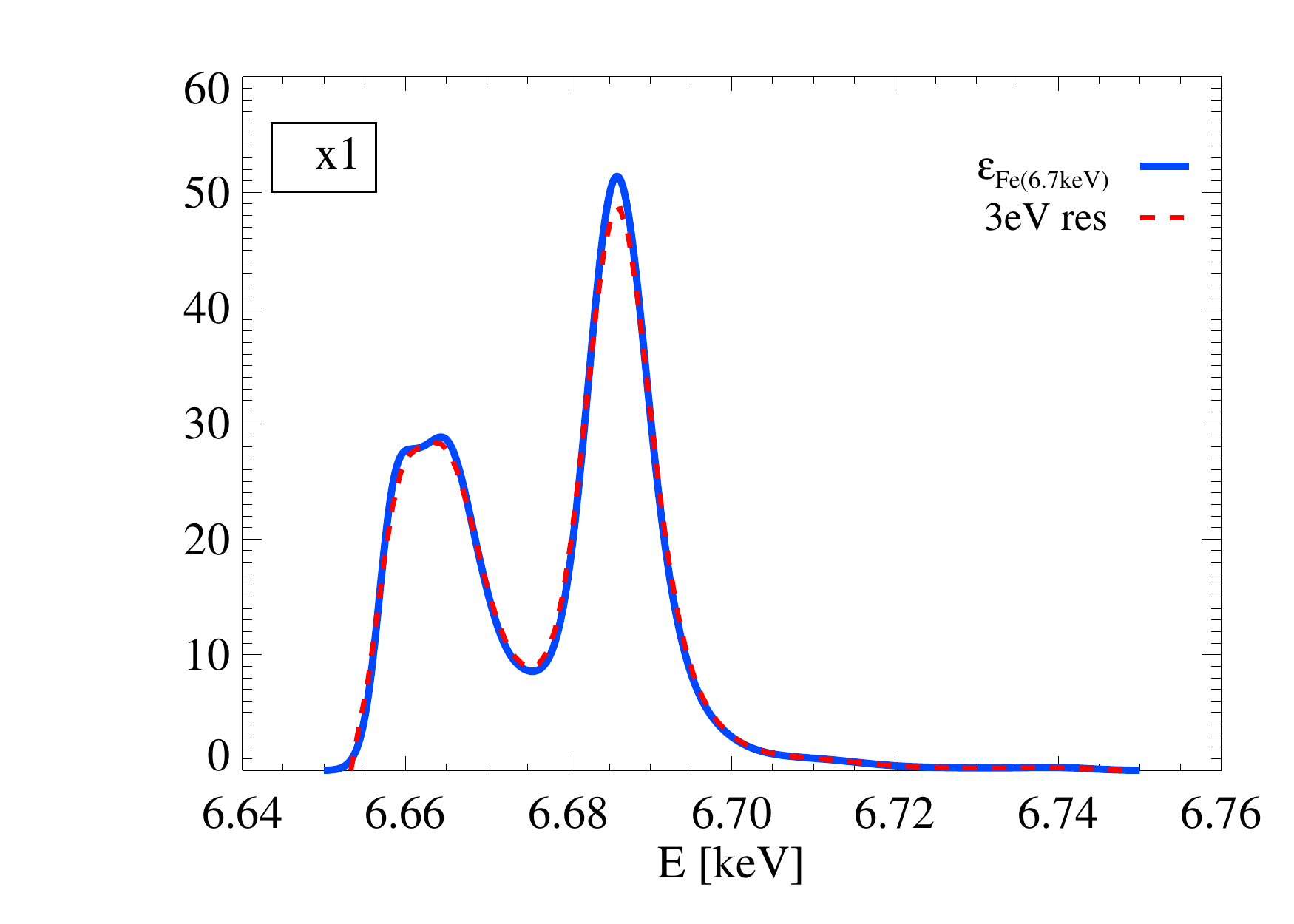}
  \\
  \includegraphics[width=0.34\textwidth,trim=0 10 20 25,clip]{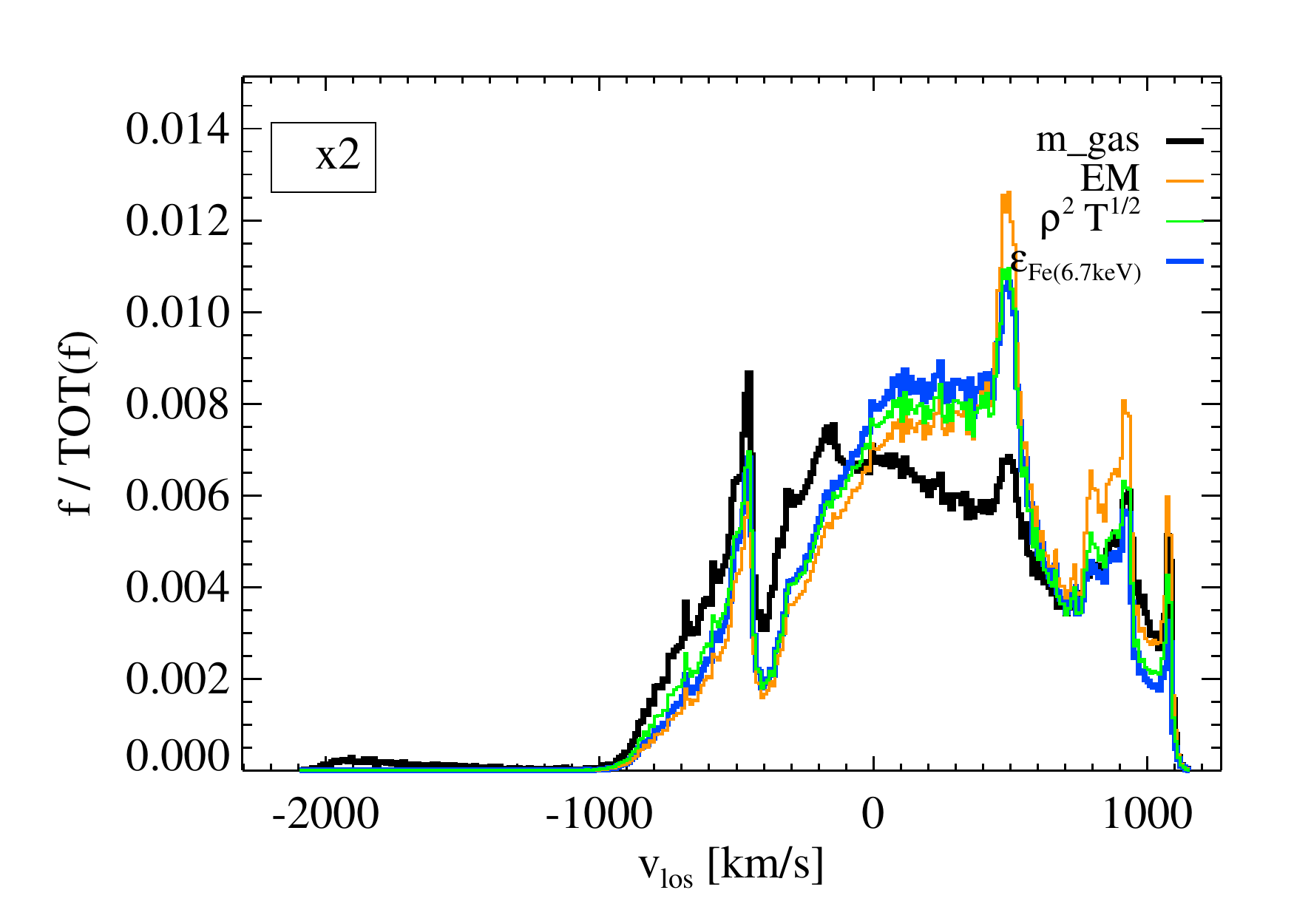}
  \includegraphics[width=0.32\textwidth,trim=100 0 20 10,clip]{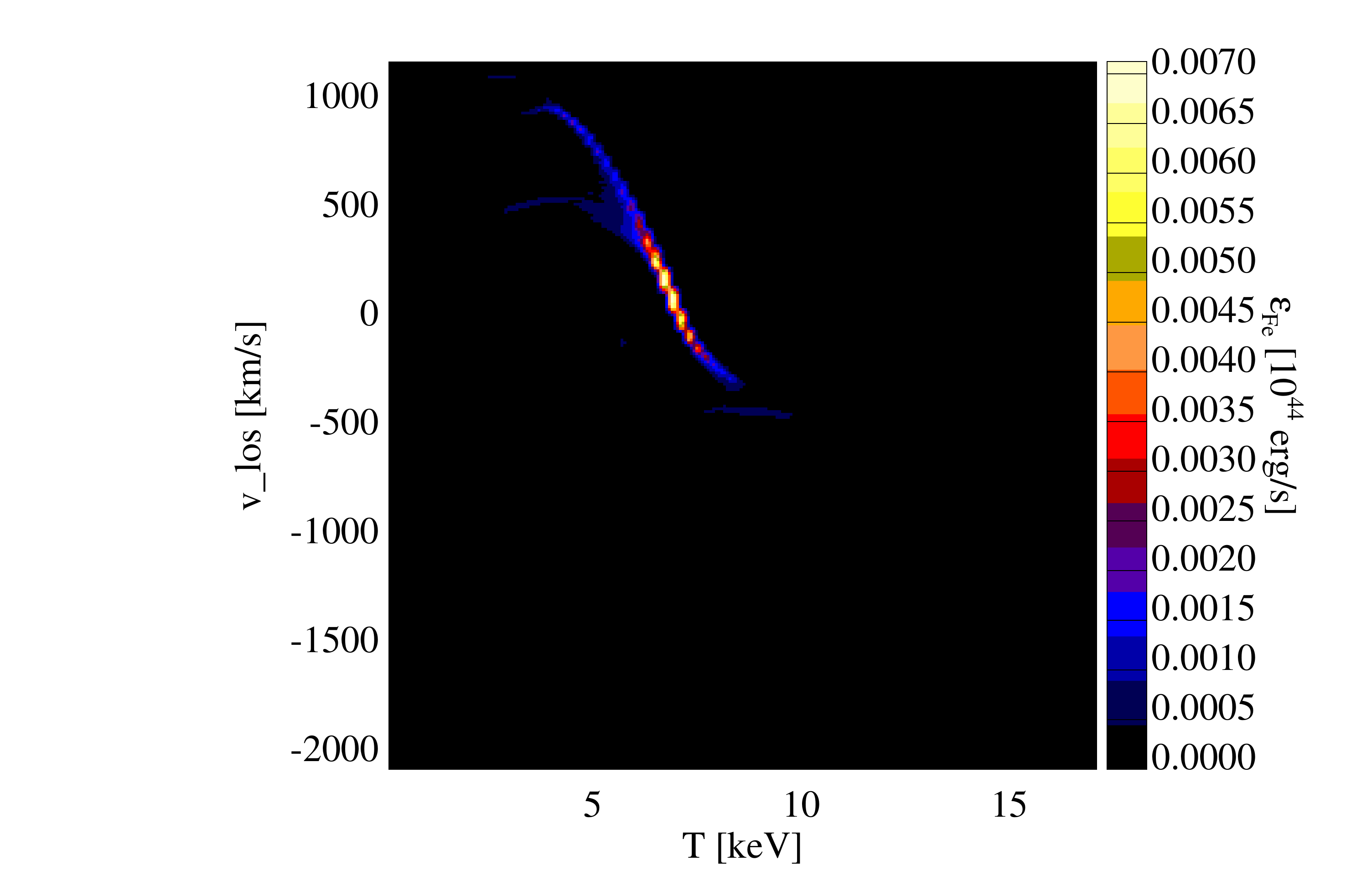}
  \includegraphics[width=0.33\textwidth,trim=20 10 10 20,clip]{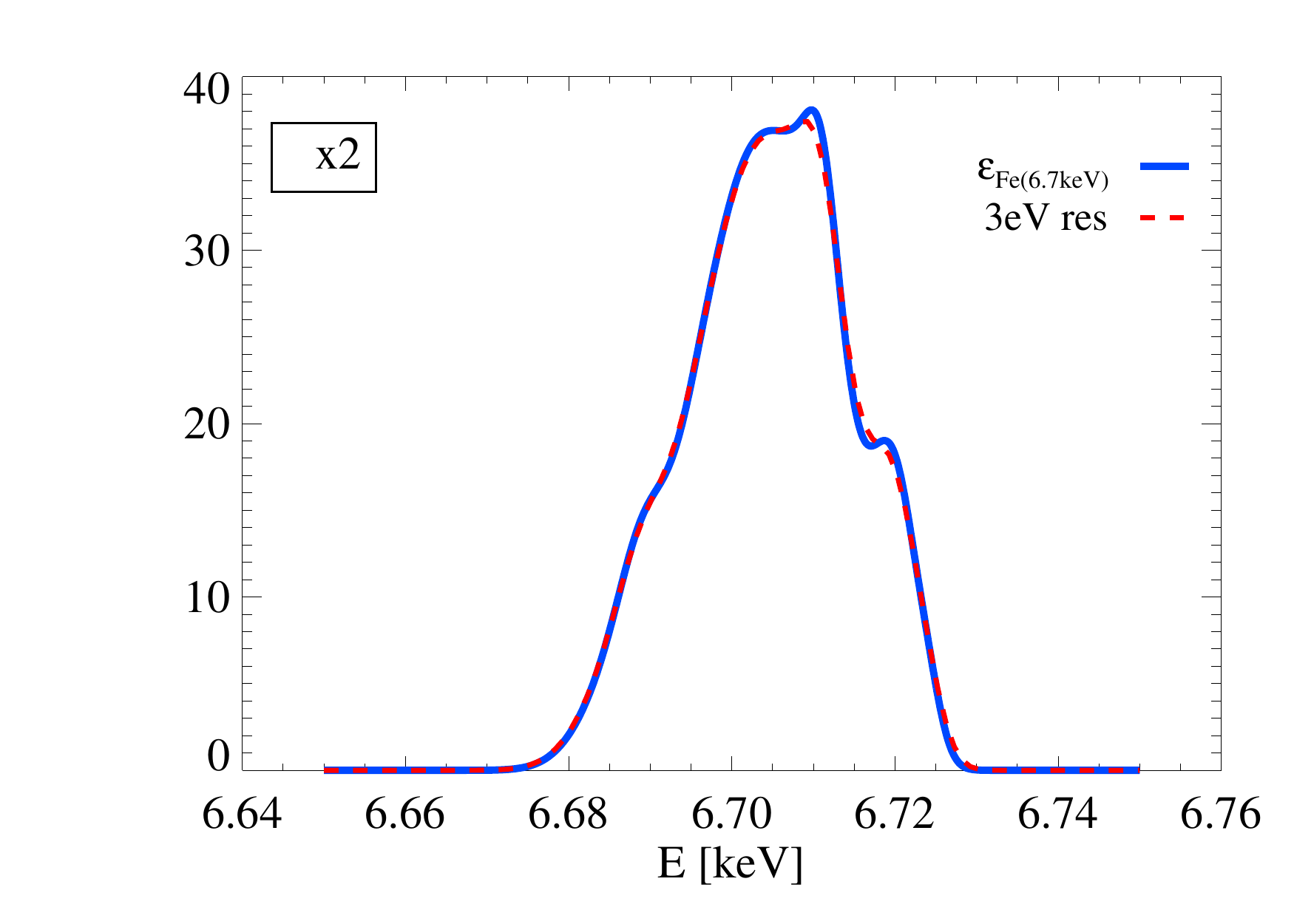}
  \caption{Same as Fig.~\ref{fig:z-pdfs1}, but for regions x1 and x2, where the l.o.s.\ is aligned with the merging axis. In the top panels (x1), one can notice the asymmetric double component distribution of ICM velocities, corresponding to the two cluster cores.
  \label{fig:x-pdfs}}
\end{figure*}

The intrinsic velocity and thermal structure of the gas along the l.o.s.\ in x1 and x2 are reported in Fig.~\ref{fig:x-pdfs} (upper and lower panels, respectively).
Similarly to Figs.~\ref{fig:z-pdfs1}--\ref{fig:z-pdfs2}, we show the l.o.s.\ velocity distribution (left) of the gas elements, comparing different weights, the Fe-emissivity of the gas in the temperature-velocity space (middle), and the theoretical shape of the toy Fe line convolved with the l.o.s.\ velocity distribution (right).

From the inspection of the theoretical gas distributions, the two merging components are clearly visible and separate in the x1 case.
As for the fainter x2 pointing, the gas velocity presents a broader, albeit not Gaussian, distribution, originating from the complex superposition of regions with different motions along the l.o.s.

\begin{figure*}
  \centering
  \includegraphics[width=0.4\textwidth,trim=0 0 0 10,clip]{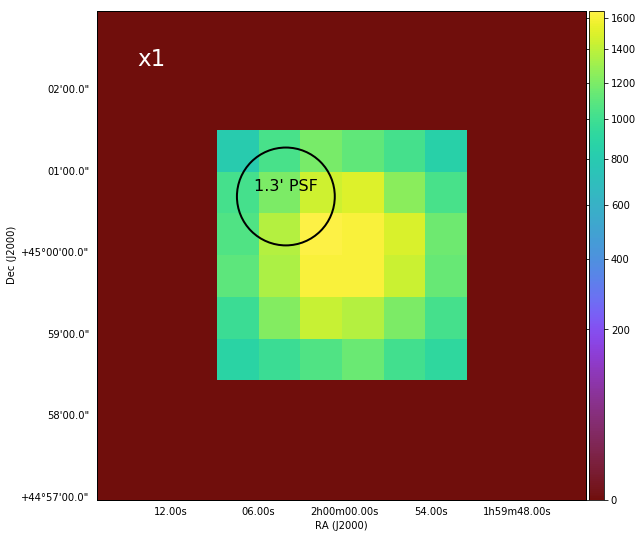}\qquad
  \includegraphics[width=0.45\textwidth,trim=10 40 40 50,clip]{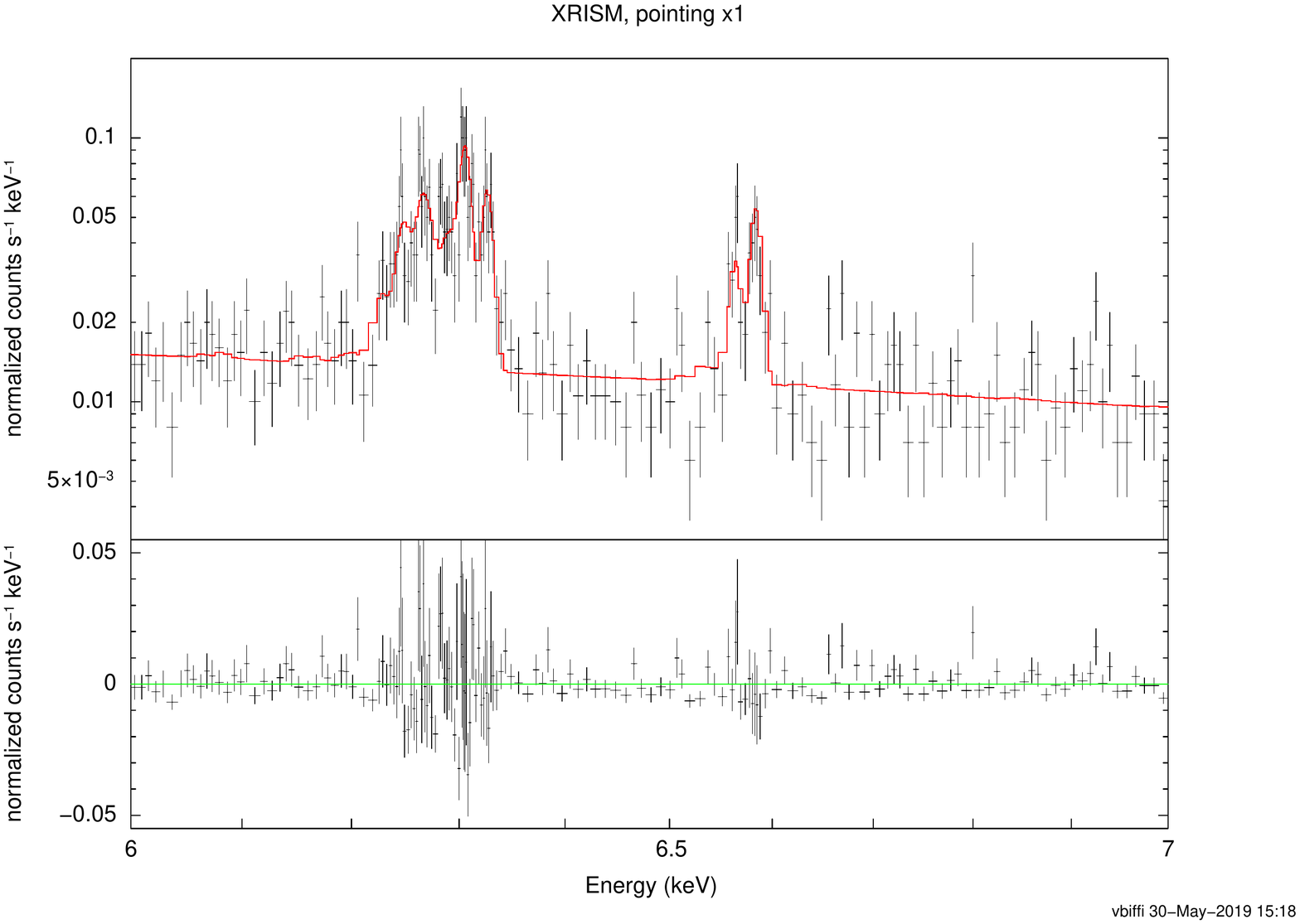}\\
  \includegraphics[width=0.4\textwidth,trim=0 0 0 10,clip]{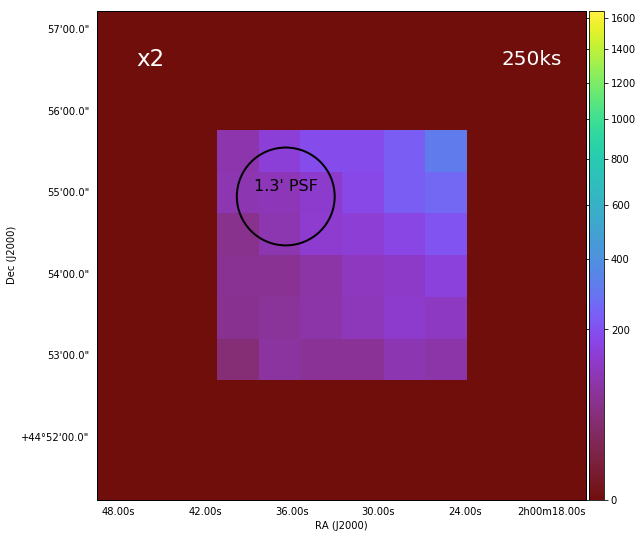}\qquad
    \includegraphics[width=0.45\textwidth,trim=10 40 40 50,clip]{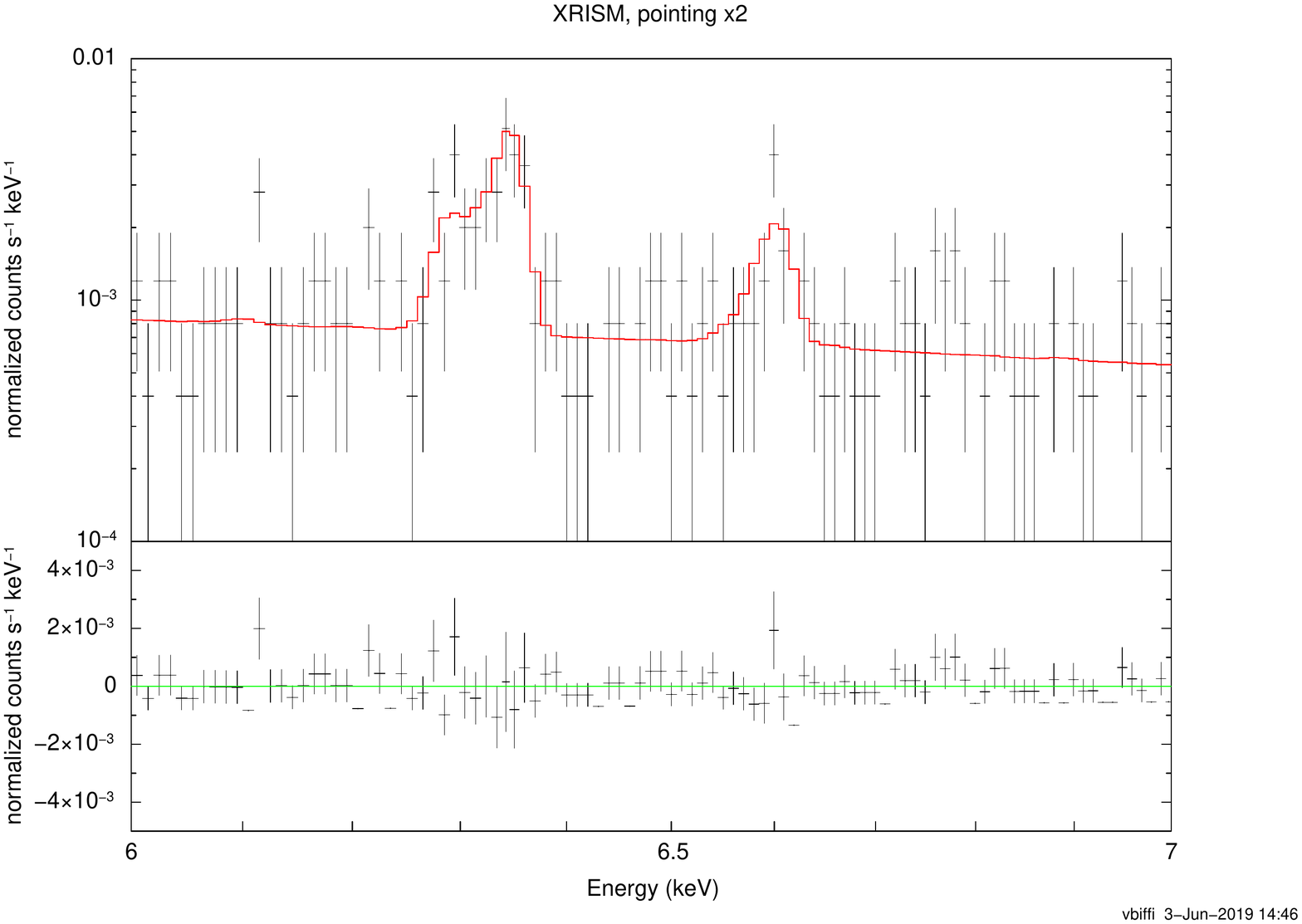}
  \caption{{\it Left:} \xrism Resolve images ($0.5$--$2$\,keV band) for the x1 (top) and x2 (bottom) pointings shown in Fig.~\ref{fig:x-pdfs}. The images (color coded with the same scale for comparison) refer to observations of 100\,ks and 250\,ks esposure, for x1 and x2 respectively. The physical size of each pointing is $200$\,kpc per side and the \xrism Resolve PSF is $1.3'$. {\it Right:} Spectrum, best-fit model and residuals in the $6$--$7$\,keV band around the Fe-K line complex,
  for the \xrism Resolve 100\,ks and 250\,ks mock observations of the x1 and x2 regions (upper and lower panel, respectively).\label{fig:xrism_xproj}}
\end{figure*}

In both cases, we use \xrism Resolve mock observations to derive spectroscopic constraints on the gas l.o.s.\ velocity shift and dispersion.
The difference in the brightness of the two regions is remarkable --- while a $100$\,ks exposure for the bright central region x1 ensures more than $1500$ counts in the 6--7\,keV band, $250$\,ks are necessary to obtain $\gtrsim 200$\,ks counts in the same energy range in x2. \xrism Resolve images (in the soft X-ray band 0.5--2\,keV) and spectra (zoomed-in onto the 6--7\,keV band), with best-fit model and residuals, are shown in Fig.~\ref{fig:xrism_xproj}.
Table~\ref{tab:xrism-x} reports the quantitative comparison between \xrism best-fit values and average estimates computed from the simulated distributions,\footnote{Also in this case, we perform 1000 realizations of the mock \xrism observations, to robustly derive the spectroscopic estimates.} for the velocity shift and dispersion in x1 and x2.
In particular, we note that the brightness of x1 allows for a double-BAPEC spectral fit, leading to a very well constrained reconstruction of the two velocity components, compared to a double Gaussian fit of the theoretical l.o.s.\ velocity distribution. Only the velocity dispersion of the secondary component ($\sigma_2$) is overestimated with respect to the simulation Gaussian-fit result, although still consistent within the $2$-$\sigma$ uncertainty. This indicates that the two Gaussian functions can capture the complex asymmetrical shape of the velocity distribution.
Interestingly, the velocity features of the fainter x2 region are well reconstructed from the spectral fit, despite the non-Gaussian shape of the velocity distribution. We expect, nevertheless, that in this fainter region the background emission --- not included here --- could compromise this reconstruction.

\begin{table*}
  \caption{Comparison between theoretical estimates and spectral best-fit values of velocity shift and dispersion, for the x1 and x2 pointings in the $x$ projection. For x1 only, the Gaussian fit of the simulation velocity distribution and the XSPEC fit are both performed with 2 Gaussian components, in order to approximate the 2-component velocity distribution shown in Fig.~\ref{fig:x-pdfs} (upper panels). 
  \label{tab:xrism-x}}
  \centering \renewcommand{\arraystretch}{1.2}
    \scriptsize \begin{tabular}{c|cc|cccc|ccccc}
    \hline
    & \multicolumn{2}{c|}{Theo. estimate} & \multicolumn{4}{c|}{Gaussian fit} & \multicolumn{5}{c}{XSPEC fit}\\
     & $\mu_{wFe}$ & $\sigma_{wFe}$  
             & $\mu_{1}$ & $\sigma_{1}$ & $\mu_{2}$ & $\sigma_{2}$  
             & $\mu_{1}$ & $\sigma_{1}$ & $\mu_{2}$ & $\sigma_{2}$  & exp
             \\
             & [km/s] & [km/s] & [km/s] & [km/s] 
             & [km/s] & [km/s] & [km/s] & [km/s] 
             & [km/s] & [km/s] & [Ms]\\[2pt]
    \hline
    x1  & -992.3  & 591.2  
    & -1629.9 & 251.7 & -630.4 & 85.2 & 
    $ -1693.8 \pm 80.4 $ & $274.7  \pm 102.1$ &  
    $ -637.7 \pm 42.3 $ & $ 171.1 \pm 59.0$ & 0.1\\
    x2  & 206.2 & 437.4 & 231.0 & 479.4 & -- & -- & 
    $255.2 \pm 139.0$  & $425.5 \pm 137.5$ & -- & -- &0.25\\
    \hline
  \end{tabular}
\end{table*}

\subsection{Insights from the kinetic Sunyaev-Zeldovich effect}

For comparison, we include maps of the thermal and kinetic Sunyaev-Zeldovich effect, characterizing the two respectively by the Compton $y$ parameter, $y=(\sigma_\textsc{t} / m_{\rm e} c^2)\int n_{\rm e} k_\textsc{b} T_{\rm e} d\ell$, and the so-called $y_{\rm kSZ}$ parameter, $y_{\rm kSZ}=(\sigma_\textsc{t}/c) \int n_{\rm e} v_z d\ell$ \citep[e.g.][]{mroczkowski2019}, ignoring secondary scattering effects proportional to $e^{-\tau}$ for electron opacity $\tau = \sigma_\textsc{t} \int n_{\rm e} d\ell$.
As noted in the introduction (Section~\ref{sec:intro}), the kinetic Sunyaev-Zeldovich effect can provide additional insight into the internal and bulk motions of the ICM.  It can also be seen in the definition of $y_{\rm kSZ}$ that the kinetic SZ traces the momentum of the gas along the line of sight. 
The thermal and kinetic SZ signals can be further expressed in terms of the effective CMB temperature~\cite[see e.g. the review by][]{mroczkowski2019}.
In particular, the change in $T_\textsc{cmb}$ due to the thermal SZ effect can be computed from the $y$ parameter as
$dT_{\rm tSZ}/T_\textsc{cmb} = f(x) \, y$, where classically $f(x)= x (e^x +1)/(e^x - 1) - 4$ and $x = h \nu / k_\textsc{b} T_{\rm e}$ is the dimensionless frequency.
Similarly, the strength of the kSZ can be computed knowing $y_{\rm kSZ}$ as $dT_{\rm kSZ}/T_\textsc{cmb} = y_{\rm kSZ}$, where we note that classically the kSZ is simply a Doppler shift in the CMB.
This feature of the classical kSZ means it cannot be distinguished from temperature fluctuations in the primary CMB, which are on the order few $\mu$K for arcminute scales \citep[e.g.][]{Dunkley2013}.

\begin{figure}
    \centering
    \includegraphics[width=0.9\columnwidth,trim=0 0 10 0,clip]{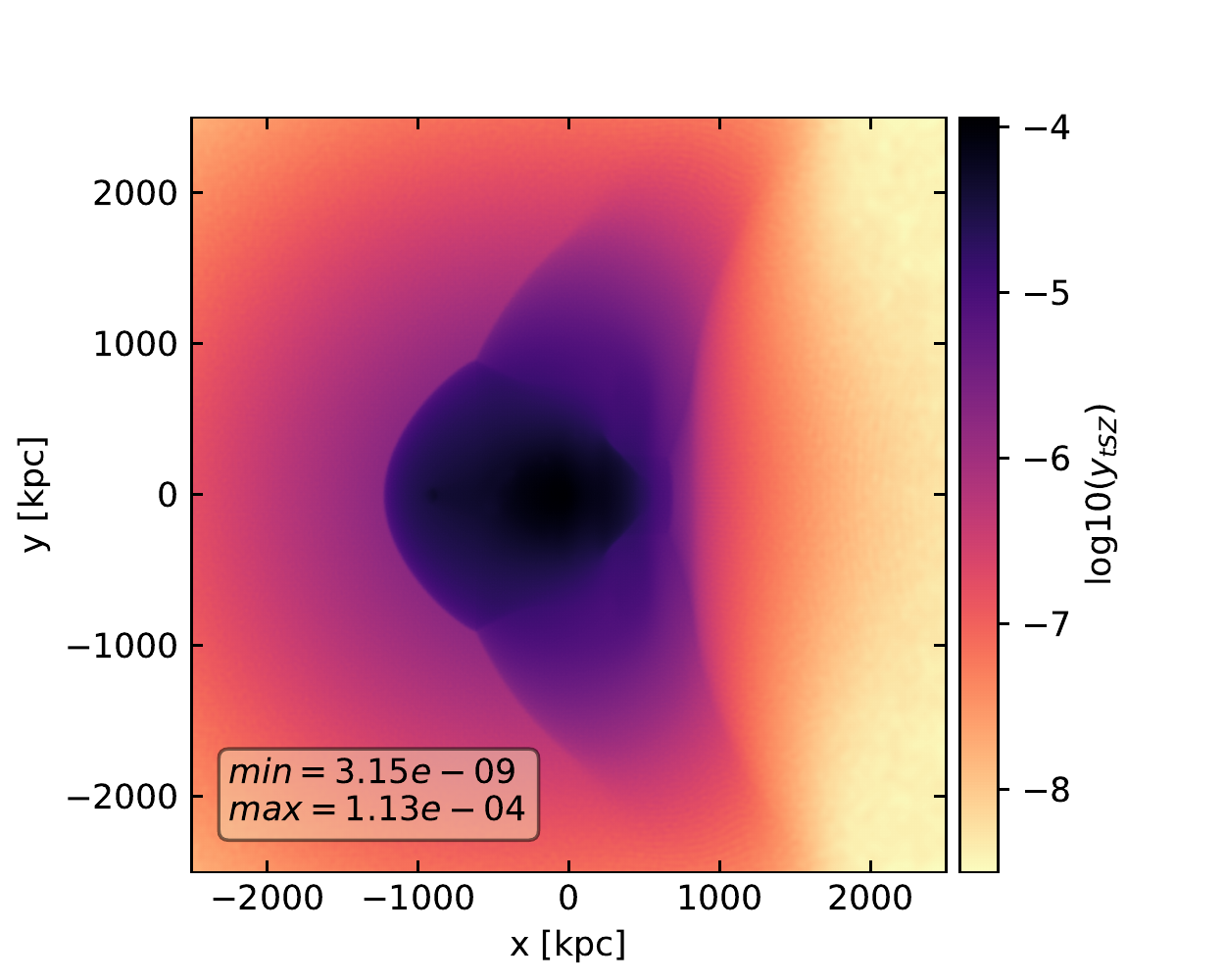}
    \includegraphics[width=0.9\columnwidth,trim=0 0 10 0,clip]{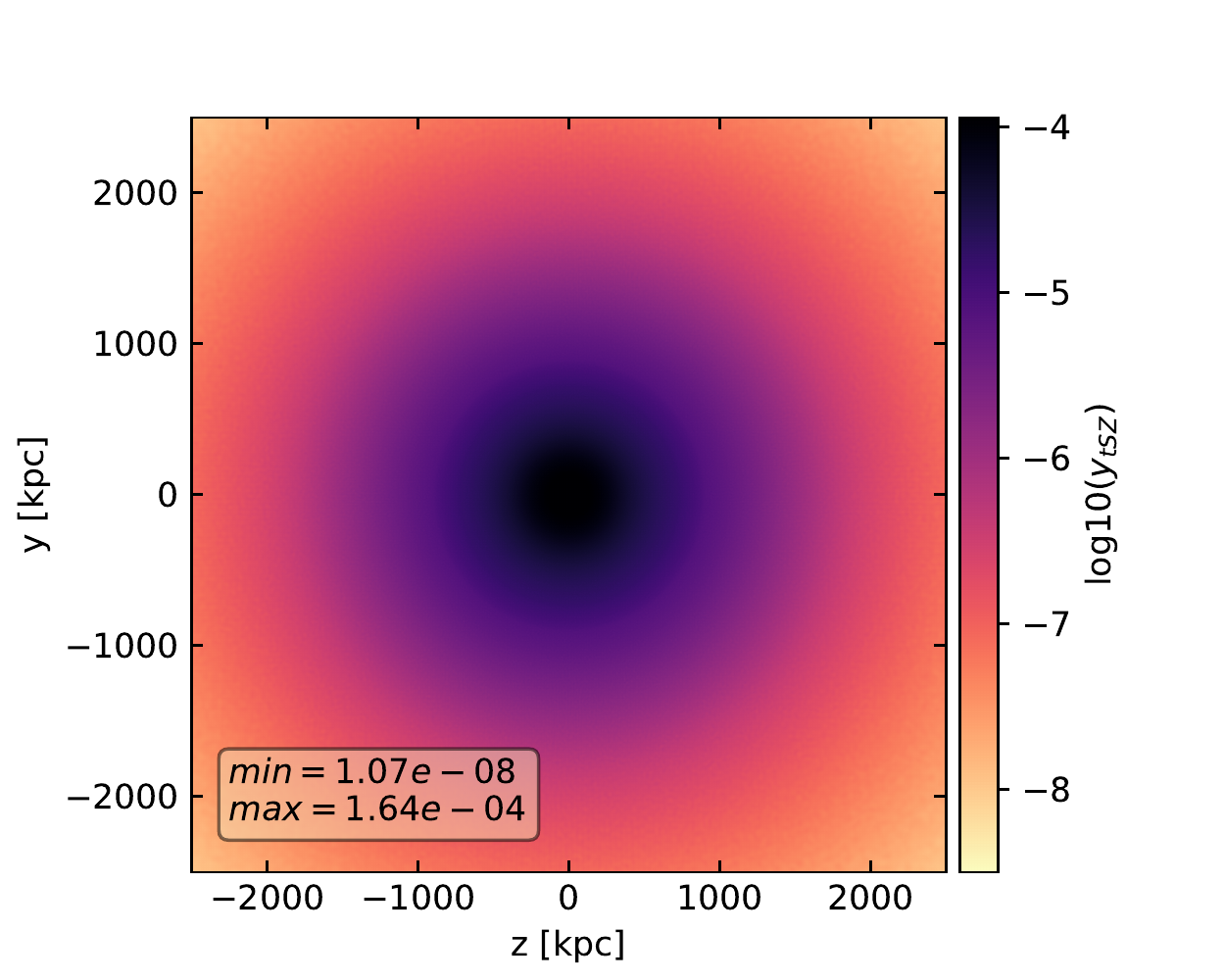}
    \caption{Noise free, native resolution thermal SZ images of the plane-of-sky (upper) and line-of-sight geometry (lower) configurations, corresponding to Figures \ref{fig:map} and \ref{fig:map.x} respectively.   }
    \label{fig:tsz}
\end{figure}

\begin{figure}
    \centering
    \includegraphics[width=0.9\columnwidth,trim=0 0 10 0,clip]{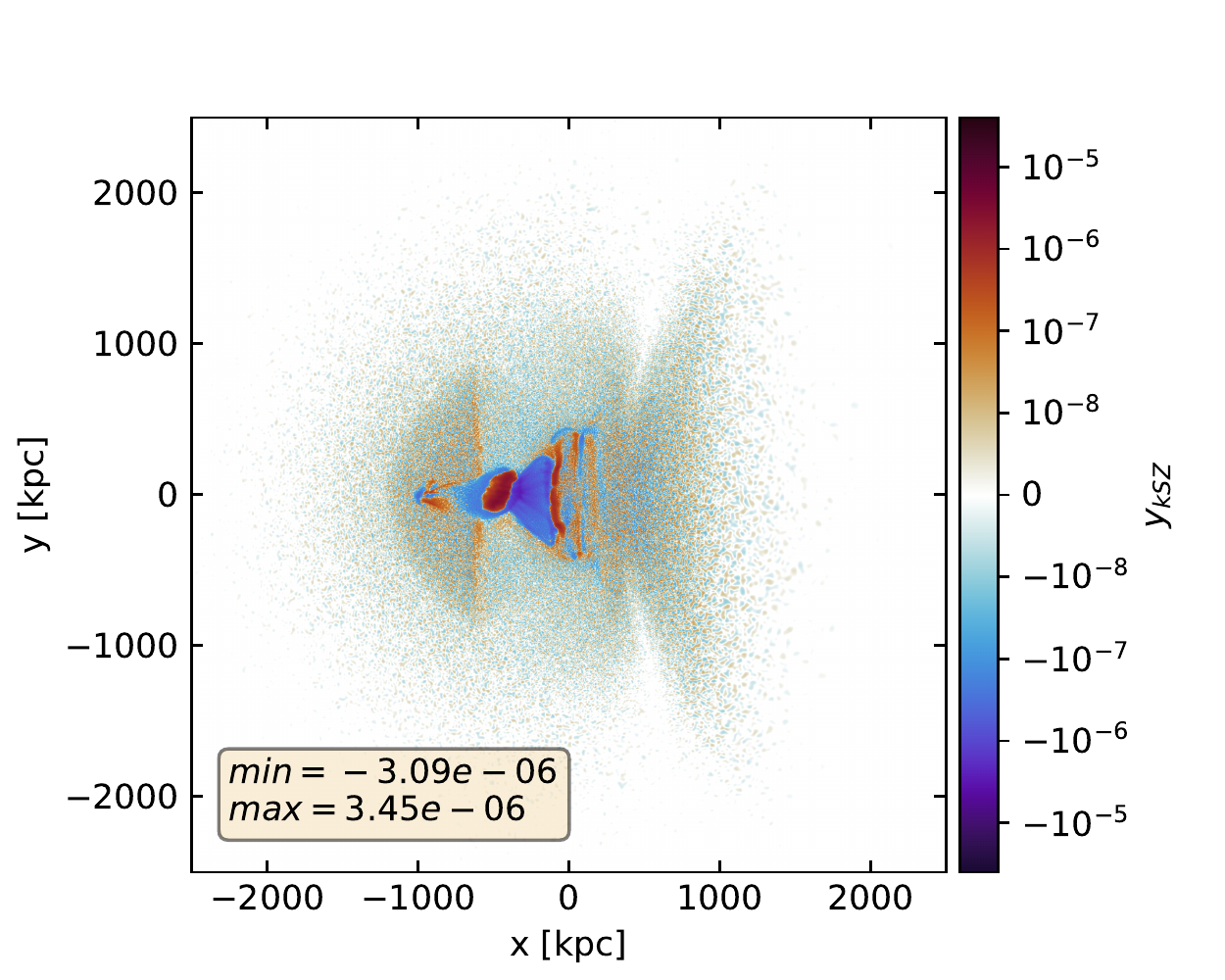}
    \includegraphics[width=0.9\columnwidth,trim=0 0 10 0,clip]{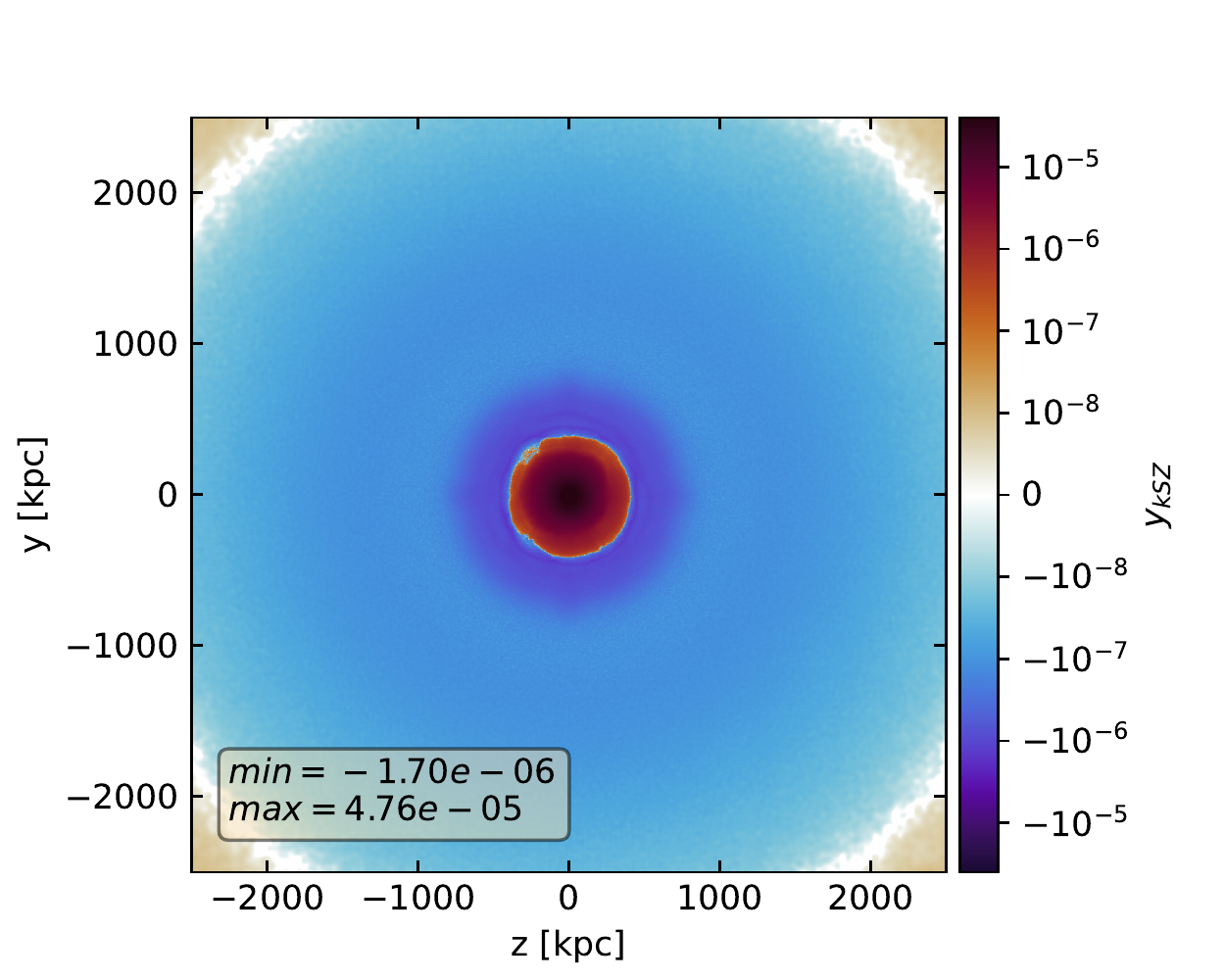}
    \caption{Noise free, native resolution kinetic SZ images of the plane-of-sky (upper) and line-of-sight geometry (lower) configurations, corresponding to Figures \ref{fig:map} and \ref{fig:map.x} respectively.}
    \label{fig:ksz}
\end{figure}

To assess the utility of the kSZ for providing complementary constraints on ICM motions to those from X-ray calorimeters, we show in Figures~\ref{fig:tsz} and~\ref{fig:ksz} projections of $y$ and $y_{\rm kSZ}$ at the resolution of our simulations.  
In the upper panel of Figure~\ref{fig:tsz}, corresponding to the plane-of-sky merger geometry, the shock front, which is fundamentally a pressure discontinuity, is readily apparent.  We note that studies such as those in~\cite{Basu2016} and~\cite{DiMascolo2019} have exploited similar geometries to constrain the properties of shocks using high resolution observations of the tSZ effect.
For comparison of the relative strengths of the tSZ and kSZ signals, we note that near 150~GHz, a frequency often chosen for tSZ surveys, $f(x)$ is nearly unity in amplitude.  This implies that in the line of sight case, the strength of tSZ is roughly one order of magnitude greater than that of the kSZ, while for the plane of the sky merger the tSZ is predicted to be two orders of magnitude greater.
For the line of sight merger geometry case, the peak kSZ signal $y_{\rm kSZ} \gtrsim 10^{-5}$ implies it would be detectable above the primary CMB.
In a future work we will explore the ability to constrain the merger inclination angle.

\begin{figure*}
    \centering
    \includegraphics[width=0.32\textwidth,trim=0 0 10 0,clip]{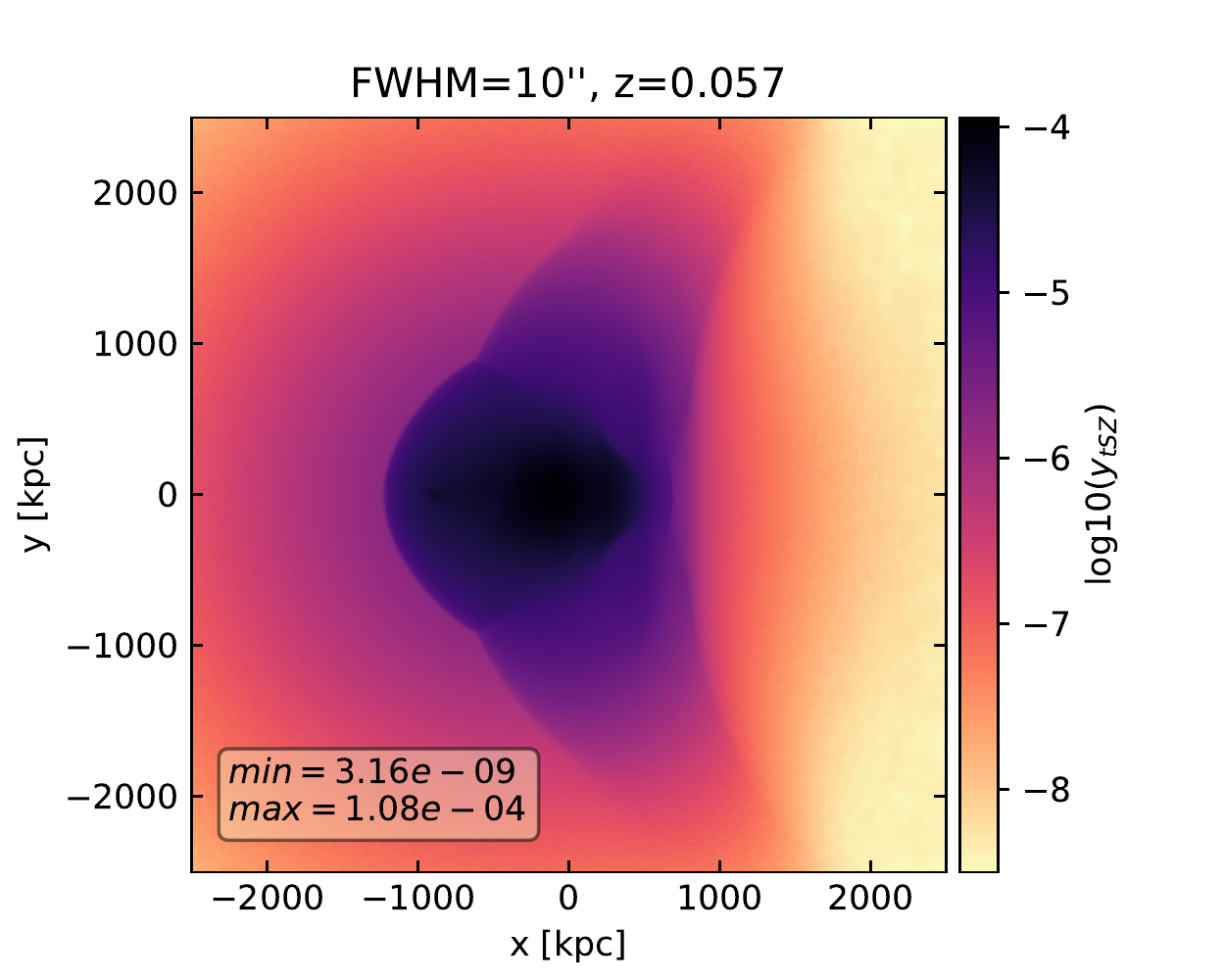}
    \includegraphics[width=0.32\textwidth,trim=0 0 10 0,clip]{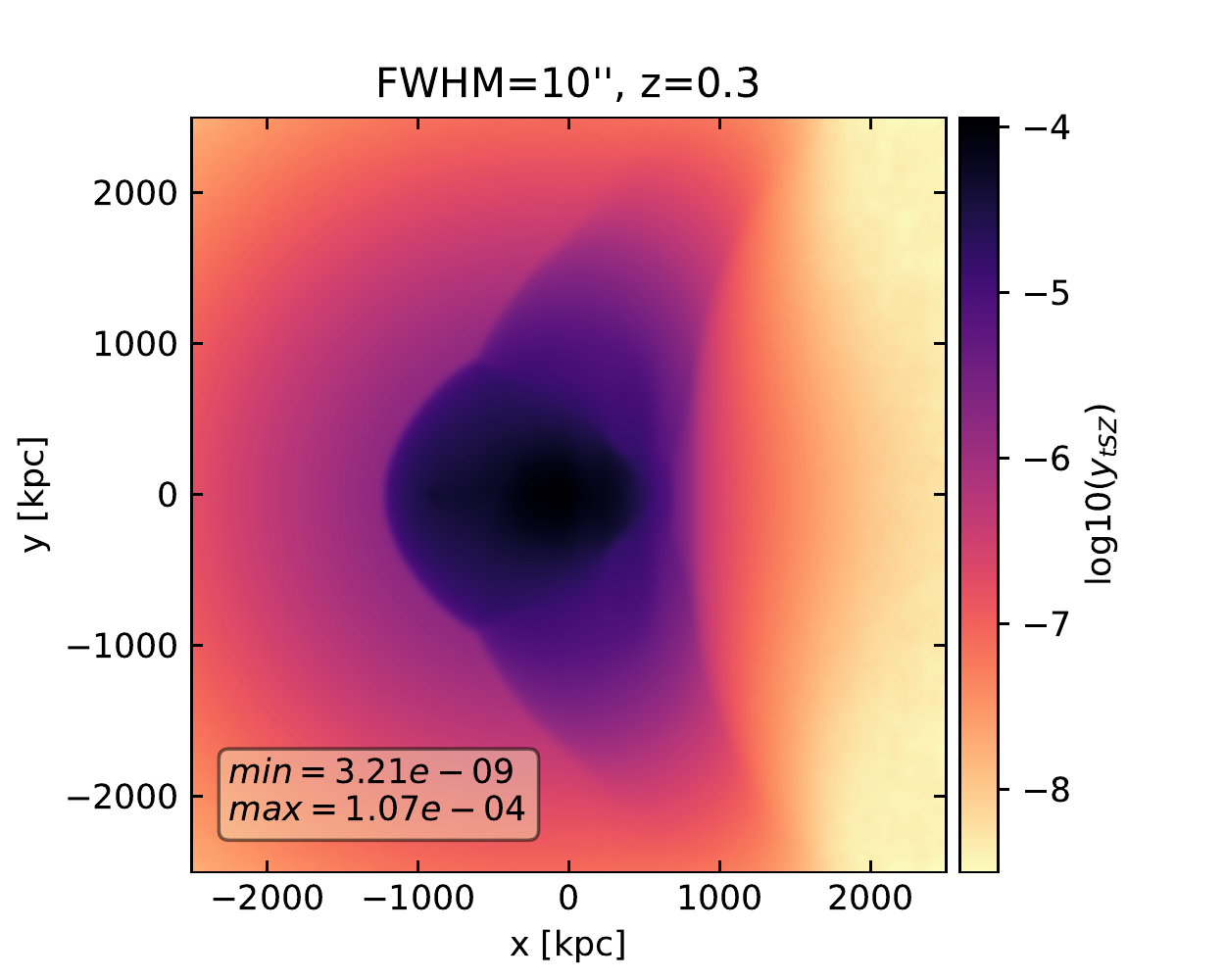}
    \includegraphics[width=0.32\textwidth,trim=0 0 10 0,clip]{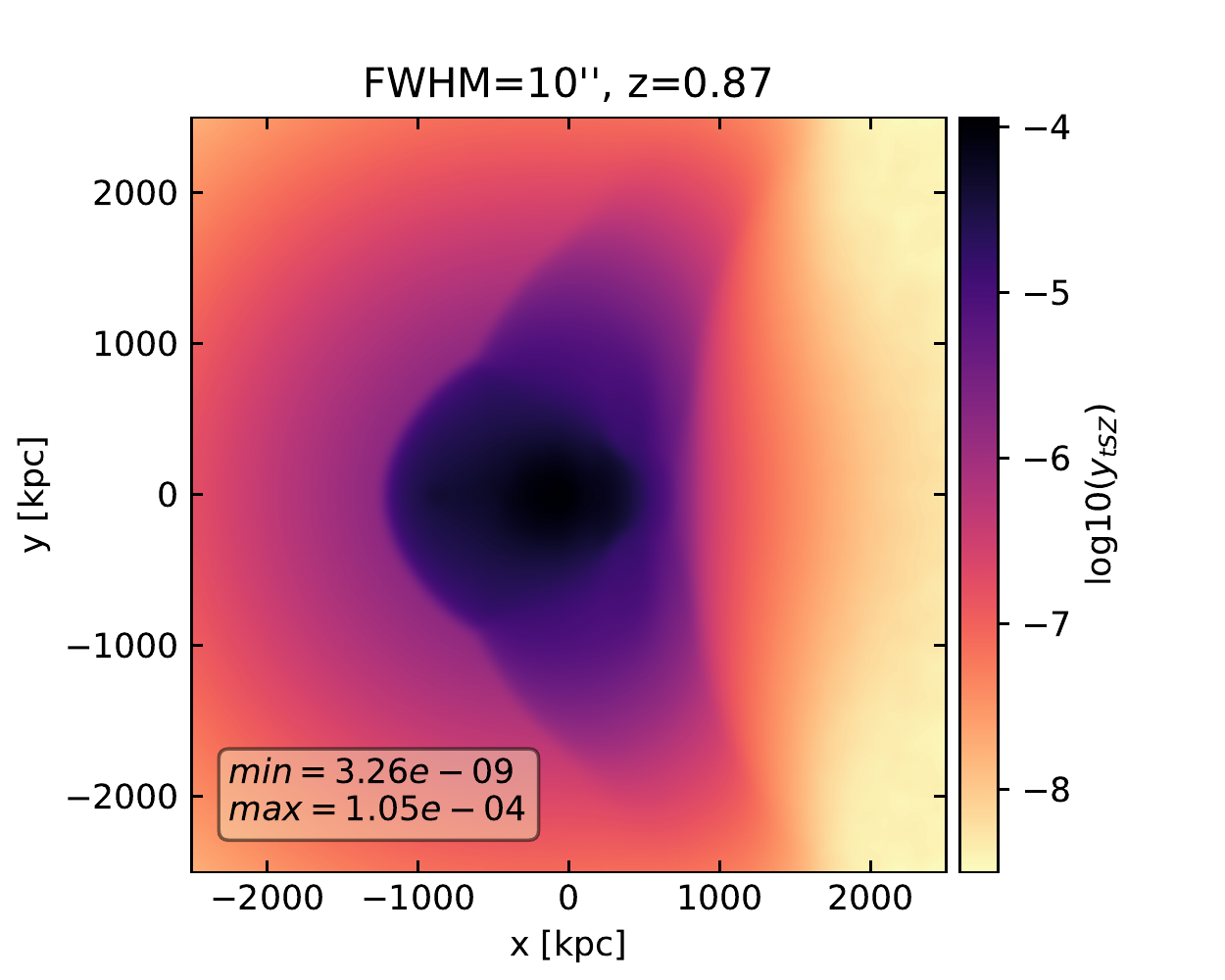}
    \\
    \includegraphics[width=0.32\textwidth,trim=0 0 10 0,clip]{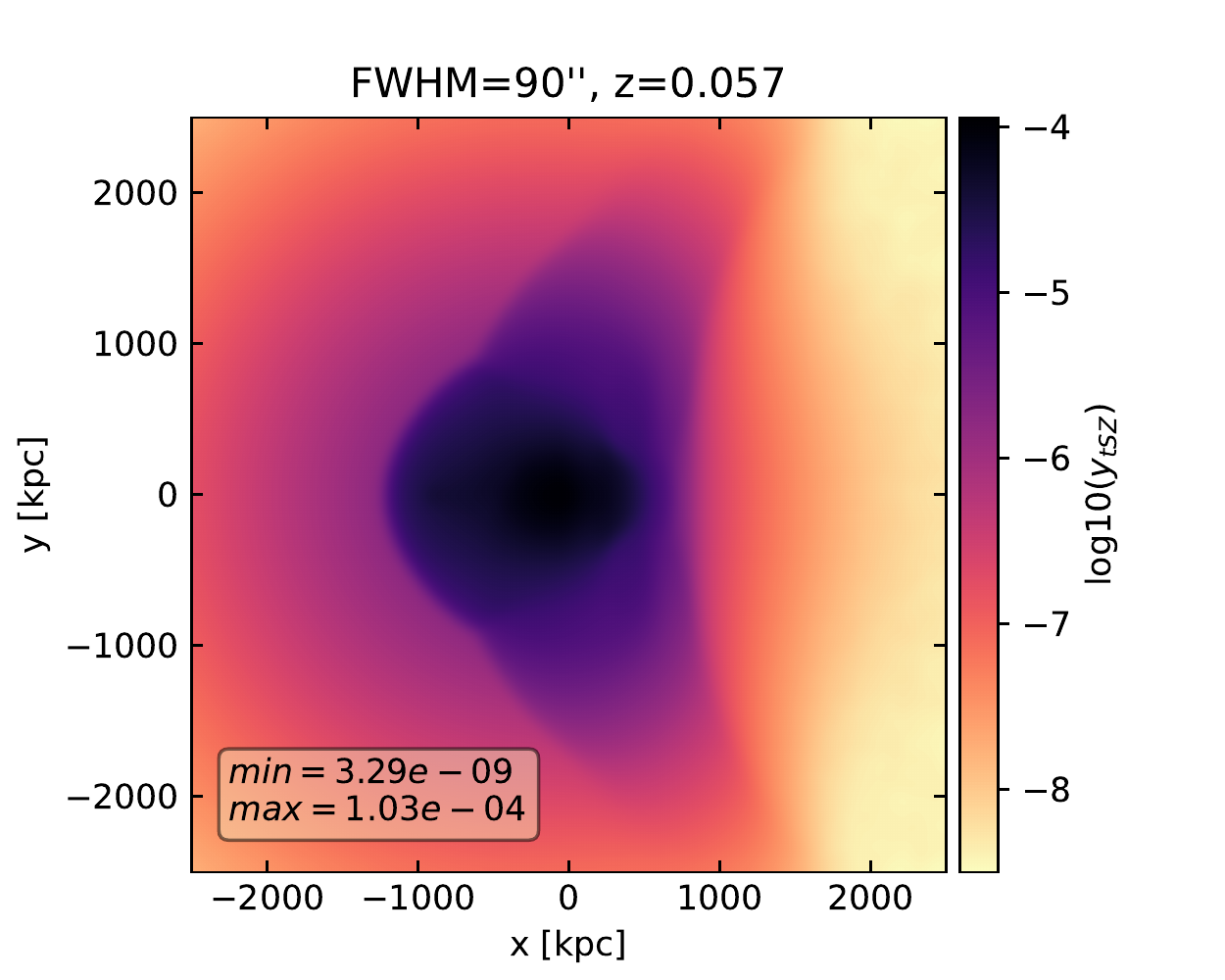}
    \includegraphics[width=0.32\textwidth,trim=0 0 10 0,clip]{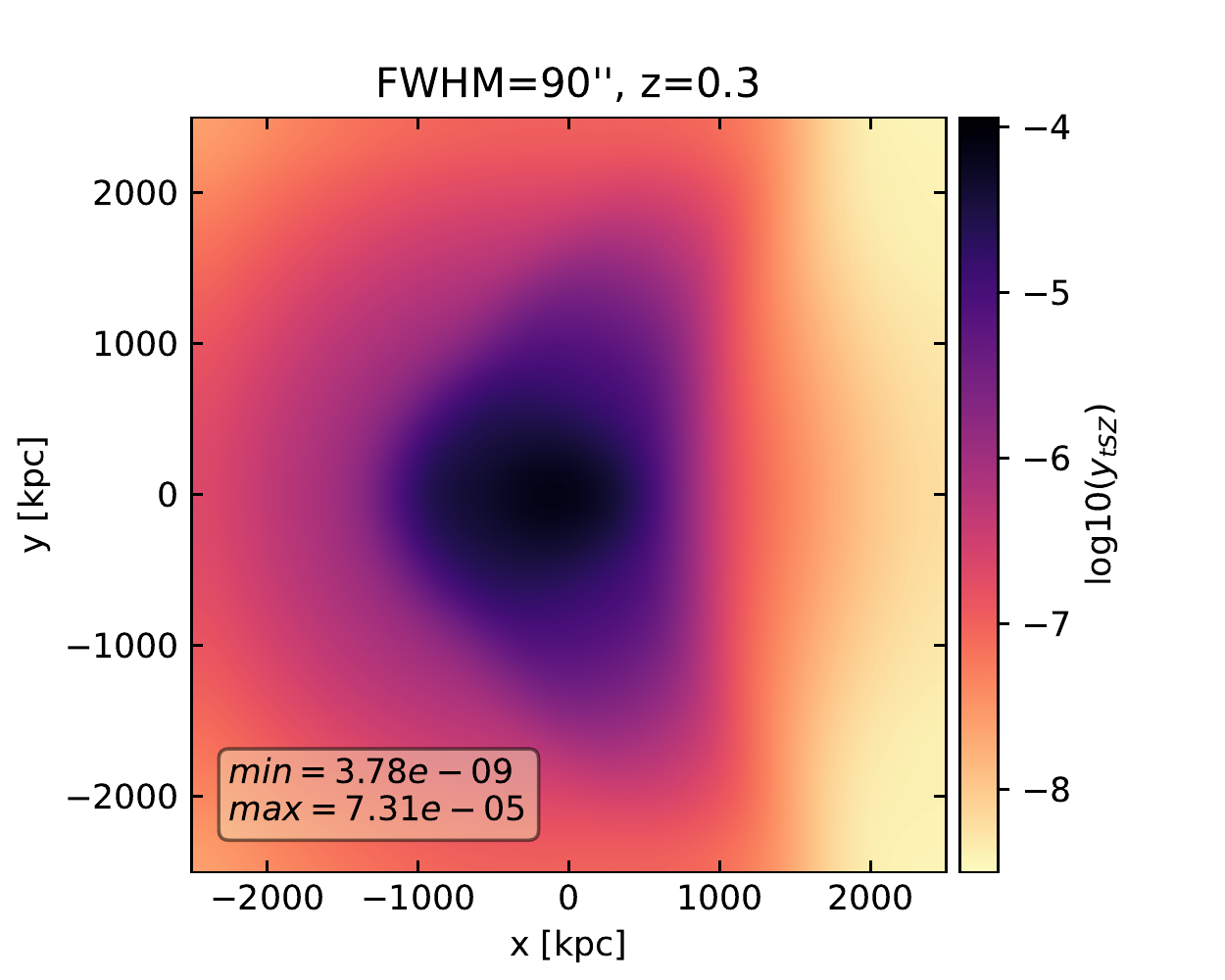}
    \includegraphics[width=0.32\textwidth,trim=0 0 10 0,clip]{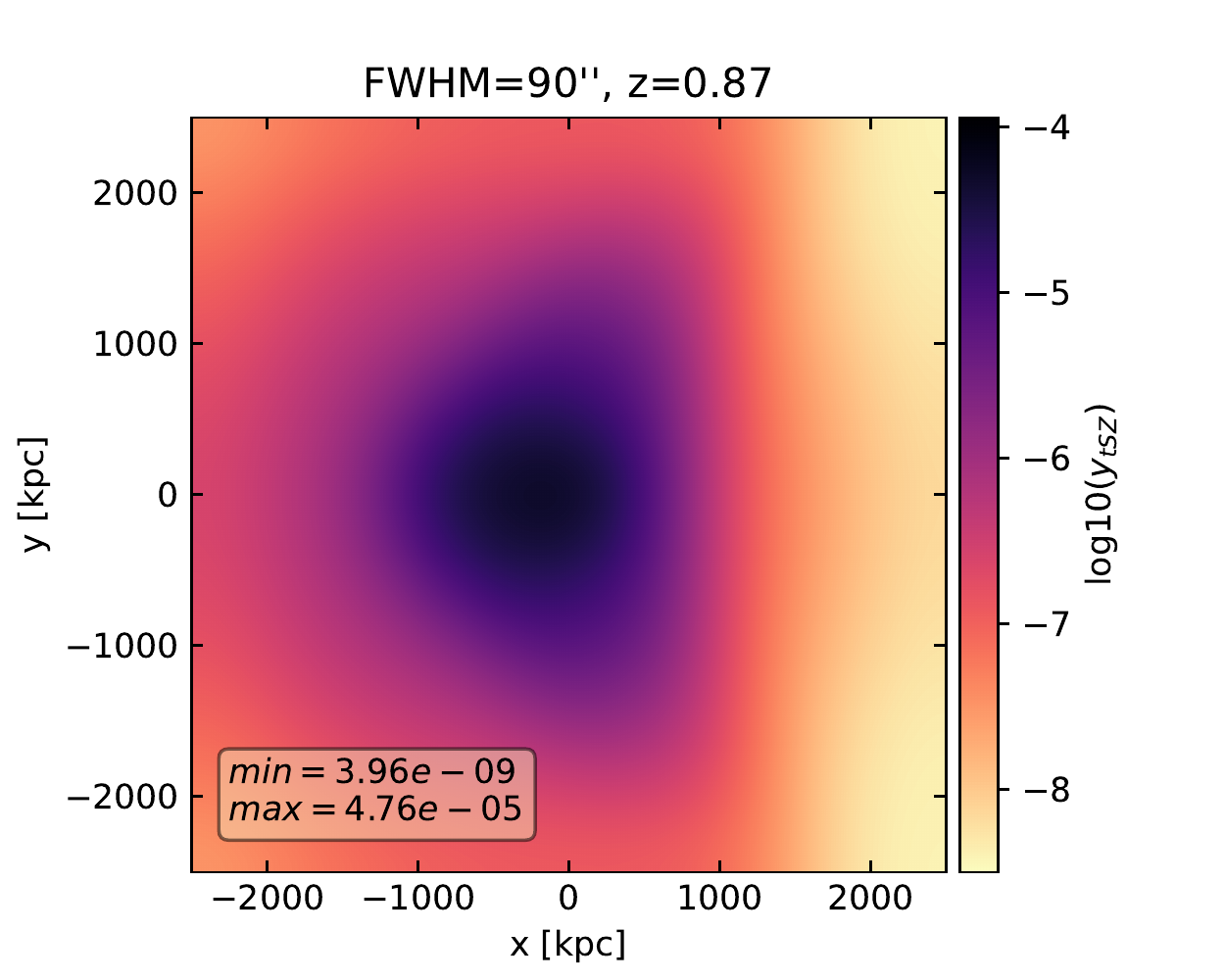}
    \\
    \includegraphics[width=0.32\textwidth,trim=0 0 10 0,clip]{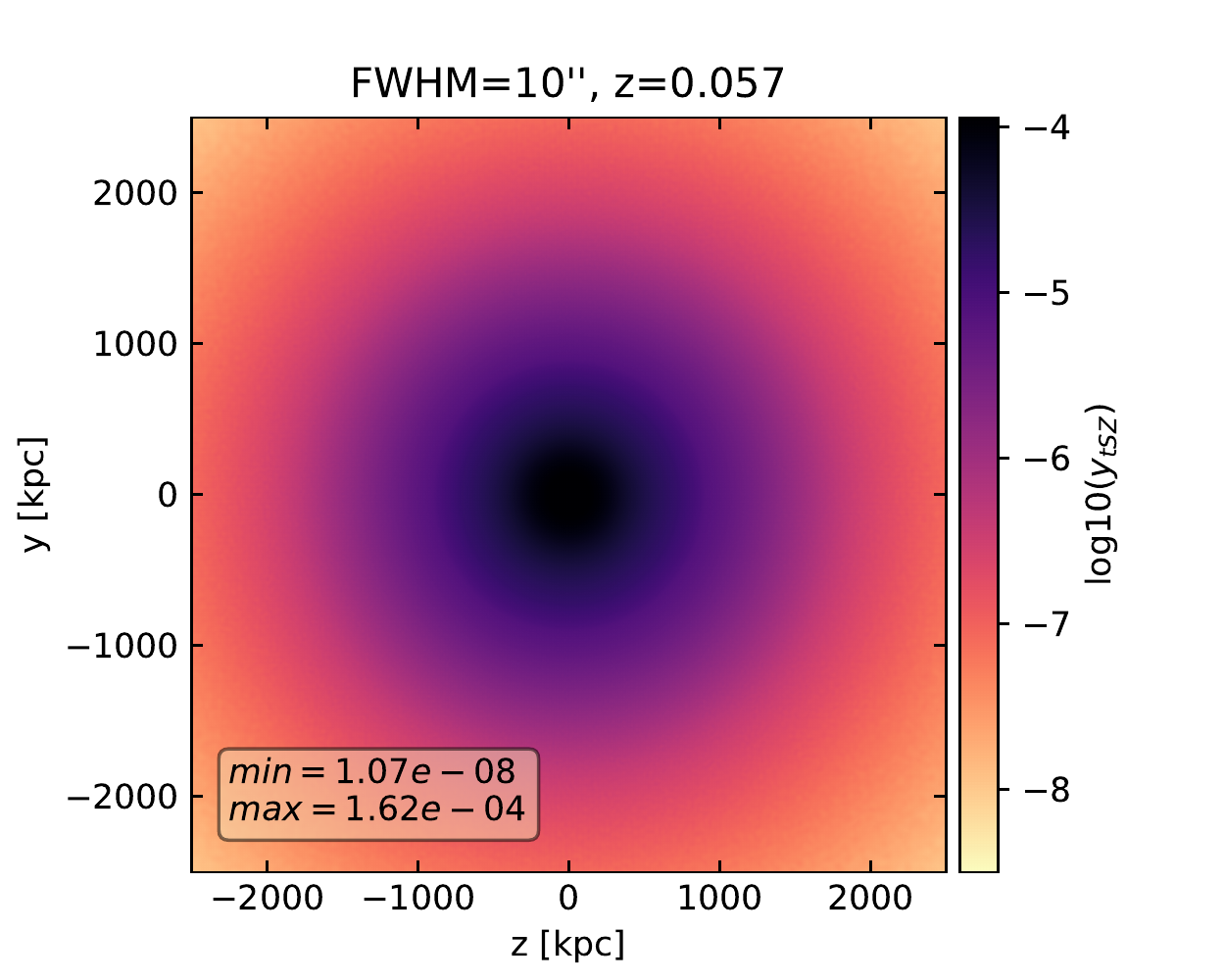}
    \includegraphics[width=0.32\textwidth,trim=0 0 10 0,clip]{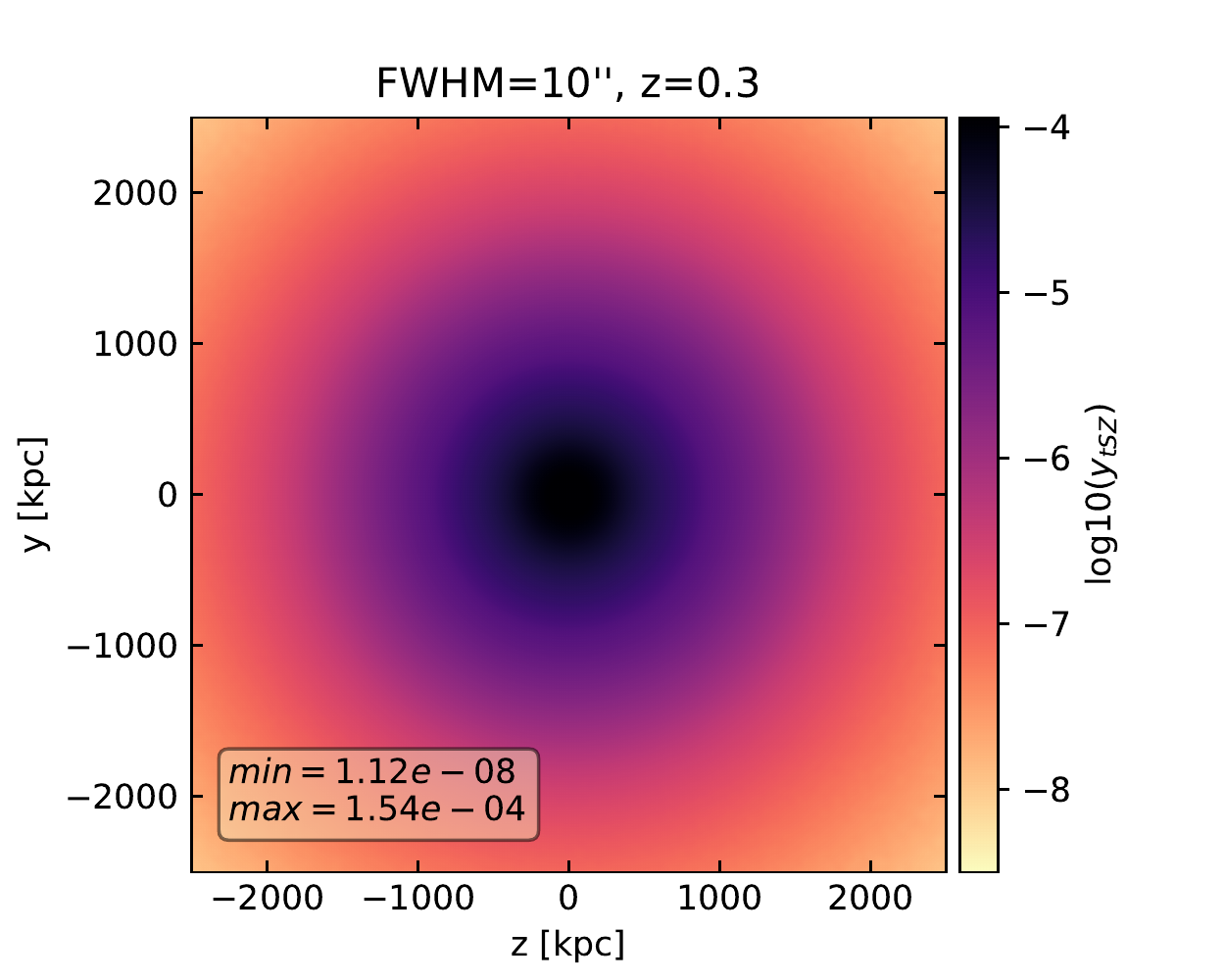}
    \includegraphics[width=0.32\textwidth,trim=0 0 10 0,clip]{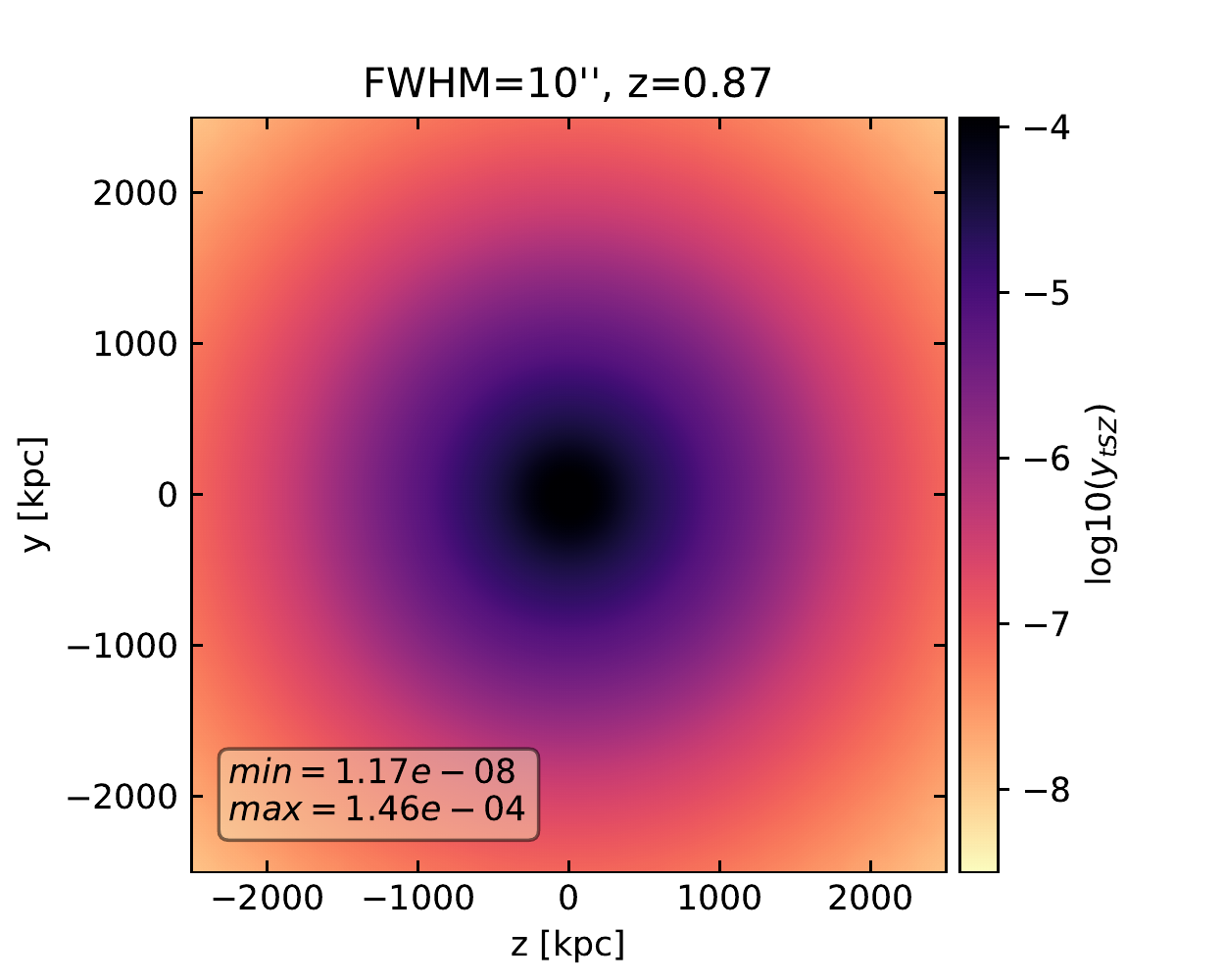}
    \\
    \includegraphics[width=0.32\textwidth,trim=0 0 10 0,clip]{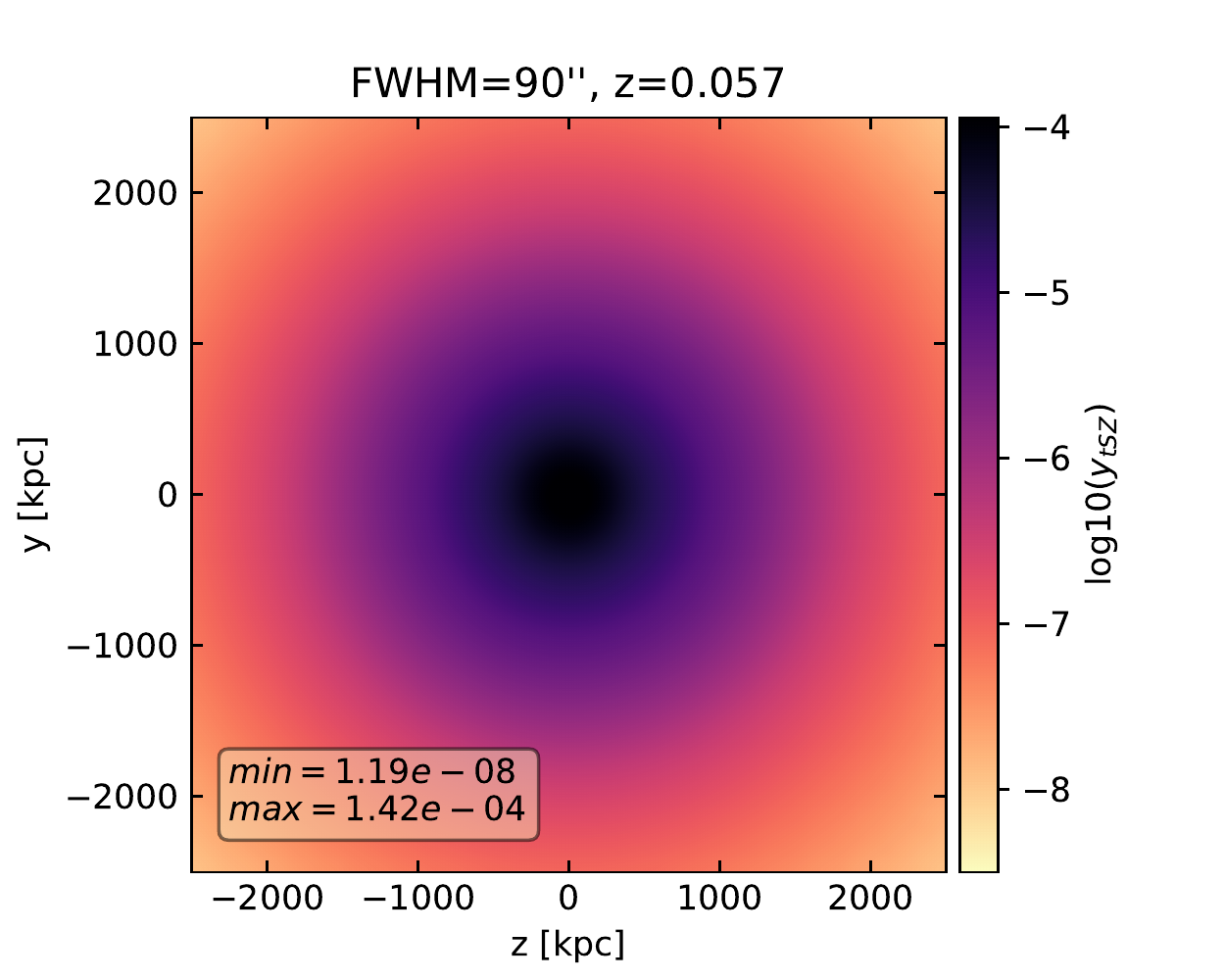}
    \includegraphics[width=0.32\textwidth,trim=0 0 10 0,clip]{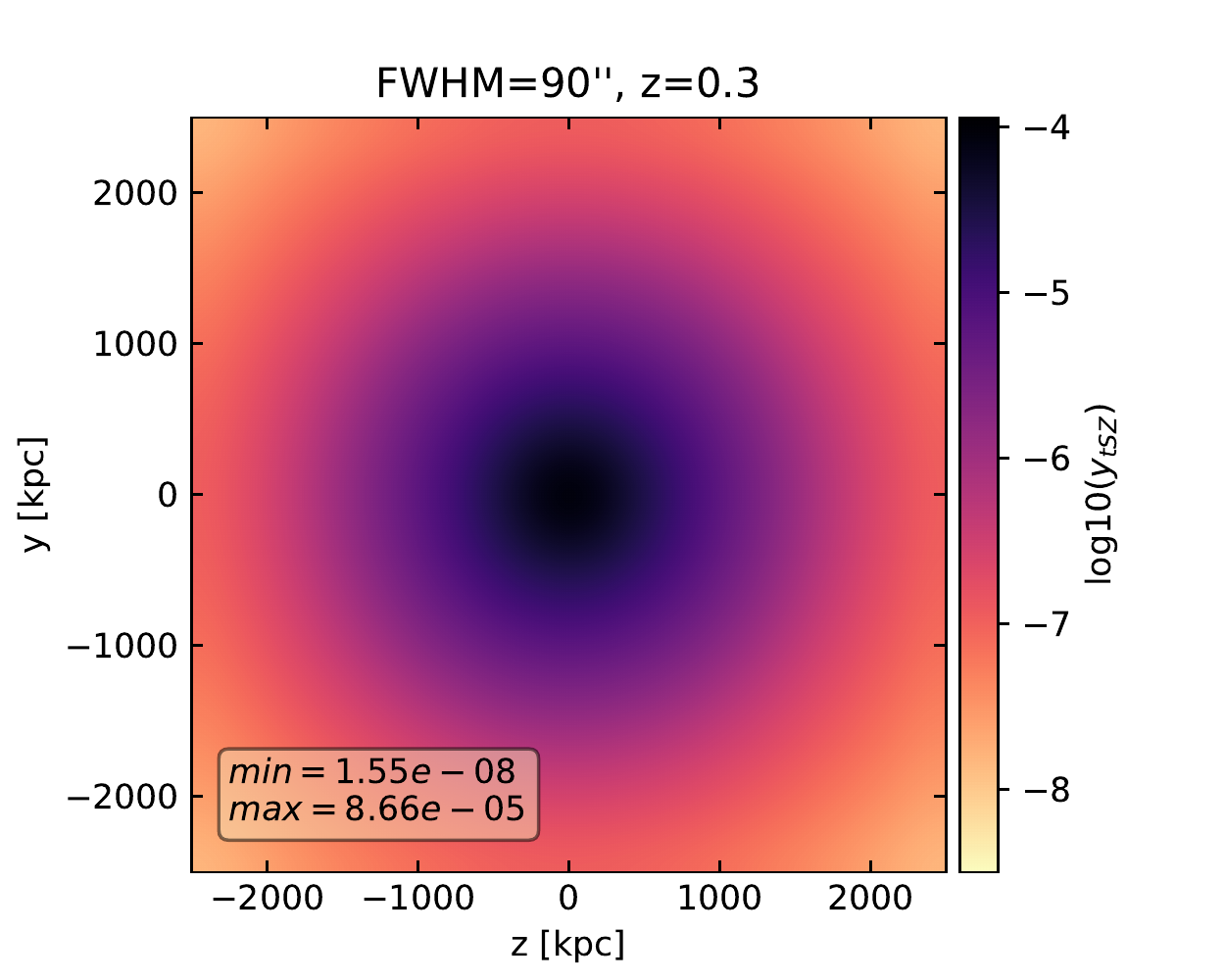}
    \includegraphics[width=0.32\textwidth,trim=0 0 10 0,clip]{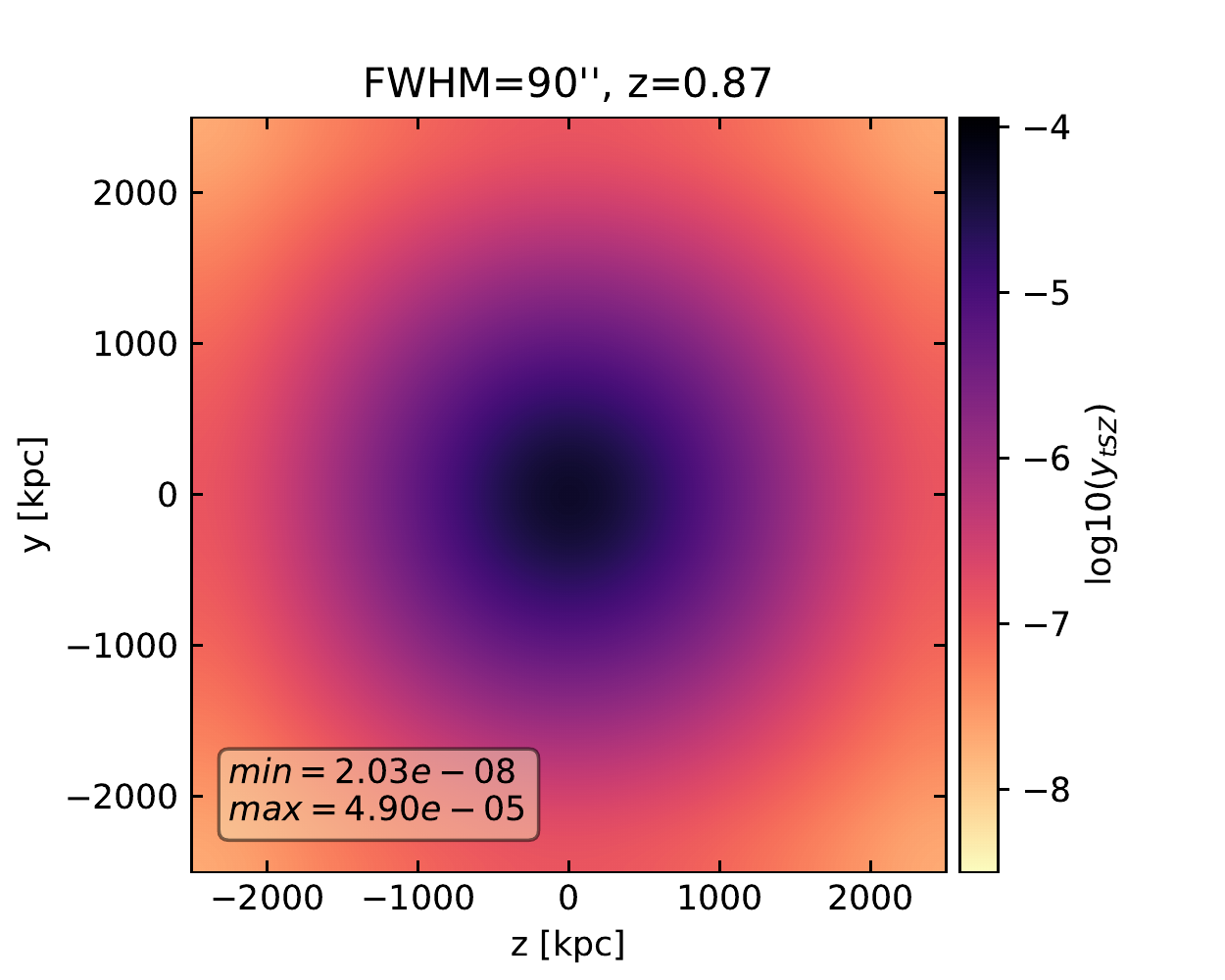}
    \caption{Noise free, smoothed thermal SZ images of the plane-of-sky (upper 6 panels) and line-of-sight geometries (lower 6 panels), corresponding to the upper and lower panels of Figure \ref{fig:tsz}.
    The first and third rows show the images for an AtLAST-like resolution at 150~GHz, while the second and third rows show the images for a CMB-S4-like resolution.  The columns correspond to the representative redshifts of $z=[0.057, 0.3, 0.87]$, chosen to correspond to the mergers in e.g. A3667, the Bullet Cluster, and El Gordo.}
    \label{fig:tsz_smooth}
\end{figure*}

\begin{figure*}
    \centering
    \includegraphics[width=0.32\textwidth,trim=0 0 10 0,clip]{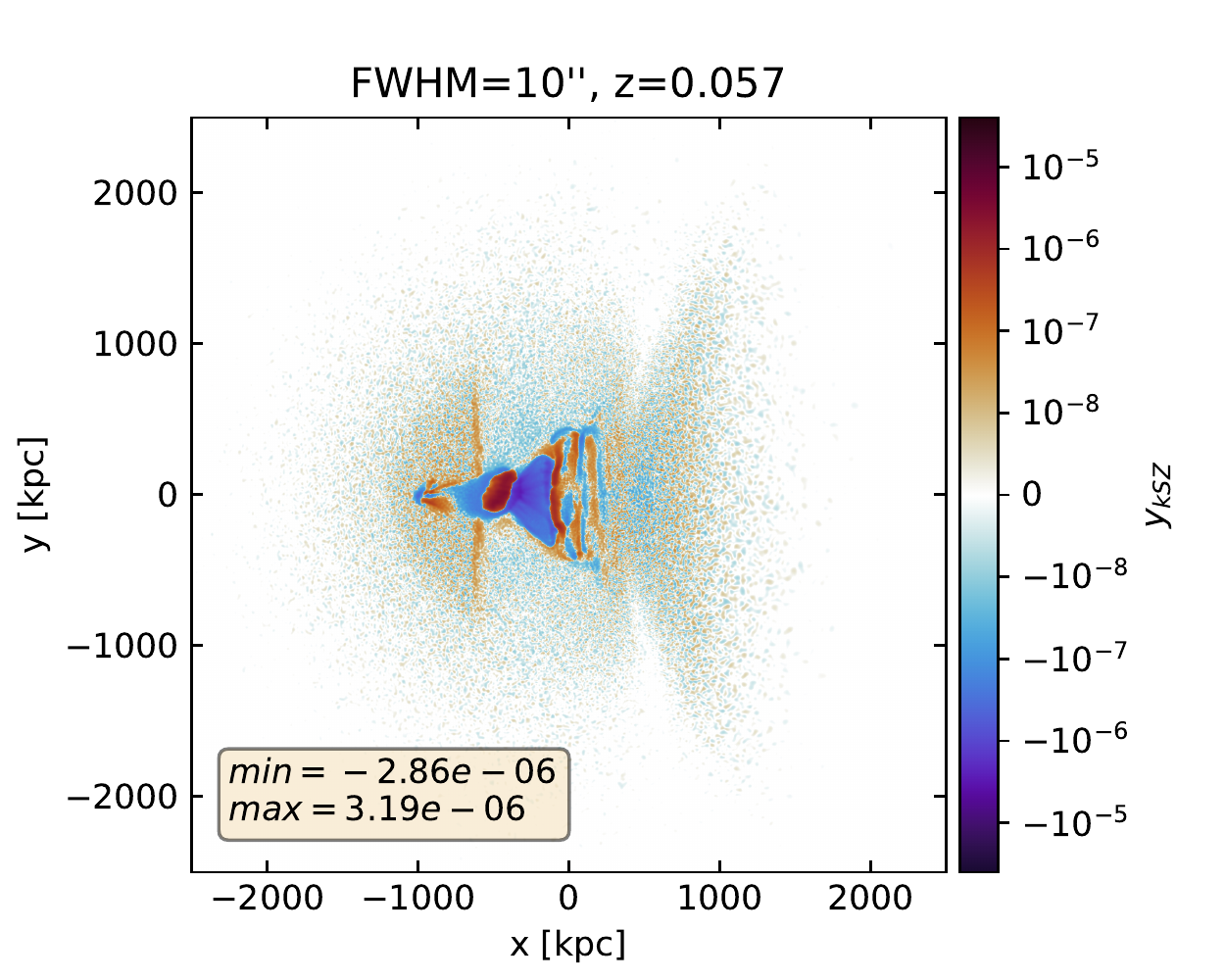}
    \includegraphics[width=0.32\textwidth,trim=0 0 10 0,clip]{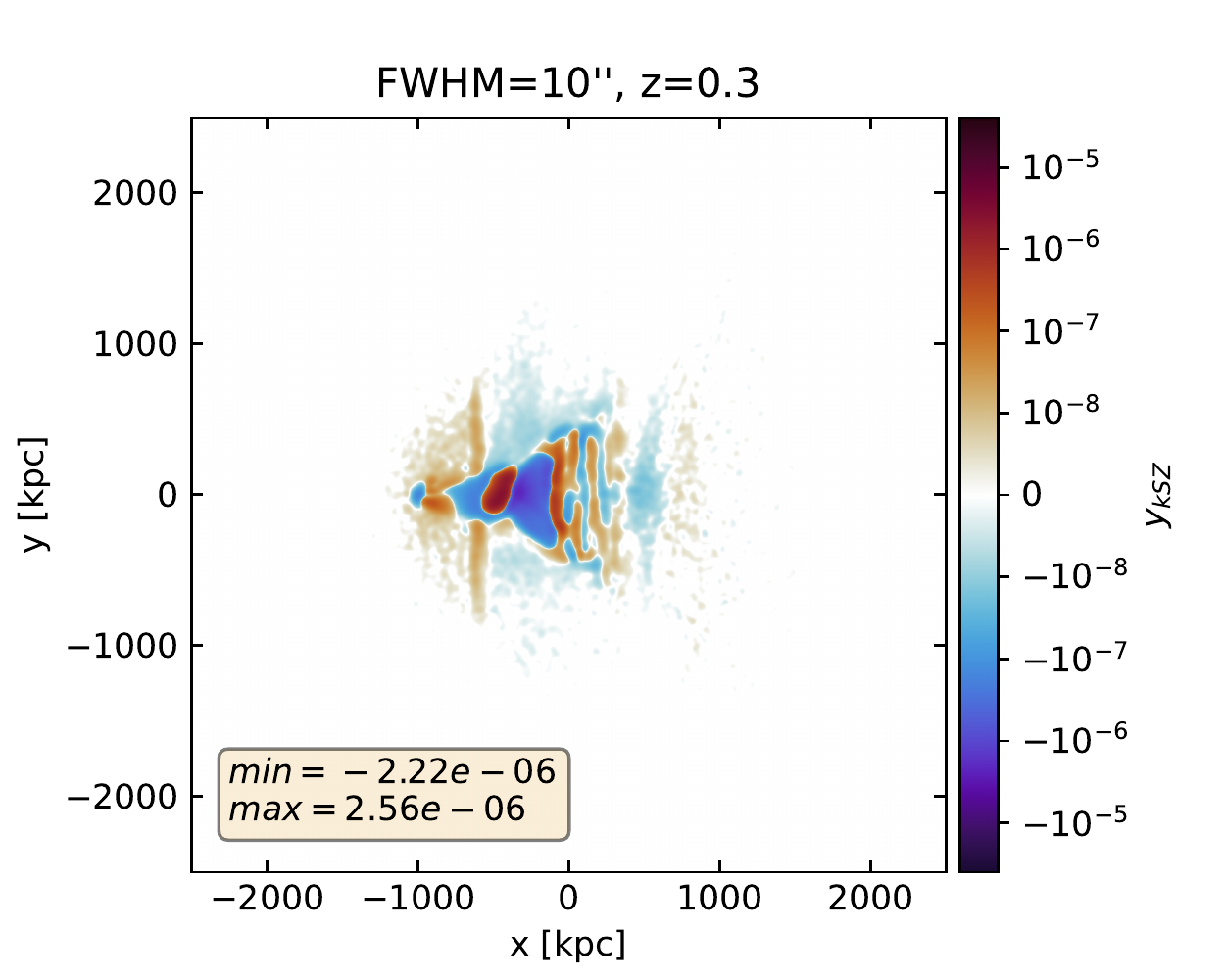}
    \includegraphics[width=0.32\textwidth,trim=0 0 10 0,clip]{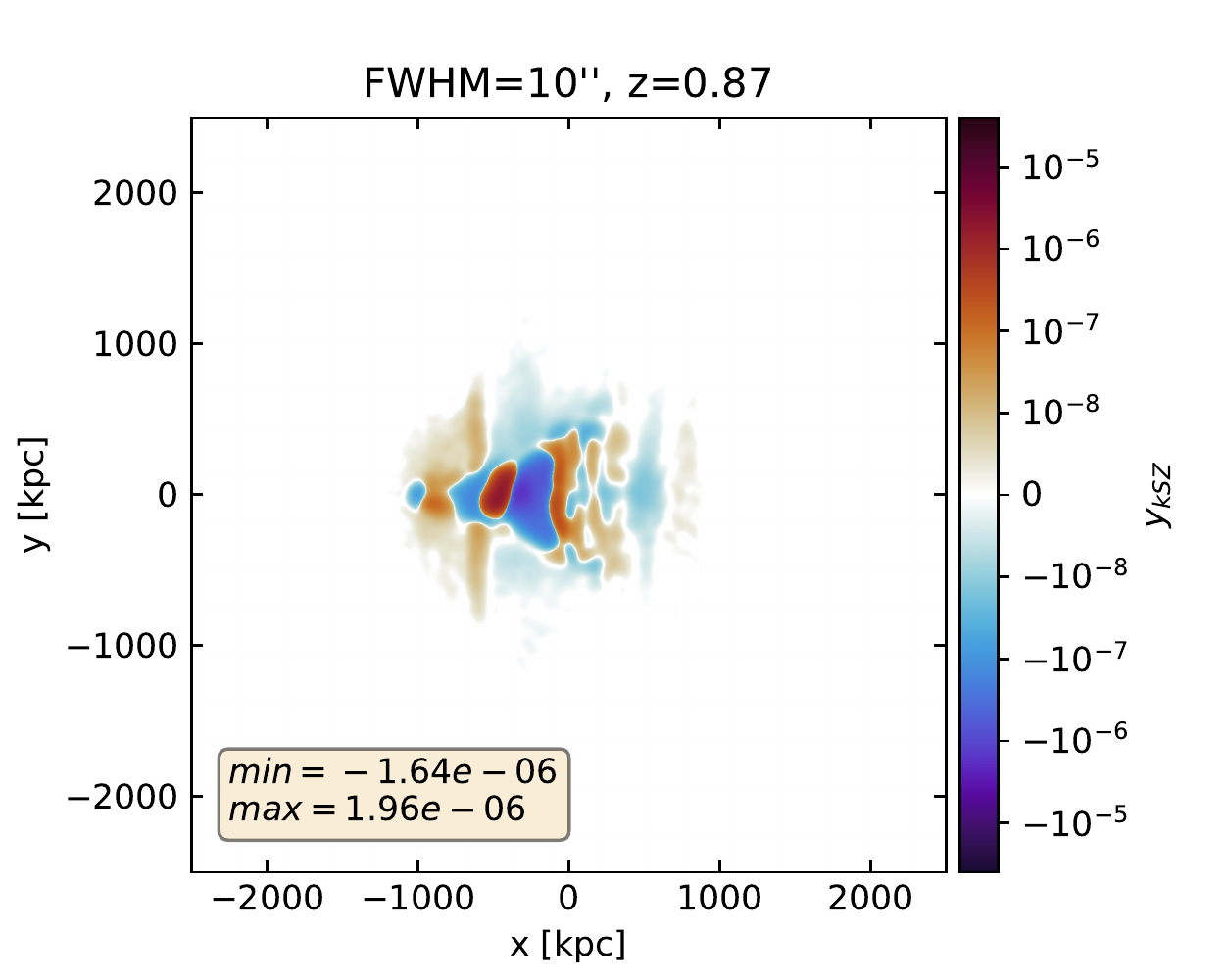}
    \\
    \includegraphics[width=0.32\textwidth,trim=0 0 10 0,clip]{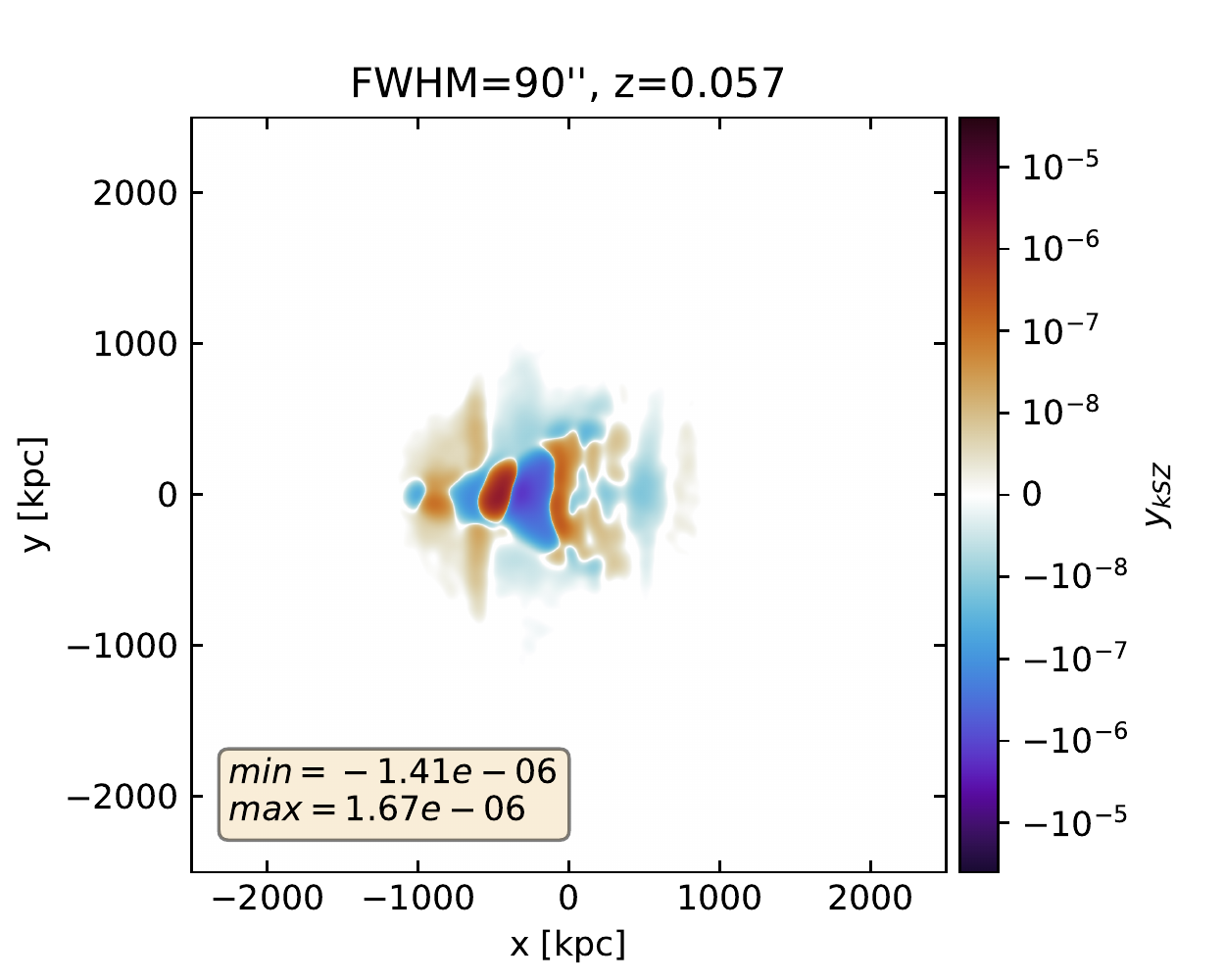}
    \includegraphics[width=0.32\textwidth,trim=0 0 10 0,clip]{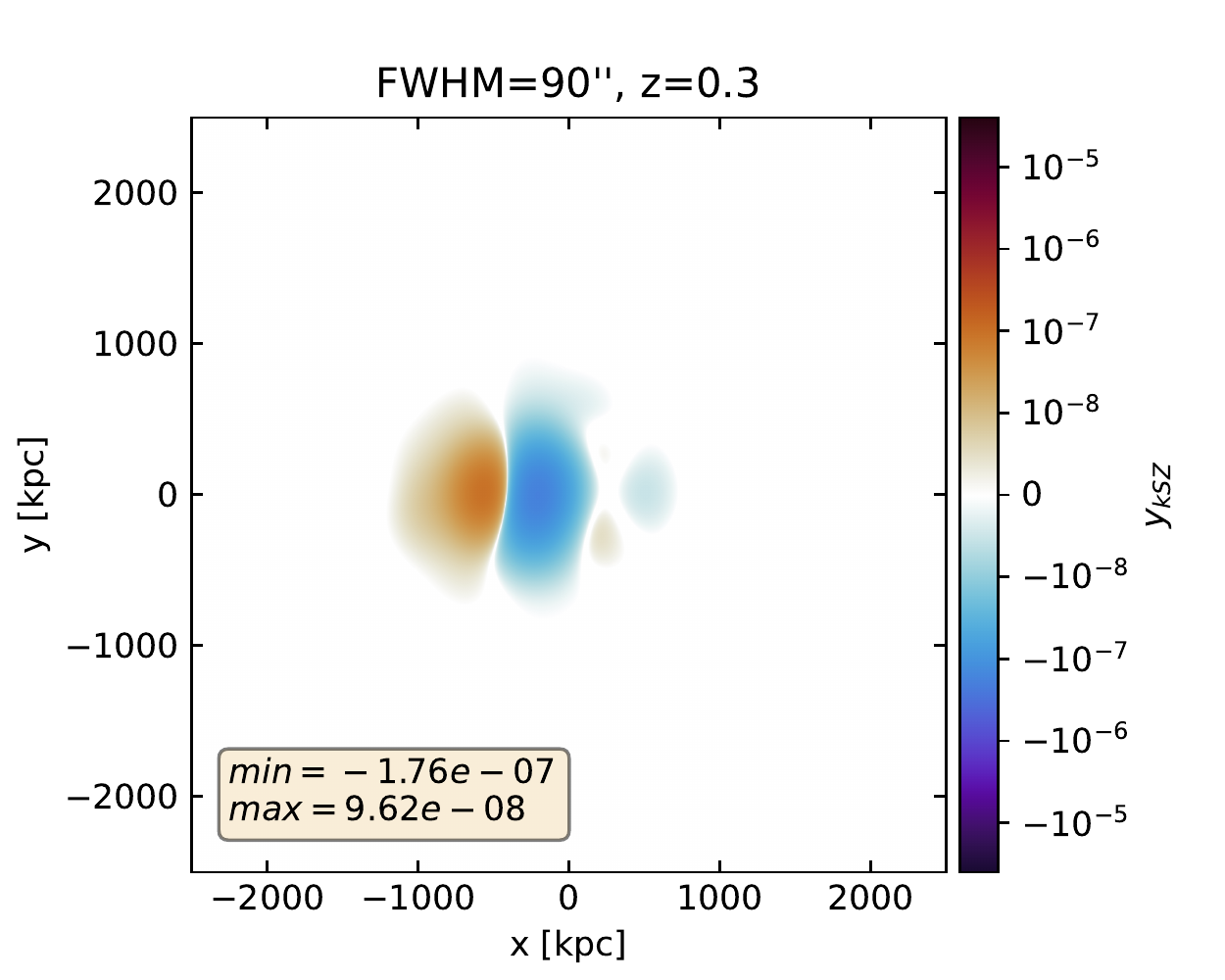}
    \includegraphics[width=0.32\textwidth,trim=0 0 10 0,clip]{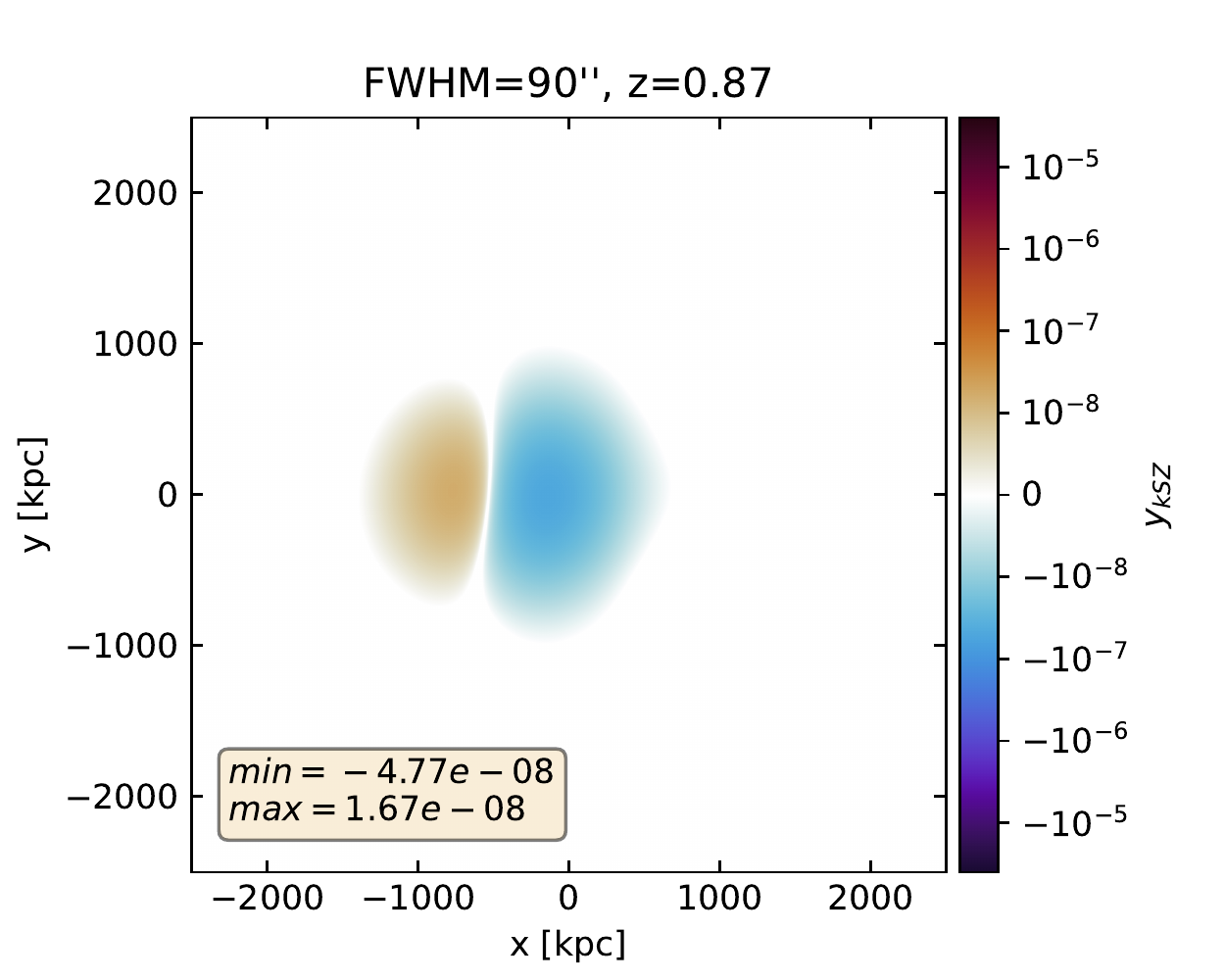}
    \\
    \includegraphics[width=0.32\textwidth,trim=0 0 10 0,clip]{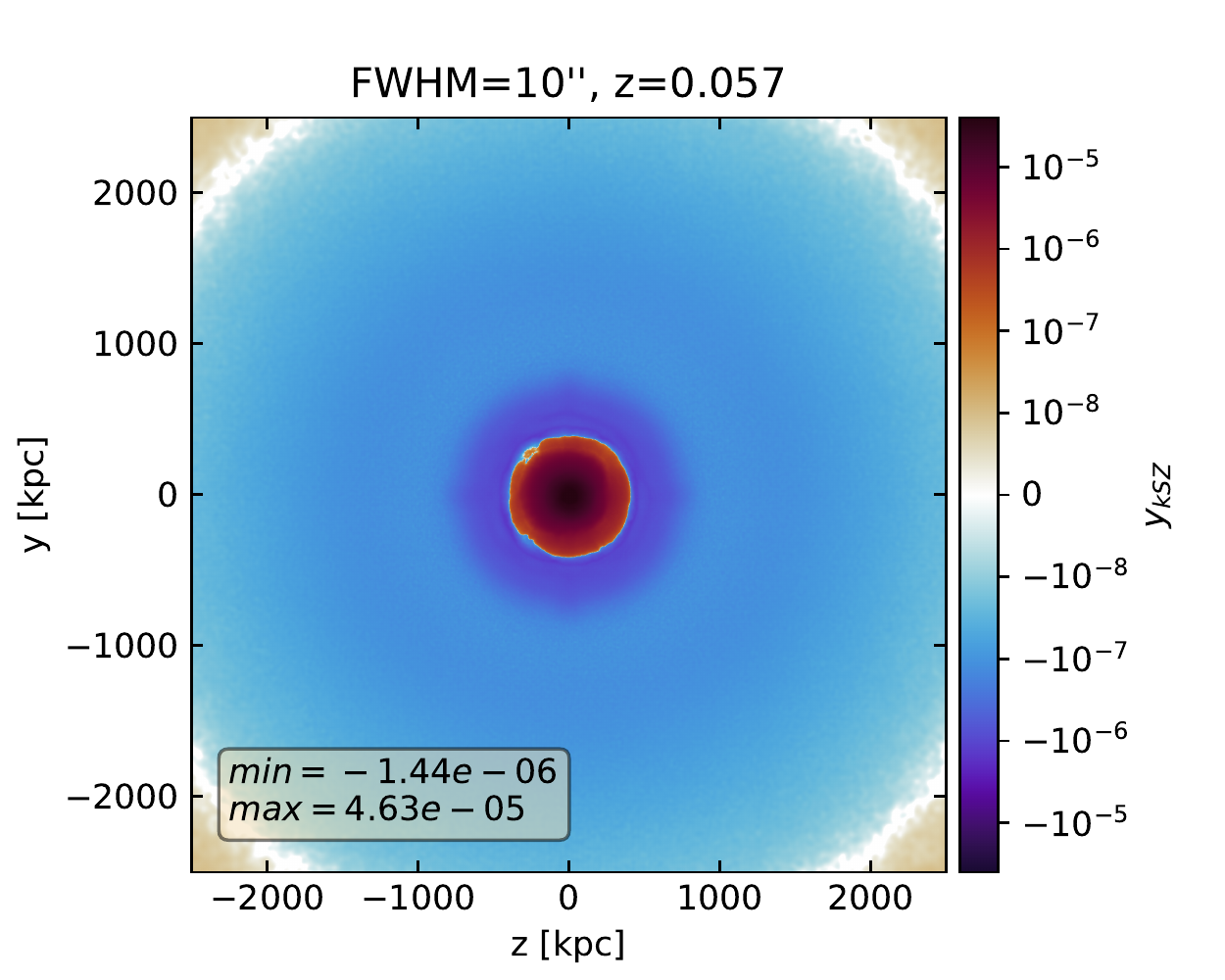}
    \includegraphics[width=0.32\textwidth,trim=0 0 10 0,clip]{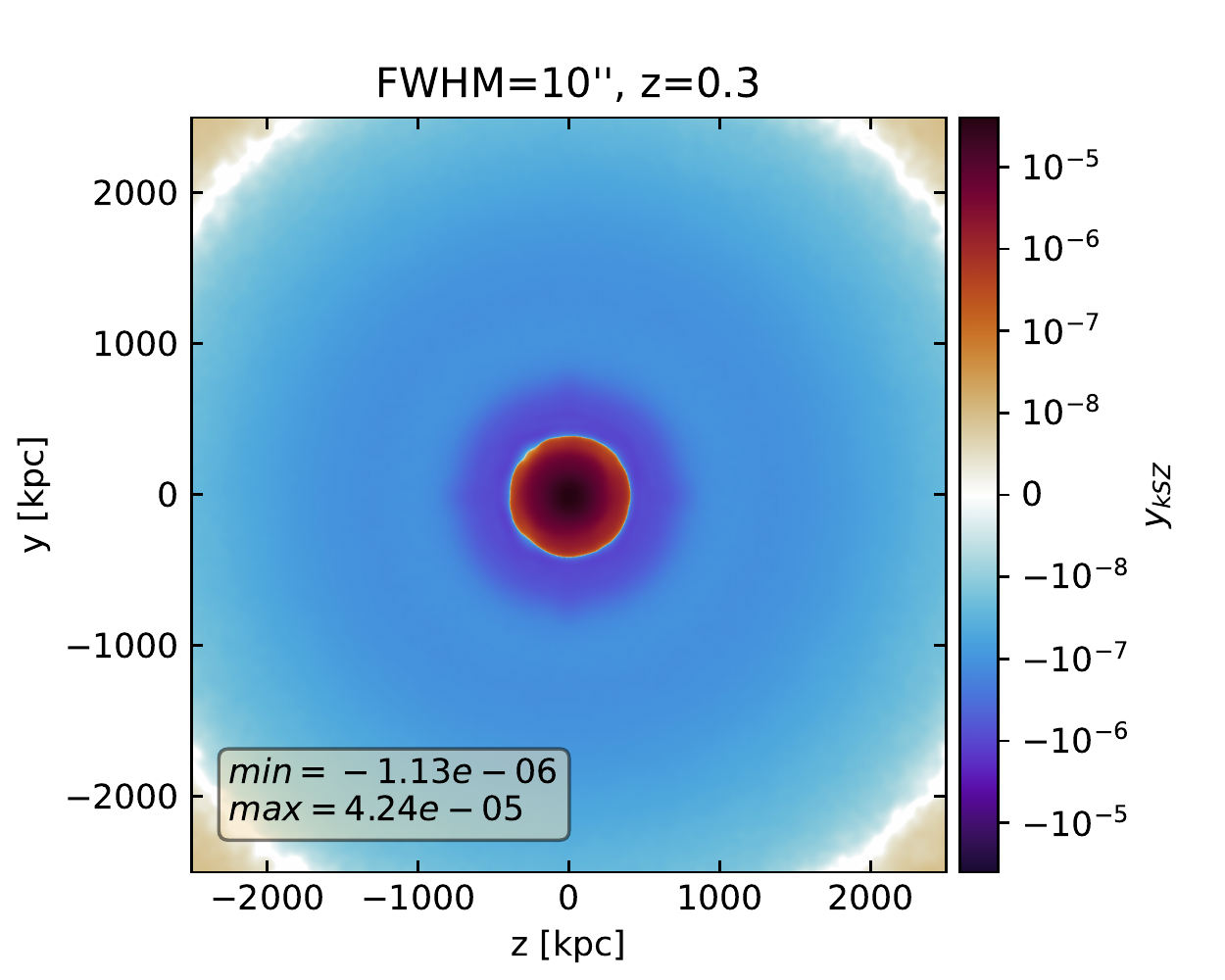}
    \includegraphics[width=0.32\textwidth,trim=0 0 10 0,clip]{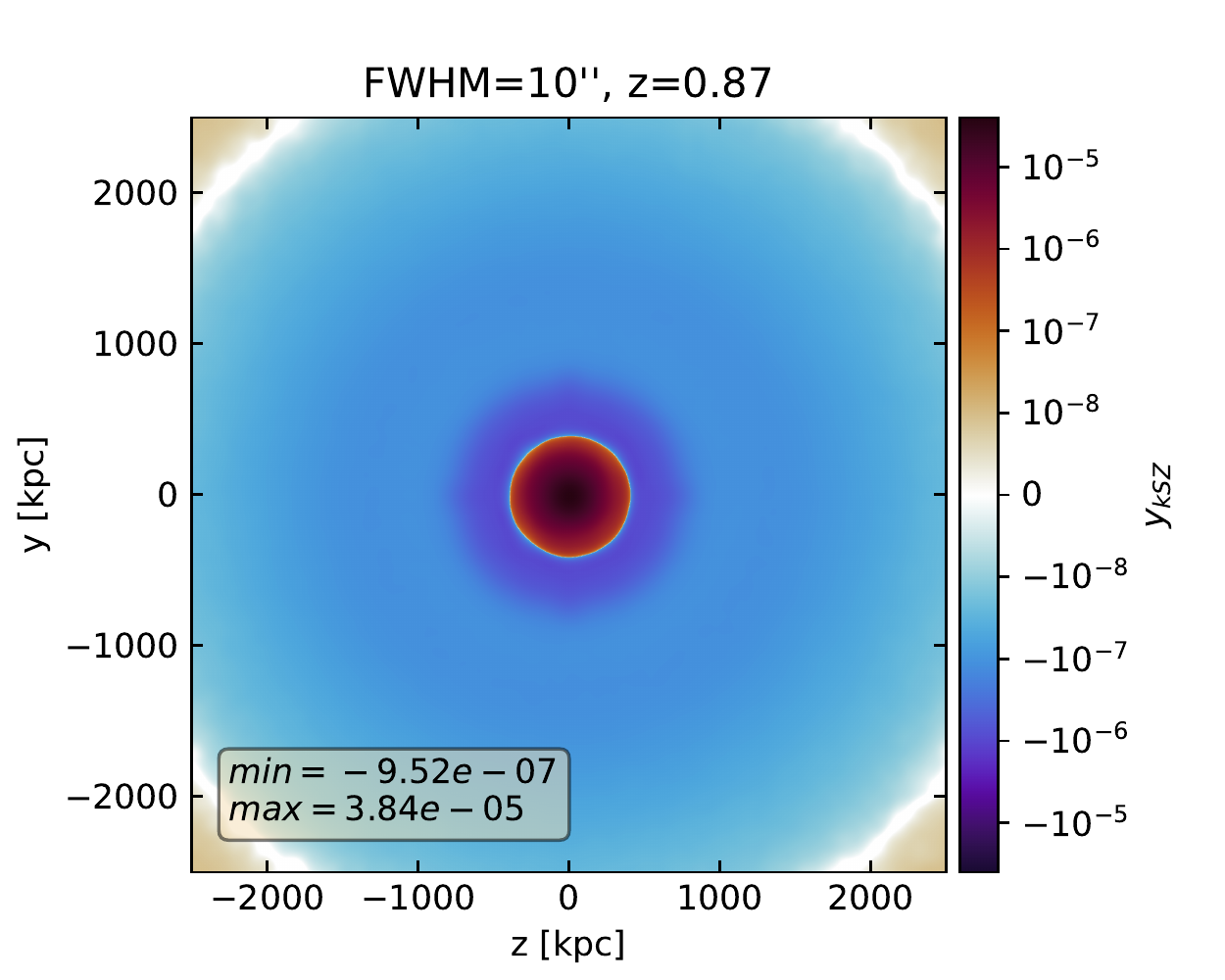}
    \\
    \includegraphics[width=0.32\textwidth,trim=0 0 10 0,clip]{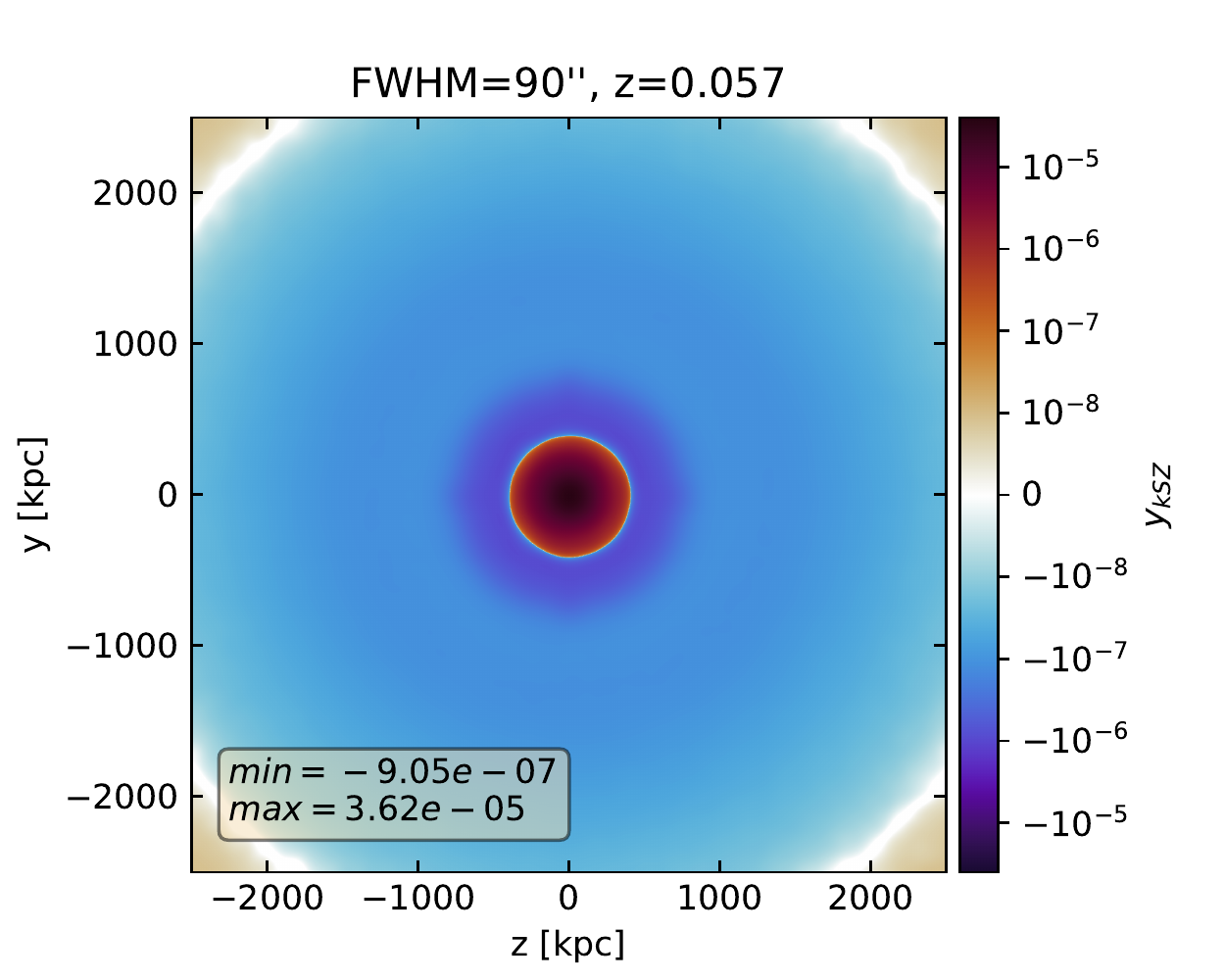}
    \includegraphics[width=0.32\textwidth,trim=0 0 10 0,clip]{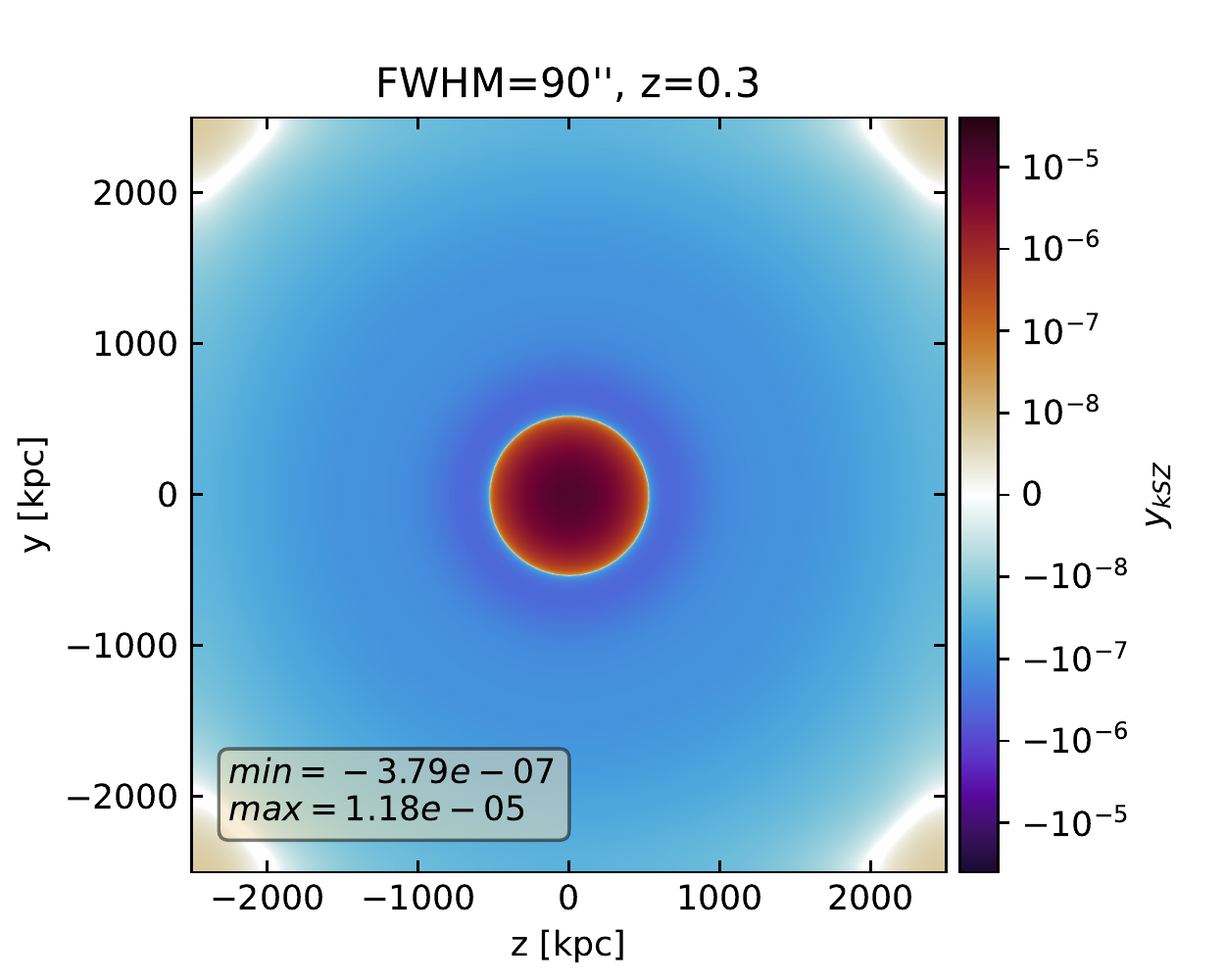}
    \includegraphics[width=0.32\textwidth,trim=0 0 10 0,clip]{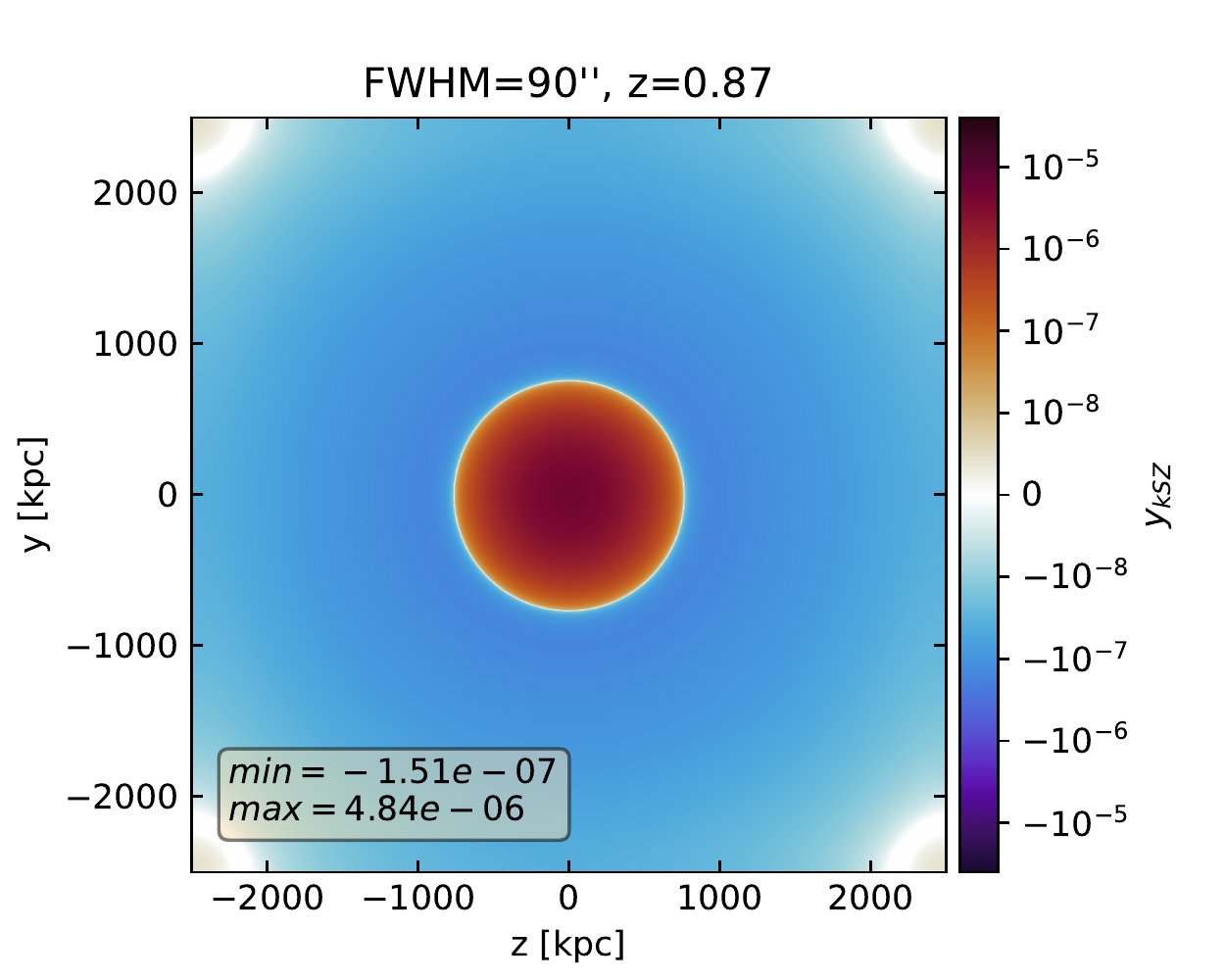}
    \caption{Noise free, smoothed SZ images of the plane-of-sky (upper 6 panels) and line-of-sight geometries (lower 6 panels), corresponding to the upper and lower panels of Figure \ref{fig:tsz}.
    The first and third rows show the images for an AtLAST-like resolution at 150~GHz, while the second and third rows show the images for a CMB-S4-like resolution.  The columns correspond to the representative redshifts of $z=[0.057, 0.3, 0.87]$, chosen to correspond to the mergers in e.g. A3667, the Bullet Cluster, and El Gordo.  
    We note that features at levels $|y_{kSZ}| < 10^{-6}$ will likely be masked by the primary CMB on arcminute scales.}
    \label{fig:ksz_smooth}
\end{figure*}

We then smooth the maps by 90\arcsec, representative of the resolutions of current and near future SZ surveys with e.g. Advanced ACTpol, Simons Observatory, and CMB-S4~\citep[see][]{Henderson2016,CMB-S4_2019,SimonsObs2019}, and by 10\arcsec, representative of larger single dish facilities such as AtLAST, the Green Bank Telescope, the Sardinia Radio Telescope, and the Large Millimeter Telescope~\citep[see respectively][]{Klaassen2020,Dicker2014,Hughes2020}, as shown in Figures~\ref{fig:tsz_smooth} and~\ref{fig:ksz_smooth} respectively.  For both resolutions, we choose representative redshifts of $z=0.057$, $z=0.3$, and $z=0.87$, corresponding approximately to the redshifts of the famous massive merging systems A3667, the Bullet cluster, and El Gordo.  

It can be seen from the Compton-$y$ images and values of the peak signal that 90\arcsec\ resolution may not present a critical challenge for tSZ measurements of plane-of-sky mergers at low redshift, where the peak in $y$ is suppressed by $9\%$ at $z=0.057$.  However, by $z=0.3$ the peak features are suppressed by $35\%$, and by $58\%$ at $z=0.87$, which may present a severe limitation.  
For systems merging along the l.o.s., the suppression of the peak signal is also similar, albeit slightly larger.
By contrast, the $\sim$10\arcsec\ or better resolution afforded by larger aperture telescopes shows a much smaller suppression of the peak Compton $y$,
which never exceeds $11\%$ in all cases explored, leaving the exciting possibility that next generation high-resolution millimetric observatories may be able to measure it above the primary CMB.
We note however that these simulations do not include turbulent subgrid physics akin to that in e.g.\ \cite{Bennett2020}, where even 10\arcsec\ resolution could limit the observational sensitivity to pressure fluctuations on the smallest scales.

For the kSZ signal, the opposite signs of the gas momentum (c.f.\ Figures~\ref{fig:map} \&~\ref{fig:map.x}) exacerbate the situation.  Figure~\ref{fig:ksz_smooth} shows the signal can be suppressed by over 2 orders of magnitude at high redshift when observed at 90\arcsec\ resolution, while smoothing to 10\arcsec\ resolution still suppresses the signal by $17\%$, $36\%$, and $43\%$ respectively at $z=0.057,~0.3$, and $0.87$.  Recovering the signal at higher resolutions would likely require interferometers like ALMA, though as noted in e.g.\ \cite{mroczkowski2019}, such observations generally require complementary information from instruments on large aperture single dish telescopes to provide constraints on scales larger than $\sim$30\arcsec\ near the null in the tSZ.
Future work will explore the ability to constrain the merger inclination angle using high resolution measurements of the SZ effects.

\section{Summary and conclusions}\label{sec:concl}

In this work, we explored the possibility to constrain the ICM velocity field in a major cluster merger by means of X-ray microcalorimeter observations, anticipating the expected performance of \xrism Resolve and \athena X-IFU.
Especially for complicated structures of the velocity field along the l.o.s.\ due to extreme geometrical configurations, X-ray spectra with an energy resolution of a few eV from upcoming microcalorimeters will be essential.
We analyse the three-dimensional thermal and velocity structure of the ICM in a hydrodynamical simulation of a major merger between two clusters, with zero impact parameter and a mass ratio of 1:3. We focus on a snapshot in time soon after the first core passage.
We also provide predictions on the SZ view of these same merger geometries.

Our main findings can be summarized as follows:

\begin{itemize}

    \item From the direct inspection of the simulated system we find significant non-thermal velocity dispersion at various locations, reaching several hundreds km/s, both in the projection perpendicular to the merger direction and along the l.o.s.\ aligned with it. 
    
    \item By inspecting the simulated l.o.s.\ velocity distribution, we notice that large velocity dispersion values can often be ascribed to multiple velocity components with large velocity differences, in line with observational constraints from redshift mapping of merging clusters~\cite[see][]{liu2016}. Thus, the detailed line shape is expected to be manifestly non-Gaussian in such cases.

    \item While substantial net l.o.s.\ velocities of order $\sim 1000$\,km/s develop when the merger direction is aligned with the l.o.s., the plane-of-sky merger shows negligible shifts of the emission line centers, corresponding to no net bulk l.o.s.\ motions~\cite[][]{sayers2019}. This is consistent with estimates for the Bullet Cluster obtained by~\cite{liu2015}. 
    
    We have shown that negligible values of the net l.o.s.\ velocities, can be due to a symmetrical, albeit broadened, distribution of the l.o.s.\ gas velocities or to a multi-component distribution with roughly opposite line shifts, which nevertheless does not imply an intrinsically low level of non-thermal or bulk motions.
    This finding suggests caution in reconstructing gas motions from the Doppler shift of emission lines, which can misleadingly under-estimate the amount of non-thermal motion in plane-of-sky merging systems.

    \item \xrism Resolve observations of bright regions can successfully constrain the velocity dispersion from the iron K$\alpha$ complexes in the 6--7\,keV band with 100\,ks exposures. 
    Some discrepancy with respect to the expected value derived from the direct simulation analysis can be ascribed to the non-Gaussian shape of the iron line, due to the complexity of the velocity distribution along the l.o.s.\ 
    For the l.o.s.\ merger, the two components associated with the two system cores can be reconstructed (both shift and broadening) from the double-component spectral fitting of the 6--7\,keV band, assuming a typical $100$\,ks observation of the central bright region.

    \item The velocity broadening in fainter regions, e.g. beyond the shock edge of the plane-of-sky merger, cannot be successfully constrained from \xrism spectra, despite the large values, due to poor photon statistics.

    \item The larger effective area of \athena X-IFU allows for consistent estimates of ICM velocities compared to the simulation values in relatively faint regions. This result is obtained using a reasonable 100\,ks exposure, and accounting for instrumental and physical background as well. A detailed reconstruction of the two symmetrical velocity components by means of the superposition of two Gaussian terms was not possible, due to the poor photon statistics.

    \item A velocity dispersion gradient downstream of the shock is developed in the plane-of-sky projection of the merger. By analysing \athena X-IFU spectra extracted from five $20''$-wide annuli across the shock, we reconstruct temperature and velocity dispersion values that are roughly consistent with the estimates computed directly from the gas-cell distributions. While the temperature gradient is indeed recovered, the moderate velocity dispersion gradient cannot be well constrained, given the large uncertainties on the best-fit values of velocity broadening. The expected velocity dispersion values are indeed relatively low, decreasing from $\sim 400$\,km/s to $\sim 200$\,km/s from the innermost annulus to outermost one.

    \item From the inspection of the expected SZ thermal and kinetic signals derived for the resolution and redshift ($z=0.057$) of our simulation, we find that the strength of tSZ is one and two orders of magnitude greater than the strength of kSZ, in the l.o.s.\ case and plane-of-sky merger respectively.
    
    \item By assuming typical resolutions of current surveys ($90''$) and large single dish facilities ($10''$) at low, intermediate and high redshifts, we note that the suppression of thermal SZ signal is not as severe as in the case of kSZ, especially at high resolution. For plane-of-sky mergers at low redshifts, measurements of the thermal SZ can provide informative constraints also at low resolution. More severe limitations are expected in the case of the kSZ signal, for which high resolution facilities like AtLAST are definitely required to constrain the ICM velocity features even in the case of mergers aligned along the line of sight.
    
\end{itemize}

We have shown that ICM motions, developing after a major merging event, can be in many cases well constrained with X-ray microcalorimeter observations. 
While \xrism can be safely employed in bright cluster regions, fainter ones --- e.g. post-shock pointings --- require larger effective areas, like those of the \athena X-IFU, or exposure times larger than $1\,$Ms.

The cases discussed in this work present some similarities with well-known observed systems, such as the disturbed clusters Abell~754 or Abell~3667. In these cases, the shock front, in the plane-of-sky merger configuration, constitutes a real challenge to constrain gas motions in the shock region with \xrism Resolve~\cite[consistent with predictions in][]{astroh2014}, especially for the moderate turbulent velocities expected there --- $\lesssim 300$\,km/s, from our simulation. Longer \xrism exposures on fainter off-set regions can only provide realistic constraints when the turbulent broadening is larger (e.g.\ of order of $400$--$500$\,km/s, for our z1 pointing).
Microcalorimeters like \athena X-IFU, with a larger effective area and smaller PSF, will greatly improve the reconstruction of both streaming and turbulent motions in fainter regions.

The most challenging aspect still remains the detailed reconstruction of the line shape, which is nevertheless crucial to distinguish between true turbulent broadening and superposition of l.o.s. streaming motions.
\cite{zuhone2016b}, focusing on \xrism-like mock observations of bright central cluster regions in a relatively relaxed cluster with sloshing motions, also found that large number counts ($10^3$ or more) in the iron complex band are required to fully model the multi-component line shape.
This is consistent with expectations obtained for observed systems in the preparatory studies for \textit{ASTRO-H}~\cite[][]{astroh2014}.

Even for the \athena X-IFU, \cite{Roncarelli:2018} showed that very large exposures, of order of several Ms, would be required to fully reconstruct the velocity structure function and velocity field, and to account for a full thermodynamical description of the ICM.
\cite{zhuravleva2013} constructed model spectra, assuming a large number of photons, to provide templates for the expected distortions of emission lines due to resonant scattering. In their theoretical study, they predicted the possibility to constrain such signatures with 100\,ks \textit{ASTRO-H} (\xrism-like) observations of the Perseus cluster core, considering a Gaussian shape of the lines~\cite[see also predictions from][]{zuhone2016a}. \cite{Cucchetti:2019} also investigated the possibility to statistically map the ICM velocity field through studies of the structure function from simulated \athena X-IFU observations with typical exposures of 100\,ks. They also extracted the spectra from binned regions to reach high signal to noise ratios (with $\sim 40\,000$ counts).

Despite the idealized set-up of our simulated merger, we showed that the low number counts in the iron 6--7\,keV band, combined with a complex thermal gas structure along the l.o.s., can prevent a detailed reconstruction of complicated line shapes, for moderate exposures.
A successful multi-Gaussian modelling, similarly to that proposed by~\cite{shang2012}, requires very large photon counts ($\gtrsim 1000$ in the Fe-K band). 
As we discussed, this can be definitely achieved with a single $100$\,ks \athena X-IFU observation for the bright central region of a l.o.s.\ merger (see Sect.~\ref{sec:x-proj}), where the multiple velocity structures can be well constrained.

The possibility to combine observations with X-ray calorimeters and SZ data can further provide valuable insights on the ICM velocity field in merging systems. Current and upcoming SZ surveys, despite their lower resolution, can in fact successfully probe mergers at low redshift, aiding in the measure of shocks in the plane of the sky merger case. Measurements of the kinetic SZ signal, directly related to the ICM velocity field along the l.o.s., are definitely more challenging to obtain. Nonetheless, l.o.s.\ cluster mergers can develop significant gas velocities, that will be eventually detectable above the primary CMB component with next generation high-resolution millimetric telescopes like AtLAST.

\begin{acknowledgements}
The authors would like to thank the anonymous referee and the Editor for providing constructive comments to improve the presentation of the results.
VB is particularly thankful to E.~Cucchetti, the SIXTE support Team and V.~Ghirardini for helpful advice on the set up of SIXTE and \athena X-IFU background.
VB acknowledges support from the Smithsonian Institution Scholarly Studies award "Probing the Velocity of the Hot and Dark Universe with Simulations of Galaxy Cluster Mergers".
This research was also partially supported by the Deutsche Forschungsgemeinschaft (DFG, German Research Foundation) --- 415510302, and by the Excellence Cluster ORIGINS which is funded by the DFG under Germany's Excellence Strategy – EXC-2094 – 390783311.
JAZ acknowledges support from the Chandra X-ray Center, which is operated by the Smithsonian Astrophysical Observatory for and on behalf of NASA under contract NAS8-03060.
WF acknowledges support from the Smithsonian Institution, the Chandra High Resolution Camera Project through NASA contract NAS8-03060, and NASA Grants 80NSSC19K0116, GO1-22132X, and GO9-20109X.
The simulations used in this work have been performed on the Pleiades supercomputer (NASA/Ames Research Center).
\end{acknowledgements}

\bibliographystyle{aa}
\bibliography{bibl.bib}

\end{document}